\documentclass[12pt, a4paper]{article}
\pdfoutput=1
\usepackage[utf8]{inputenc}
\usepackage{fancybox}
\usepackage[T1]{fontenc}
\usepackage{lmodern, textcomp}
\usepackage{bm}
\usepackage[american]{babel}
\usepackage{csquotes}

\usepackage{eurosym}
\usepackage[nottoc]{tocbibind}
\usepackage{authblk}
\usepackage{fancyhdr}
\usepackage{xspace}

\usepackage{graphicx}
\usepackage{subcaption}
\usepackage{float}
\usepackage{makecell}
\usepackage{multirow}
\usepackage{multicol} 
\usepackage{marginnote}
\usepackage[multiple]{footmisc}

\usepackage[usenames, dvipsnames, svgnames, table]{xcolor}

\usepackage[backend=bibtex8, citestyle=numeric-comp, maxbibnames=99,
			sorting=none, sortcase=false, sortcites=true, giveninits=true]{biblatex}

\usepackage{amsmath, amsfonts, amssymb}
\usepackage{mathtools}
\usepackage{mathrsfs}
\usepackage{cancel}
\usepackage{braket}
\usepackage{tensor}
\usepackage{slashed}
\usepackage{siunitx}
\usepackage{listings}
\usepackage{stmaryrd}
\usepackage{caption}
\usepackage{relsize,exscale}
\usepackage{array}
\usepackage[toc,page]{appendix}

\usepackage{enumerate}

\DeclareSymbolFont{largesymbols}{OMX}{cmex}{m}{n}

\setcellgapes{1pt}
\makegapedcells
\newcolumntype{R}[1]{>{\raggedleft\arraybackslash }b{#1}}
\newcolumntype{L}[1]{>{\raggedright\arraybackslash }b{#1}}
\newcolumntype{C}[1]{>{\centering\arraybackslash }b{#1}}

\newcommand{\Tr}{\mathrm{Tr}}

\newtheorem{remark}{Remark}

\newcommand{\beq}{\begin{equation}}
\newcommand{\eeq}{\end{equation}}
\newcommand{\bea}{\begin{eqnarray}}
\newcommand{\eea}{\end{eqnarray}}
\definecolor{mygray}{gray}{0.3}

\newcommand{\bes}{\begin{eqnarray}}
\newcommand{\ees}{\end{eqnarray}}

\newcommand\restr[2]{{% we make the whole thing an ordinary symbol
  \left.\kern-\nulldelimiterspace % automatically resize the bar with \right
  #1 % the function
  \vphantom{\big|} % pretend it's a little taller at normal size
  \right|_{#2} % this is the delimiter
  }}

\usepackage{multicol}
\usepackage{amssymb}

\usepackage[heightrounded, top=3.5cm, bottom=3.5cm, left=3cm, right=3cm, headheight=14pt]{geometry}

\usepackage[pdftex, breaklinks=true, linktocpage,
	colorlinks=true, urlcolor=blue, linkcolor=blue, citecolor=red
]{hyperref}

\newcommand{\email}[1]{\href{mailto:#1}{\nolinkurl{#1}}}
\newcommand{\emailfoot}[1]{\thanks{\email{#1}}}

\usepackage{marginnote}
\newcounter{draftcommentcnt}
\NewDocumentCommand{\draftcomment}{s O{red} m}{%
	\def\margnote{\IfBooleanTF{#1}{\marginnote}{\marginpar}}%
	\stepcounter{draftcommentcnt}%
	\textcolor{#2}{#3}%
	\margnote{\textcolor{#2}{$\Leftarrow$ \arabic{draftcommentcnt}}}%
}

\fancypagestyle{preprint}{
	\fancyhf{}

	\fancyhead[R]{\textsf{\preprint}}
}

\numberwithin{equation}{section}

\hypersetup{
	pdftitle={Functional renormalization group for multilinear
disordered Langevin dynamics II},
}

\title{Functional renormalization group for multilinear
disordered Langevin dynamics II:\\
\begin{Large}
Revisiting the $p=2$ spin dynamics for Wigner and Wishart ensembles
\end{Large}}

\author[1]{Vincent Lahoche\emailfoot{vincent.lahoche@cea.fr}}

\author[1,2]{Dine Ousmane Samary\emailfoot{dine.ousmanesamary@cipma.uac.bj}}

\author[1]{Mohamed Tamaazousti\emailfoot{mohamed.tamaazousti@cea.fr}}

\affil[1]{%
	Université Paris Saclay, \textsc{Cea}, \textsc{List}, Gif-sur-Yvette, F-91191, France
}

\affil[2]{%
	Faculté des Sciences et Techniques (ICMPA-UNESCO Chair)
	\protect\\
	Université d'Abomey-Calavi, 072 BP 50, Bénin
}

\begin{document}
\maketitle

\begin{abstract}
In this paper, we investigate the large-time behavior for a slightly modified version of the standard $p=2$ soft spins dynamics model, including a quartic or higher potential. The equilibrium states of such a model correspond to an effective field theory, which has been recently considered as a novel paradigm for signal detection in data science based on the renormalization group argument. We consider a Langevin-like equation, including a disorder term that leaves in the Wigner or Wishart ensemble. Then we construct a nonperturbative renormalization group formalism valid in the large $N$ limit, where eigenvalues distributions for the disorder can be replaced by their analytic limits, namely the Wigner and Marchenko–Pastur laws. One of the main advantages of this approach is that the interactions remain local in time, avoiding the non-locality arising from the approaches that integrate out the disorder at the partition function level. 
\end{abstract}

\newpage

\hrule
\pdfbookmark[1]{\contentsname}{toc}
\tableofcontents
\bigskip
\hrule

\newpage

\section{Introduction}
%[arXiv:2005.05118,0905.4878,0510731,2208.07309,2001.07682]
Dynamical aspects of glassy systems are essentially characterized by their out-of-equilibrium properties, and especially regarding the so-called “aging” phenomenon in the non-ergodic phase. In the vicinity of the glassy transition temperature, the relaxation time increases until it diverges, and the correlation function fails to have an exponential decay \cite{Cugliandolo1,Cugliandolo2,Cugliandolo3,Castellani1,Bouchaud1}. In the static limit, the glassy systems are essentially characterized by the existence of a \textit{replica-symmetry breaking} (RSB) for the overlap matrix $Q_{ab}$ below the critical temperature $T_c$. The RSB arises as we consider $n$ copies of the system and when we average over the disorder effect before taking the $n\to 0$ limit for replica number. At the critical temperature, the replica symmetric solution for the free energy exhibits instabilities along some eigen-directions at the saddle point and predicts negative entropy states. This bad behavior is generally improved by ansatz, which explicitly breaks the replica symmetry. The entries $Q_{ab}$ quantify the average correlation between spins, namely $Q_{ab}:=\frac{1}{N} \sum_{i=1}^N \langle S_i^{a}S_i^{b} \rangle$, where $S_i^{a}$ denotes the spin of site $1\leq i \leq N$ for the replica number $0\leq a \leq n$ \cite{Castellani1,Sherrington,Nishimori,Contuccibook,mezard1984nature,derrida1980random}. One of the most popular theoretical models for glassy systems is the well-known $p$-spin model, whose disorder effect is quantified by a random tensor of rank $p$, $J_{i_1i_2\cdots i_p}$ that couple spins with Hamiltonian $H_J[S]:=\sum_{i_1,\cdots i_p} J_{i_1i_2\cdots i_p} S_{i_1}\cdots S_{i_p}$. Depending on the model and the nature of the spins, this Hamiltonian can be completed by a deterministic component $V[S]$ regarding some symmetries. For soft spins models, $S_i \in \mathbb{R}$, and the Hamiltonian $H_j$ is usually constraint by the spherical condition $\sum_i S_i^2=N$ defining the so-called spherical $p$-spin model. Alternatively, the glassy transition can be investigated from a dynamic point of view regarding the apparition of metastable states with infinite lifetimes. Due to these metastable states, ergodicity must be broken, and equilibrium can never be reached for a temperature small enough. In the equilibrium dynamics setting, ergodicity is expected at first, and the phase transition looks like an ergodicity breaking, where the large-time two-point correlation function fails to vanish. Formally, the dynamics are described by a Langevin-like equation analogous to a coarse-grained Glauber dynamics for Ising spins, and for the spherical model reads \cite{DeDominicisbook,BenArous2,Sommers1,Kim2001}:
\begin{equation}
\frac{d q_i}{dt}=- \frac{\partial}{\partial q_i} \left(\frac{1}{p}\sum_{i_1,\cdots,i_p} J_{i_1 i_2\cdots i_p}\, q_{i_1}(t)\cdots q_{i_p}(t)\right)-\ell(t) q_i(t) +\eta_i(t)\,,
\end{equation}
where $\ell(t)$ is a Lagrange multiplier ensuring the spherical constraint and $\eta$ some white noise. In this paper, we focus on the $p=2$ dynamical spin model. Still, with a non-trivial interacting potential instead of a Lagrange multiplier, that suppresses configurations with large values of the soft spins $q_i$. The $p=2$ soft spin dynamics which has the characteristic to be investigated and fully understood analytically \cite{Cugliandolo2,Cugliandolo3,Caiazzo1}, including the static limit, despite the randomness of the disorder coupling $J_{ij}$. This is essentially a consequence of the large $N$ properties of the random matrix spectra, allowing to replacement of the randomness of the matrix $J$ with deterministic density spectra as the Wigner law. In particular, solving the static limit does not require a replica method, and using it may allow showing that no replica symmetry breaking is expected. Hence, the model has no true spin-glass phase but is rather a ferromagnet in disguise \cite{van2010second}, exhibiting a second-order phase transition at finite temperature $T_c$ where the component $q_\lambda$ for the largest eigenvalue $\lambda_*$ has a macroscopic occupation number $q_{\lambda_*}^2 \propto N$. Surprisingly, however, the dynamic aspects of the model have many interesting features close to what we expect for a true spin-glass and capture some relevant experimental scenarios. In particular, it exhibits aging effect and weak ergodicity breaking that almost characterize true glassy transitions. To put it in a nutshell, the spherical $p=2$ spin dynamics never reaches equilibrium, except for very particular initial conditions called ‘‘staggered states" \cite{DeDominicisbook,Cugliandolo3}. Interestingly, the behavior of the system is reminiscent of the behavior of a ferromagnet in $\mathbb{R}^d$, described by the $N$-component field $\phi_a(\vec{x},t)$ which obeys the Langevin equation,
\begin{equation}
\frac{d\phi_a(\vec x,t)}{dt}=\Delta \phi_a(\vec x,t)+\phi_a(\vec x,t) \left((T_c-T)-\frac{g}{2N} \vec{\phi}\,^2 \right)+\eta_a(t)\,, \label{domco}
\end{equation}
where $\vec{x}\in\mathbb{R}^d$, $a\in \llbracket1,N \rrbracket$, $\Delta$ is the standard Laplacian over $\mathbb{R}^d$ and $\vec \phi=(\phi_1,\cdots,\phi_N)$. This system exhibits for large $N$ a phenomenon called \textit{coarsening} \cite{DeDominicisbook,bray2002theory,newman1990dynamic} and domains with positive and negative magnetization increase their size without thermalization (the correlation length behaves as $\sqrt{t}$ for the quartic model), except if the initial condition is of staggered type, with a macroscopic occupation number in the ground state of the potential. The correspondence between the two models is obvious for $d=3$, where the energy states distribution $\rho(p^2) \sim (p^2)^{\frac{d-2}{2}}$ of the ferromagnet behaves as the density states of the Wigner distribution for large eigenvalues. 
\medskip

The purpose of this paper is to address the issue of the $p=2$ spin dynamics with the so-called \textit{functional renormalization group} (FRG) formalism \cite{Dupuis_2021,Wett1,Wett2,MORRIS_1994,Morris_19942}, including especially the case of non-trivial confining potential. We consider the  Wetterich-Morris incarnation of the FRG that focus on the effective average action $\Gamma_k$, which smoothly interpolates between ultraviolet scales (large $k$) and infrared scales (small $k$). The Wetterich-Morris approach has many advantages, among them the ability to deal with theory having strong couplings, and will be of great interest to address the issue of confining potentials with arbitrary shapes, not achievable with perturbative approaches of RG. Furthermore, for polynomial potential, and especially for the quartic case, we recover the analytical statement that the system fails to reach equilibrium for generic initial conditions, and the equilibrium flow diverges at a finite timescale for all the RG trajectories. Moreover, for the “staggered-like” initial vacuum, we recover that the system goes toward equilibrium. This paper follows the previous work \cite{Lahoche:2021tyc} of the authors, which investigated the frequency RG flow of the bi-local Lagrangian obtained after averaging over a disorder of rank $p$. Our originality in this work is that we perform a coarse-graining both in time and eigenvalues of the random disorder, whose spectra are assumed to converge toward an analytic law for $N$ large enough. One of our goals with this study is to propose a framework that can be used for example to address the issue of signal detection for nearly continuous spectra with an RG point of view, in the continuation of our works in \cite{lahoche2020field2,lahoche2021signal,lahoche2020field,lahoche2020generalized,LahocheSignal2022}, restricted to the equilibrium theory. In this paper, we follow the efforts of \cite{Lahoche:2021tyc} where FRG was considered as a powerful tool to address with a great computational efficiency some issues in spin glass dynamics and signal detection as for tensorial principal component analysis (PCA) \cite{BenArous1,BenArous2, ouerfelli2021selective, ouerfelli2022random}. The main purpose of this paper is then: Can we construct a reliable exact renormalization group flow solution that can be used to investigate such a kind of issue?
\medskip

Let us conclude this introduction with a general comment. We focus essentially on the equilibrium dynamics in this paper, hence we expect to reach the out-of-equilibrium transition from above, as $T\to T_c^+$. This restriction to equilibrium implies the validity of a number of relations, such as the fluctuation-dissipation theorem, which are consequences of time reversal invariance. Note that, the renormalization group has a certain ability to “see” beyond the limits imposed by the assumptions of its derivation. An example is provided by the ordinary $d<4$ theory, where the vertex development, which assumes that we are in the symmetric phase, reveals however the presence of a fixed point of negative mass as well as a region of the phase space where the zero vacuum is unstable \cite{defenu2015truncation}. Another non-trivial example is given in the context of two-dimensional quantum gravity. The renormalization group has indeed been considered to study phase transitions in discrete models of random geometry, such as matrices and random tensors \cite{zinn2014random,brezin1992renormalization,lahoche2020reliability,lahoche2020revisited,eichhorn2014towards,eichhorn2020universal}. In the case of matrices, for instance, it has been shown that, in the large $\textit{N}$ limit, the flow equations predict a fixed point whose critical exponents match the predictions of the well known \textit{double scaling limit} \cite{DiFrancesco1}, although the direct construction of this limit requires not only the leading sector but also all the sub-leading sectors in the neighborhood of the critical point.  Note that  we also addressed the general issue of $p=2$ soft spin dynamics for arbitrary confining potential in low temperature (i.e., below the critical temperature $T_c$) analytically, see \cite{Lahoche:2022lmf}.
\medskip

\textbf{Outline.} The paper is organized as follows. In section \ref{preliminaries} we present the model as well as useful definitions and the construction of the equilibrium path integral using Martin-Siggia-Rose (MSR) formalism, and we conclude with a short review of some basics about the dynamics of the spherical $p=2$ spin dynamics (more details should be found in Appendix \ref{appendix1} and references therein). We furthermore investigate the large $N$ behavior of the solutions using Schwinger-Dyson equations,  In section \ref{FRG} we present the FRG formalism, and in sections \ref{vertexexp} and \ref{LPAformalism} we consider vertex expansion and local potential approximation to solve the exact RG equation, and numerically investigate them. We conclude with a presentation of the two-particle irreducible (2PI) RG equations in section \ref{section2PI}, whose hierarchy closes in the large $N$ limit, and which will allow us to discuss ergodicity breaking from the point of view of the 2-point correlation function. Section \ref{conclusion} summarizes the results of the paper and provides some open issues that can be addressed in a future works. Furthermore, additional material, including analytical insights, should be found in the Appendices \ref{appendix1}--\ref{appendix5}. 
\pagebreak

\section{Preliminaries}\label{preliminaries}

\subsection{The model}

\paragraph{From Langevin equation to field theory.} We consider the $p=2$ soft spin model whose dynamics are described by a Langevin-like equation where the disorder is materialized by a random matrix $J_{ij}$ with eigenvalues $\{\lambda\}$ and eigenvectors $\{u_i^{(\lambda)}\}$. In the diagonal basis for a given sample, the Langevin dynamics equation reads:
\begin{equation}
\boxed{\frac{dq_\lambda(t)}{dt} =-\lambda q_\lambda(t)-V^\prime(Q^2)q_\lambda(t)+\eta_\lambda(t)\,,}\label{eqLangevin}
\end{equation}
where: 
\begin{enumerate}
\item The random field $\eta_\lambda(t)$ is Gaussian, centered, with Dirac delta correlations:
\begin{equation}
\langle \eta_\lambda(t) \eta_{\lambda^\prime}(t^\prime) \rangle = 2D \delta_{\lambda \lambda^\prime} \delta(t-t^\prime)\,,\label{stateta}
\end{equation}
\item For $J$ of Wigner or Wishart type, $\lambda$ is assumed to be distributed accordingly with the semi-circle or Marchenko-Pastur laws respectively, in the large $N$ limit. 
\item The potential $V(Q^2)$ depending on the Euclidean norm $Q^2:=\sum_{\lambda}q_\lambda^2$ avoids configurations with large $\vert q_\lambda \vert$. 
\end{enumerate}
Physically, $D$ is the \textit{temperature} of the system at equilibrium, and we will occasionally note this coefficient by $T$. One of the simplest potentials one can think of is the quartic one, $V^\prime(Q^2)=\kappa_1+\frac{\kappa_2}{N} Q^2$, and we expect this model to be equivalent to the spherical one in the large $N$ limit, provided that $\kappa_2 \to \infty$, $\kappa_1\to -\infty$ but $\kappa_1/\kappa_2=-\mathcal{O}(1)$. Indeed, quantities like $Q^2$ self averages in the large $N$ limit, and the relaxation time is larger for $V^\prime(\langle Q^2 \rangle)=0$, where $Q^2$ fluctuates shortly around the average $\langle Q^2 \rangle = -N \kappa_1/\kappa_2$ \cite{van2010second}. Finally, note that the non-linear effects due to the local potential could induce interesting non-Gaussian effects, see for instance \cite{carreras1996fluctuation,kim2009probability,kimanderson2008nonperturbative}.
\medskip

The equation \eqref{eqLangevin} can be rewritten as  a path integral, following the Martin-Siggia-Rose strategy \cite{ZinnJustinBook1,ZinnJustinBook2,DeDominicisbook}. Explicitly, introducing the so-called \textit{response field} $\varphi$, the generating functional $Z[j,\tilde{j}]$ reads for \textit{equilibrium dynamics}\footnote{Note that this implicitly assumes that the minimum of the action is unique, and this is why we explicitly refer as the ‘‘equilibrium solution” for this model.}:
\begin{equation}
Z[j,\tilde{j}]= \int  d\phi d\bar{\varphi} \, e^{-S[\phi,\bar{\varphi}]+\int dt\,(j(t)\cdot\phi(t)+\tilde{j}(t)\cdot\bar{\varphi}(t))}\,,\label{expressionZ}
\end{equation}
where the dot is the standard inner product $j(t)\cdot\phi(t):=\sum_\lambda\, j_\lambda(t) \phi_\lambda(t)$ and:
\begin{equation}
\boxed{S[\phi,\bar{\varphi}]=\int dt \sum_\lambda\, \left( \frac{\bar{\varphi}^2_\lambda}{2}+i \bar{\varphi}_\lambda \left( \dot{\phi}_\lambda+\frac{\partial W_0}{\partial \phi_\lambda}\right) \right)\,,}\label{classicalaction}
\end{equation}
the fields $\phi_\lambda$ and the Hamiltonian $W_0$ being defined as:
\begin{equation}
q_\lambda=: \sqrt{2D} \phi_\lambda\,,\qquad H_0=: 2D W_0\,.\label{rescaling}
\end{equation}
such that:
\begin{equation}
\frac{\partial W_0}{\partial \phi_\lambda}=:\lambda \phi_\lambda+U^\prime(\phi^2) \phi_\lambda\,.
\end{equation}
Note that we use the Ito prescription for the computation of path integral, imposing $\theta(0)=0$ for the Heaviside function and $\det \mathcal{M}=1$, where $\mathcal{M}$ has entries:
\begin{equation}
\mathcal{M}_{ij}(t^\prime,t):=\frac{d}{dt}\delta_{ij}\delta(t-t^\prime) +\frac{\partial^2 H_0}{\partial q_i(t) \partial q_j(t^\prime)}\,. \label{matM}
\end{equation}
Finally, due to the equilibrium dynamics assumption, the action \eqref{classicalaction} should be invariant under time reversal, which corresponds to the field transformation:
\begin{equation}
\phi^\prime_\lambda(t):=\phi_\lambda(-t)\,,\qquad \bar{\varphi}_\lambda^\prime(t):=\bar{\varphi}_\lambda(-t)+2i\dot{\phi}_\lambda(-t)\,.\label{fieldtransform}
\end{equation}

\begin{remark}
In contrast with \cite{Lahoche:2021tyc}, there is no integration of the disorder here. The effect of the disorder is blind in the additional kinetic contribution $\lambda\bar{\varphi}_\lambda  \dot{\phi}_\lambda$. Moreover, the local potential $U(\phi)$ is \textit{local} in time! Finally, to derive the expression \eqref{expressionZ} we assumed that the vacuum of the action is non-degenerate. Hence, if an ergodicity breaking occurs, it is reached “from above” as for instance in \cite{DeDominicis1978}. 
\end{remark}

\paragraph{Equilibrium state.} If we denote as $P[\phi(t)]$ the probability to obtain the configuration $\phi_i(t)$ to the time $t>0$ from some initial condition $\phi_i(t=0)=:c_i$ $\forall\, i$, one can show that the functional $\Psi[\phi(t)]:=e^{W_0[\phi]} P[\phi(t)]$ obeys the  Schrödinger like equation $\dot \Psi = -\hat{H} \Psi$, where $\hat{H}$ is the positive definite operator:
\begin{equation}
\hat{H}:= \frac{1}{2} \sum_i \left(-\frac{\partial}{\partial \phi_i}+ \frac{\partial W_0}{\partial \phi_i}\right)\left(\frac{\partial}{\partial \phi_i}+ \frac{\partial W_0}{\partial \phi_i}\right)\,.
\end{equation}
The ground state $\Psi_0$, solving $\hat{H} \Psi_0=0$ is formal of the form $\Psi_0 \sim e^{-W_0[\phi]}$, and corresponds to the equilibrium configuration of the system, namely:
\begin{equation}
\int d\phi \, e^{-2W_0[\phi]} < \infty\,.\label{normcondeq}
\end{equation}
Hence, the partition function of the equilibrium states reads explicitly:
\begin{equation}
Z[j]:=\int d\phi \,e^{-S_{\text{eq}}[\phi]+ \sum_\lambda j_\lambda \phi_\lambda}\,,
\end{equation}
where:
\begin{equation}
S_{\text{eq}}[\phi]=2 W_0[\phi]\equiv  \left( \sum_\lambda\lambda \phi_\lambda^2+U(\phi^2) \right)\,.\label{eqmodel}
\end{equation}

\paragraph{Large-N limit distributions.} In many cases, in the large $N$ limit, discrete sums over eigenvalues of a random matrix converge toward an integral over a bounded domain $\mathcal{D}$ \cite{Benaych1,Wigner1,Potters1}:
\begin{equation}
\frac{1}{N}\sum_\lambda f(\lambda) \to \int_{\mathcal{D}} \,d\lambda \, \mu(\lambda) f(\lambda)\,,\label{integralapp}
\end{equation}
where the shape of the distribution depends on the statistical ensemble that we consider for $J_{ij}$. Assuming $J$ to be centered random matrices with independent and identically distributed (i.i.d) Gaussian entries with variance $\sigma^2/N$, namely the Gaussian orthogonal ensemble (GOE), the distribution $\mu(\lambda)$ converges weakly toward the Wigner semi-circle law $\mu_W$
\begin{equation}
\mu_W(x):=\frac{1}{2 \pi\sigma^2}\sqrt{4\sigma^2-x^2}\,\,\mathrm{I}_{[-2\sigma,2\sigma]}\,,
\end{equation}
where $\mathrm{I}_{[-2\sigma,2\sigma]}$ vanishes outside the real interval $\mathcal{D}\equiv [-2\sigma, 2\sigma]$. 
\medskip

In this paper, we also consider the Wishart ensemble. We recall that a Wishart matrix is a random matrix taking the following form:
\begin{equation}
J= \frac{X X^T}{P} \,,\label{MPMAT}
\end{equation}
where $X^T$ is the transpose of the matrix $X$. The matrix $X$ is furthermore assumed to be a random matrix with i.i.d entries having standard deviation $\sqrt{\sigma}$ and finite higher momenta. Up to these assumptions, the eigenvalue distribution converges toward the Marchenko-Pastur (MP) distribution $\mu_{MP}$, provided that $N\to \infty, \, P\to \infty$ such that $N/P\equiv c$ remains finite. For $c\leq 1$, $\mu_{MP}$ reads:
\begin{equation}\label{mp}
\mu_{MP}(x)=\frac{1}{2\pi \sigma}\frac{\sqrt{(x-\lambda_-)(\lambda_+-x)}}{cx}\,,
\end{equation}
where: $\lambda_{\pm}=\sigma(1\pm \sqrt{c})^2$. 
\medskip

\paragraph{Time scales and relaxation toward equilibrium.} In the quartic regime, for a potential written as $V(Q^2)=\kappa_1 Q^2$, the system relaxes toward equilibrium with the typical relaxation time $\tau(\lambda)=(\lambda+\kappa_1)^{-1}$ for the mode $\lambda$. This relaxation time, however, may diverge up to a critical value $\kappa_1^*=-\min (\mathcal{D})$. For the Wigner ensemble, for instance, $\kappa_1^*=2\sigma$, and the mode $\lambda=2\sigma$ reach equilibrium for an arbitrarily long time. We thus introduce a new quantity $p$  called the \textit{momentum}, such that:
\begin{equation}
p:=\lambda-\min (\mathcal{D})\,,\label{equationp}
\end{equation}
such that $p$ is defined positive as $N\to \infty$. We furthermore denote as $\mu(p)$ the corresponding distribution for $p$. Interestingly, for $p$ small enough, both Wigner and MP distributions behave as $\sqrt{p}$, provided that $\lambda_- \neq 0$. For $\lambda_-=0$, $\mu(p)\sim 1/\sqrt{p}$. 

\begin{remark}\label{remarkdim}
The asymptotic behavior of the distributions has to be compared with the state distribution of an ordinary field theory in dimension $d$. Indeed, for the ordinary $\phi^4$ theory in dimension $d$ with Laplacian propagator for instance, whose action reads:
\begin{equation}
S[\phi]= \frac{1}{2}\int dx\, \phi(x) (-\Delta +m^2) \phi(x) + \frac{g}{4!} \int dx\, \phi^4(x)\,, \label{actionbare}
\end{equation} 
or for the domain coarsening model considered in the introduction (see \eqref{domco}), the distribution for the eigenvalues of the Laplace propagator $\rho(p^2)$ behaves as $\rho(p^2) \sim (p^2)^{\frac{d-2}{2}}$. Hence, for $\lambda_- > 0$ ($c\neq 1$), the behavior of Wigner and Marchenko-Pastur distribution agrees with a field theory in dimension $d=3$, and with a theory in dimension $d=1$ for $\lambda_-=0$ ($c=1$). 
\end{remark}

\subsection{Large $N$ limits from Feynman graphs}\label{largeN}
In this section, we discuss the large $N$ limit from the viewpoint of Feynman diagrams. In the first time, we sketch the existence of the $1/N$ expansion for the quartic model. In a second time, we consider the large $N$ limit of standard Schwinger-Dyson equations and show that they look like closed equations which are easy to solve in the equilibrium high-temperature regime. 

\subsubsection{$1/N$ expansion} 

\begin{figure}
\begin{center}
\includegraphics[scale=0.8]{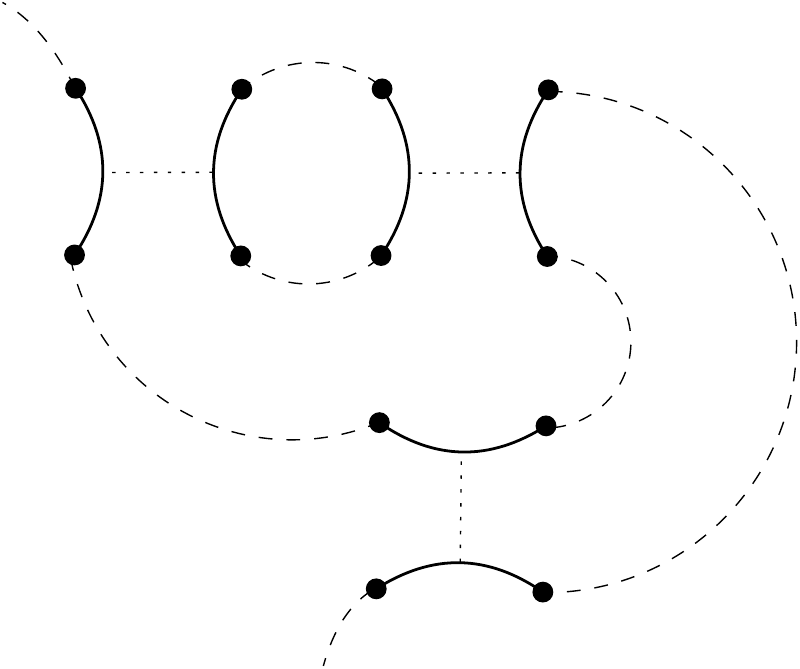}
\end{center}
\caption{A typical Feynman graph involving three vertices and contributing to the expansion of the $2$-point function. Open dashed edges materialize half external edges.}\label{TypicalFeynman}
\end{figure}

\begin{figure}
\begin{center}
\includegraphics[scale=0.8]{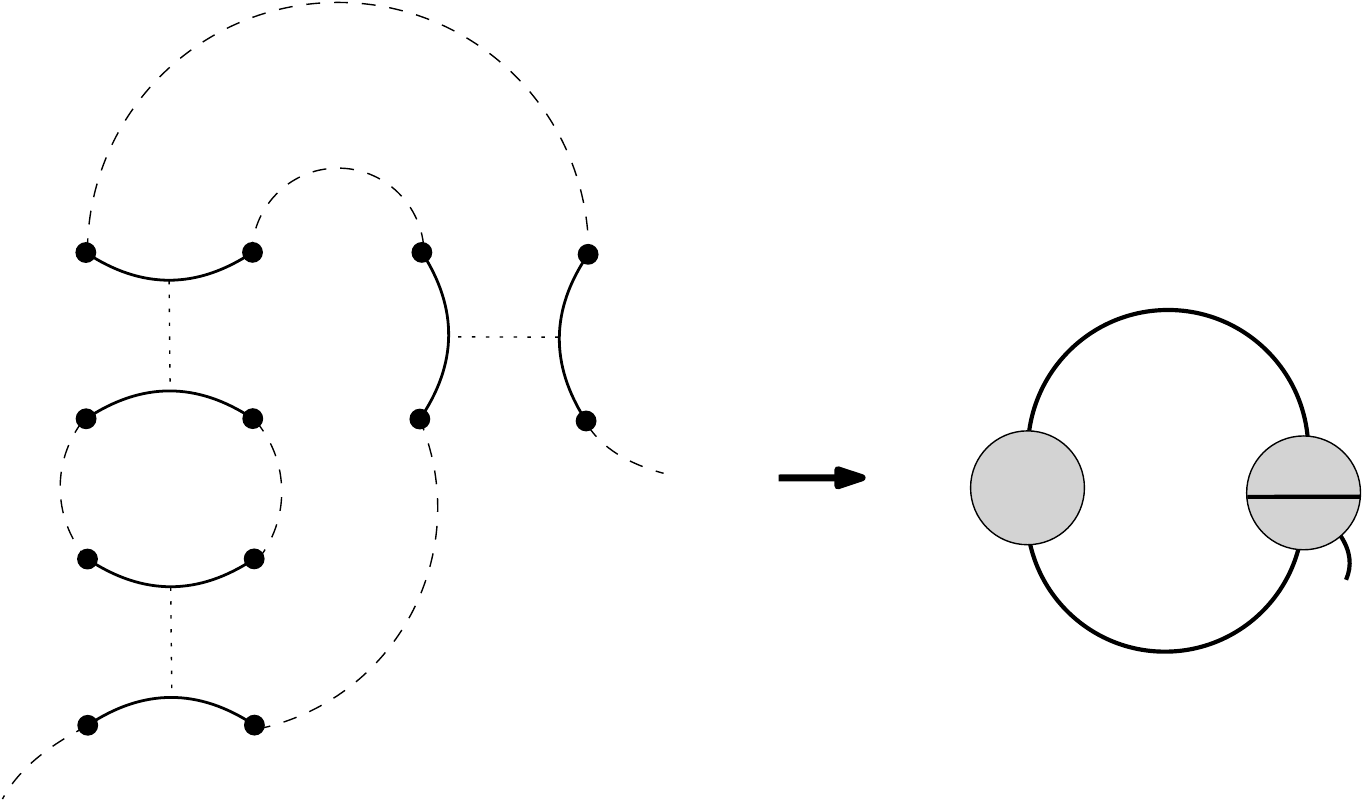}
\end{center}
\caption{Illustration of the correspondence between original and intermediate field representation. Loop vertices are gray discs and intermediate field propagators are solid edges. The “cilium” on the loop vertex on the right-hand side represents the external edges in the original representation.}\label{figintermediate}
\end{figure}

Usually, $O(N)$ vector models, that are, quantum or statistical field theories for random vectors of size $N$ with $O(N)$ invariant interactions, enjoy power counting and 1/$N$ expansion. Standard reviews exist on this topic, and we refer the reader to them, for instance, \cite{Moshe_2003,ZinnJustin4} and references therein. In \cite{Lahoche:2021tyc} the authors proposed a derivation for a model in the equilibrium limit similar to \eqref{eqmodel}, where the propagator enjoys a Wigner or MP distribution. Let us provide here a sketched discussion to make this section as self-consistent as possible. For such kind of models, the power counting is generally easier to construct using intermediate field formalism and \textit{loop-vertex representation} (LVR) \cite{rivasseau2018loop,hubbard1959calculation}. To begin, let us focus on the equilibrium model given by \eqref{eqmodel}. 
\begin{figure}
\begin{center}
\includegraphics[scale=1]{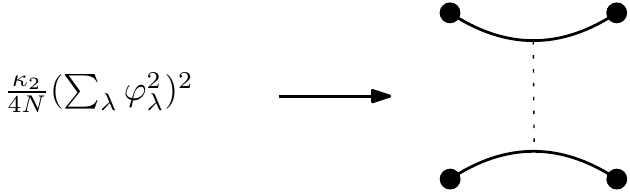}
\end{center}
\caption{Graphical convention used to materialize the quartic interaction for the equilibrium theory.}\label{convquartic}
\end{figure}
For the quartic model that we consider in this section, we have a single kind of vertex that, accordingly to the conventions of the previous section, can be pictured as in Figure \ref{convquartic}. A typical Feynman graph is then pictured on Figure \ref{TypicalFeynman}, dotted edges materializing the Gaussian propagator $C_{\lambda\lambda^\prime}=C(\lambda)\delta_{\lambda\lambda^\prime}$  with diagonal entries:
\begin{equation}
C(\lambda)=\frac{1}{\lambda-\min\mathcal{D}+\mu_1}\,.
\end{equation}
A Feynman graph like the one pictured in Figure \ref{TypicalFeynman} involves some \textit{closed loops}, i.e., closed cycles made of an alternating sequence of dashed and solid edges. The power counting can be established as follows. Each vertex shares a factor $1/N$. Moreover, each loop involves a discrete sum that should be counted as a factor $N$, a convention which is more transparent from the integral approximation \eqref{integralapp}. If we denote as $V$ and $L$ the number of vertices and loops, the global scaling with $N$ is, therefore: $N^{L-V}=N^{1-(V-L+1)}$. In the intermediate field representation, vertices look like intermediate field edges and loops are contracted as \textit{loops vertices}. The correspondence is illustrated in Figure \ref{figintermediate} for the Feynman diagram of Figure \ref{TypicalFeynman}. 
Let us consider a vacuum diagram in the original representation, and let $\mathcal{E}$ and $\mathcal{V}$ respectively the number of intermediate field edges and loop vertices. Hence, the scaling reads $N^{-\nu+1}$ with:
\begin{equation}
\nu=\mathcal{V}-\mathcal{E}+1 \geq 0\,.\label{scaling}
\end{equation}
$\nu$ is the number of loops in the intermediate field graph, and it vanishes only if the graph is a tree. Hence, the leading order graphs are trees in the intermediate field representation. Leading order non-vacuum $1$PI diagrams can be obtained from a vacuum tree by cutting some loops on the leaves of the tree. Finally, it is not hard to convince ourselves that this power counting holds for the dynamical field theory given by \eqref{expressionZ}. Formally, the only change is the introduction of square nodes at the vertex level, and of two kinds of propagator edges linking $\bar{\varphi}$ with $\phi$ and $\phi$ with themselves respectively. We call \textit{off shell,} such a diagram formally obtained from an equilibrium graph. Nevertheless, some of these off shell diagrams vanish \textit{on shell} because of causality. Indeed, it can be proved that the response field $\bar{\varphi}$ does not propagate to all orders of the perturbative expansion, ensuring that $G_{\bar{\varphi}\bar{\varphi}}$ -- the component ‘‘$\bar{\varphi}\bar{\varphi}$'' of the full propagator—is identically zero (see Appendix \ref{appendix3}).

\subsubsection{High temperature closed equations}\label{Closed}

In this section, we focus on the quartic theory in the large temperature $T>T_c$ equilibrium phase, with:
\begin{equation}
U(\phi^2):=\kappa_1 \phi^2+\frac{\kappa_2}{N} (\phi^2)^2\,.\label{potquart}
\end{equation}
In the large $N$ limit, we will show that the self-energy $\Sigma_{\bar{\varphi}\phi}$ reduces to a global translation of mass and obeys a closed “gap” equation that can be solved analytically. Furthermore, we show that the perturbative series for the $4$-point function can be formally resumed, and the effective quartic coupling analytically solved.  

\begin{figure}
\begin{center}
\includegraphics[scale=1]{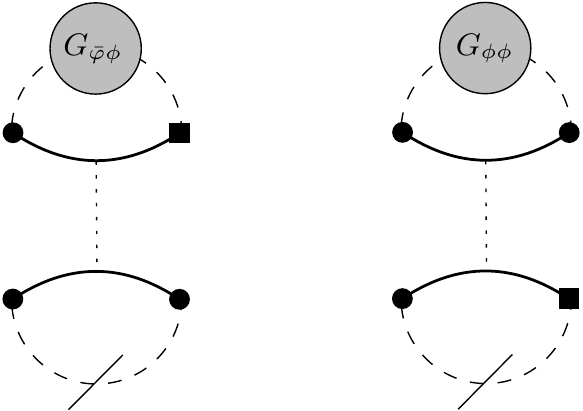}
\end{center}
\caption{The two kinds of leaves whose deletion gives a 1PI $2$-point function (off shell).}\label{figconfig2}
\end{figure}

\paragraph{Gap equation for the quartic theory.} 
In the previous section, we recalled that leading order graphs are trees in the intermediate field representation. Starting from a vacuum diagram, we obtain a $1$PI $2$-point diagram by deleting one loop from a leaf on and off shell tree. Let us denote as $v$ the vertex (in the original representation) sharing this leaf. In fact, there are two kinds of leaves, both pictured in Figure \ref{figconfig2}. 
In this figure, we formally resumed the remaining contributions of the diagrams, and it is easy to check that they correspond to the effective $2$-point functions $G_{\bar{\varphi} \phi}$ and $G_{\phi\phi}$ -- see \cite{lahoche2018nonperturbative,samary2014closed,samary2015correlation}. The diagram on the right contributes to the component $\Sigma_{\bar{\varphi} \phi}$, while the diagram on the left contributes to the component $\Sigma_{\phi\phi}$. But the condition $G_{\bar{\varphi}\bar{\varphi}}=0$ should require, on shell:
\begin{equation}
\Sigma_{\phi\phi}=0\,, 
\end{equation}
and hence:
\begin{equation}
\int dt \int \mu(\lambda) d\lambda\, G_{\bar{\varphi} \phi}(\lambda,t)=0\,.\label{conditioncausal}
\end{equation}
This condition is also true at the leading order of the perturbative expansion, and roughly speaking, the vacuum diagram on the left of Figure \ref{figconfig2} has to vanish. 
\medskip

\begin{figure}
\begin{center}
\includegraphics[scale=0.7]{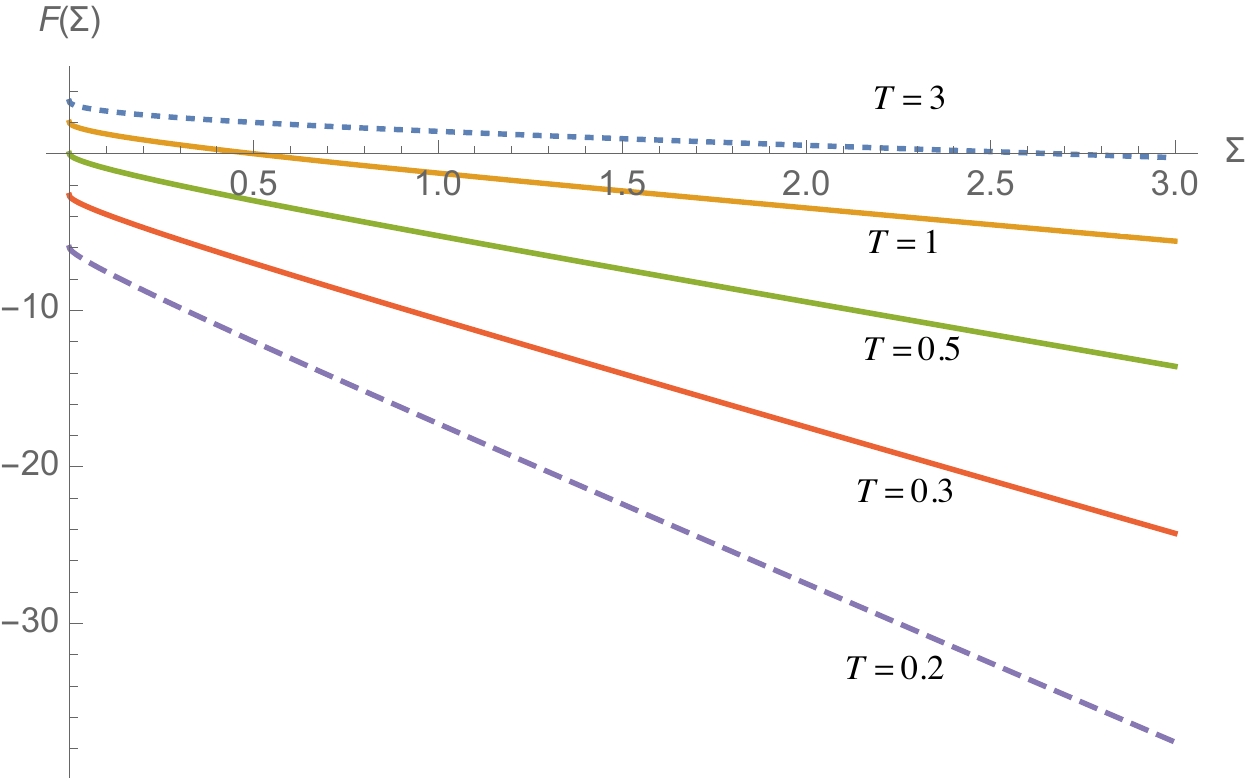}
\end{center}
\caption{Graphical solutions of the closed equation for Wigner distribution for temperatures varying from $T=3$ (dotted blue curve) to $T=0.2$ (dashed purple curve). $F(x):= \left(x -\sqrt{x  (x +4)}+2\right)-2\left( x/\kappa_2-\kappa_1/\kappa_2\right)$.}\label{figplot1}
\end{figure}

To summarize, we have the formula:
\begin{equation}
\boxed{\Sigma_{\bar{\varphi} \phi}\, \delta_{\lambda \lambda^\prime}\delta(\omega-\omega^\prime)=-\left(2\kappa_2 \, \int d\lambda\, \mu(\lambda)  G_{\phi\phi}(\lambda,0)\right) \delta_{\lambda \lambda^\prime}\delta(\omega-\omega^\prime)\,.}\label{closedSigma1}
\end{equation}
The self-energy is diagonal both in frequency and in eigenvalue spaces. Therefore, in the leading order, only the mass is shifted by the quantum effects. Note that we assumed the translation invariance of the effective propagator. The left-hand side of the equation \eqref{closedSigma1} can be computed explicitly, because:
\begin{equation}
G_{\phi\phi}(\lambda,\omega)=\frac{1}{\vert -i \omega+(\lambda-\min\mathcal{D})+\kappa_1-\Sigma_{\bar{\varphi} \phi}\vert^2}\,.
\end{equation}
Therefore, 
\begin{equation}
\Sigma_{\bar{\varphi} \phi}=-2\kappa_2 \, \int dp\, \mu(p) \int \frac{d\omega}{2\pi}\frac{1}{\vert -i \omega+p+\kappa_1-\Sigma_{\bar{\varphi} \phi}\vert^2}\,,\label{closedzero}
\end{equation}

can be computed analytically. For the Wigner distribution, and setting $\sigma=1$, we get two branches of solutions, respectively for $\Sigma>0$ and $\Sigma<-4$, where $\Sigma \equiv \kappa_1-\Sigma_{\bar{\varphi}\phi}$ is the effective mass. In the positive branch, we get:
\begin{equation}
\Sigma_+ = \kappa_1+\frac{1}{2}\kappa_2  \left(\Sigma_+ -\sqrt{\Sigma_+  (\Sigma_+ +4)}+2\right)\,.\label{sigmaclosed1}
\end{equation}
Moreover, the negative solution changes the sign of the pole of the $2$-point function and breaks causality. Hence, only the positive branch $\Sigma_+$ has a physical meaning. The equation can be solved exactly in terms of $\kappa_1$ and $\kappa_2$. However, it is convenient to recall that, with the rescaling \eqref{rescaling}, $\kappa_2\propto T$, the temperature ($T \equiv D$, see \eqref{stateta}), and we normalize such that $\kappa_2=T$. We furthermore define $\kappa_1=T-T_0$, and set $T_0=1$. Solving equation \eqref{sigmaclosed1}, we get two branches of solutions, but only one is positive for high temperature. This solution however breaks down as the discriminant becomes negative. This arises for the temperature:
\begin{equation}
T_c:=2\sqrt{3}-3\approx 0.46\,,
\end{equation}
which should be compared with the analytical value $T_c\approx 0.62$ computed in Appendix \eqref{appendix2}. For $T<T_c$, equation \eqref{sigmaclosed1} has no real solution, meaning that the equilibrium assumption does not hold. In Figure \ref{figplot1} we illustrate what happens graphically. For high temperatures (blue dotted curve with $T=3$) the curve crosses the positive abscissa axis one time. This holds for higher and smaller temperatures, as soon as $T>T_c$ where the solution for $\Sigma_+$ vanishes, and no solution exists for $T<T_c$. This is reminiscent of a second-order phase transition, where $\Sigma_+$ plays the role of an order parameter.

\begin{figure}
\begin{center}
\includegraphics[scale=0.5]{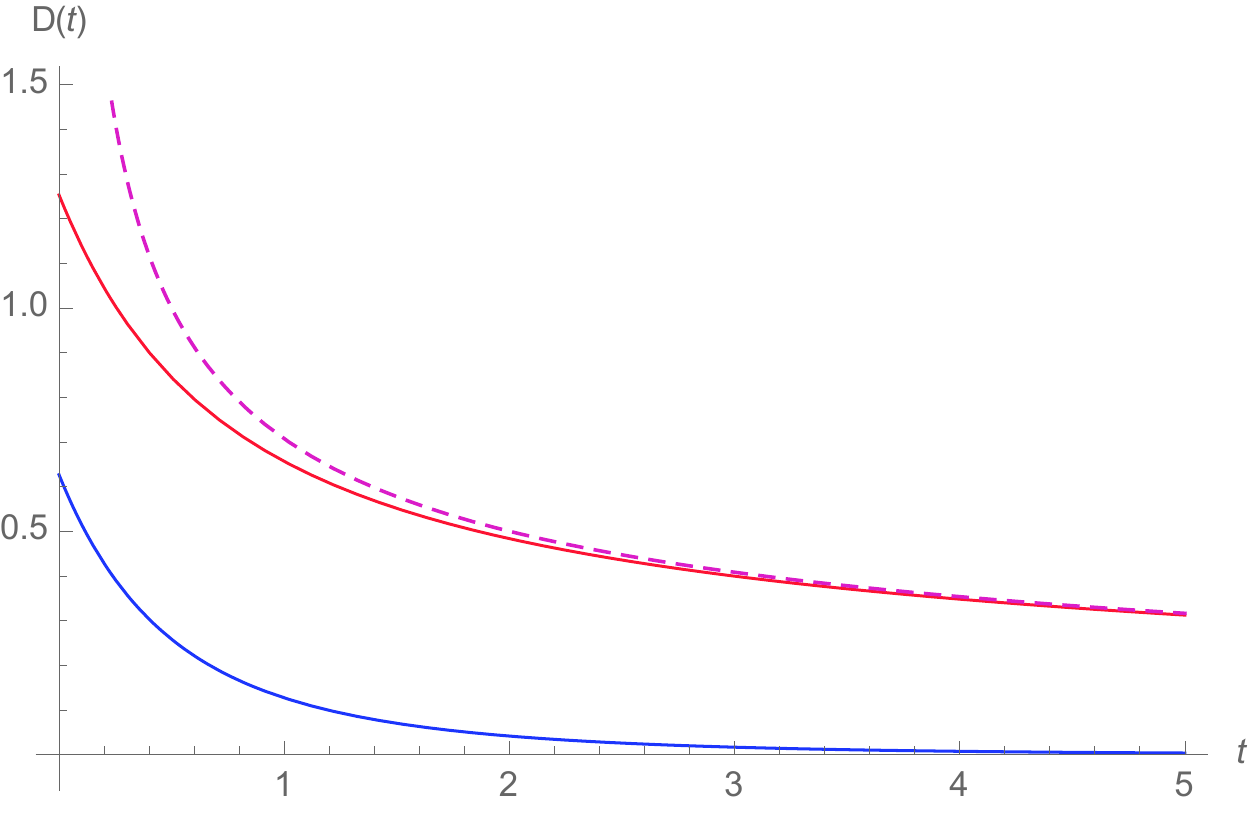}\qquad \includegraphics[scale=0.5]{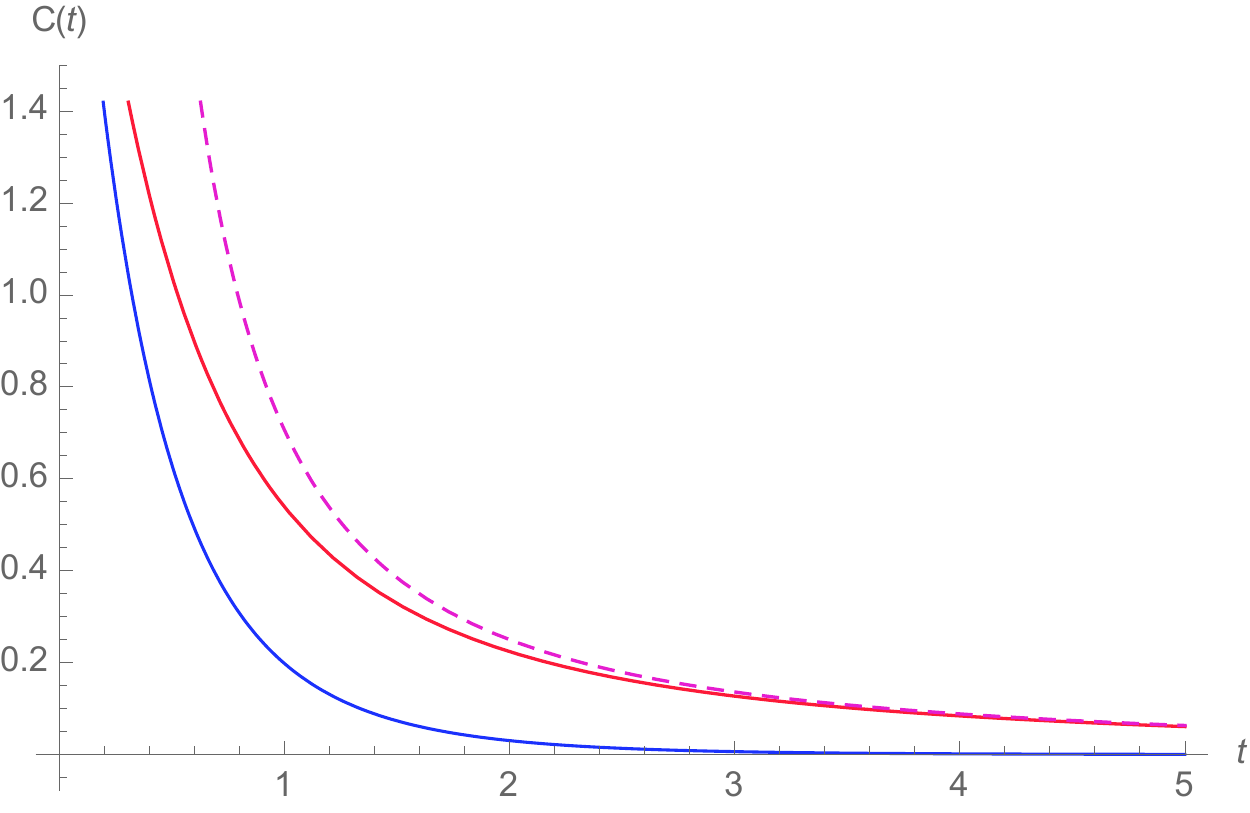}
\end{center}
\caption{Plots of functions $D(t)$ (on left) and $C(t)$ (on right). The blue curve is for $T>T_c$, and the red curve is for $T=T_c$. The dotted purple curve gives the asymptotic behavior.}\label{figCD}
\end{figure}

The nature of the phase transition can be investigated from the $2$-point correlation functions. Let us define:
\begin{equation}
C(t):=\int dp \, \mu(p) G_{\bar{\varphi}\phi}(p+\min (\mathcal{D}),t)\,,
\end{equation}
and 
\begin{equation}
D(t):=\int dp \, \mu(p) G_{\phi\phi}(p+\min (\mathcal{D}),t)\,.
\end{equation}
The typical behavior of these functions with $t$ is pictured in Figure \ref{figCD}. The solid blue curve materializes the behavior of $D(t)$ (on left) and $C(t)$ (on right) for $T>T_c$. They exhibit an exponential decay in both cases, $D(t), C(t) \sim e^{-t}$ for $t$ large enough. The red curves show the same function for $T=T_c$, and exhibit a power law relaxation with infinite relaxation time for $t$ large enough (but not too large):
\begin{equation}
C(t)\sim \frac{1}{\sqrt{2}} \frac{1}{t^{3/2}}\,,\qquad D(t)\sim \frac{1}{\sqrt{2}} \frac{1}{t^{1/2}}\,.
\end{equation}
This asymptotic behavior is materialized by the dotted purple curves in Figure \ref{figCD}. It is (at least qualitatively) in agreement with the expected behavior for the spherical model and asymptotic analysis of the quartic model, see Appendices \ref{appendix1}-\ref{appendix2} and reference \cite{Cugliandolo2}, where it is assumed that $T<T_c$ while our analysis aims to reach the transition from the high-temperature regime. 
\medskip

\begin{figure}
\begin{center}
\includegraphics[scale=0.5]{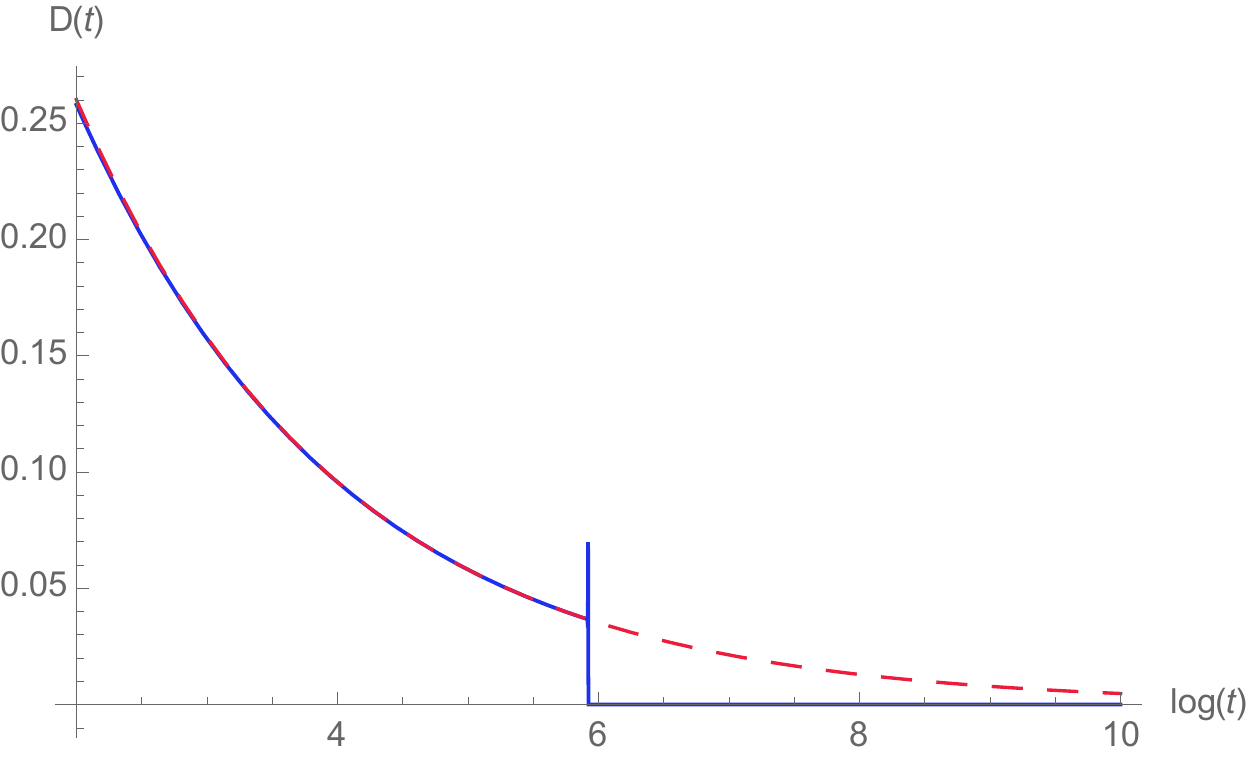}\qquad \includegraphics[scale=0.5]{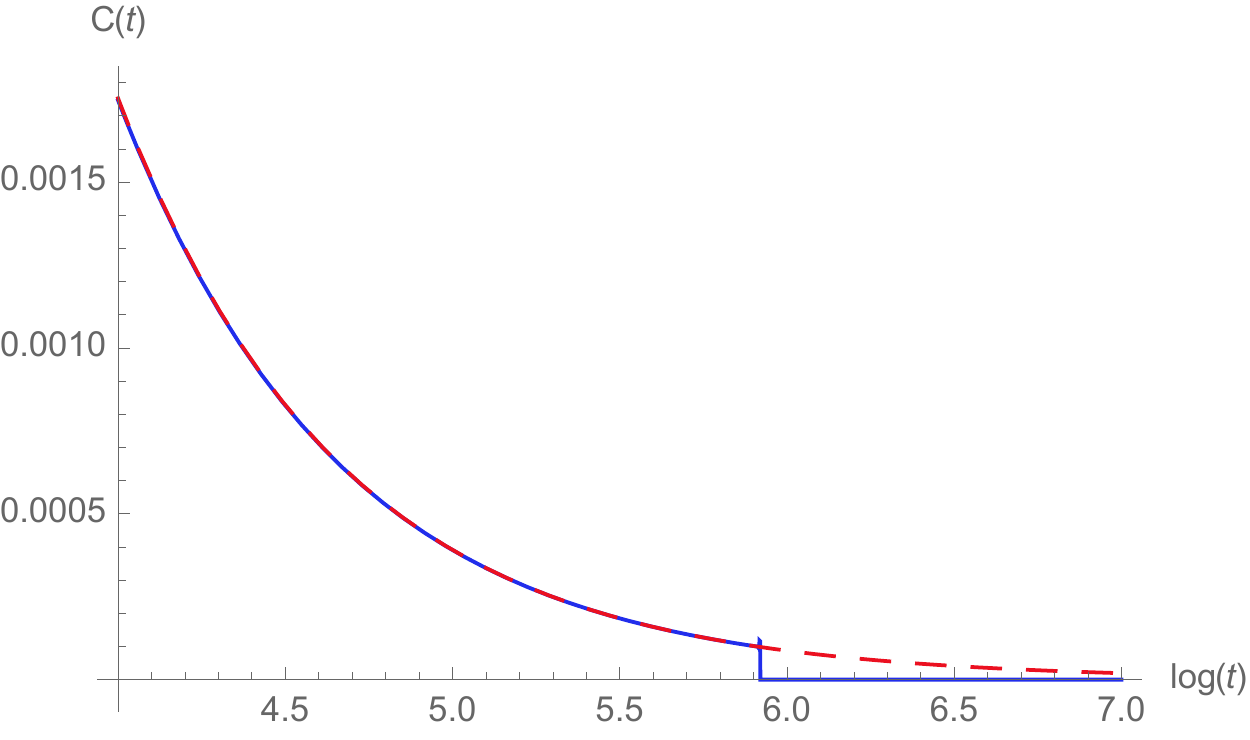}
\end{center}
\caption{Plots of functions $D(t)$ (on left) and $C(t)$ (on right) for large time using a logarithmic scale. The red curve materializes the asymptotic behavior.}\label{figCDlog}
\end{figure}

Finally, it is interesting to investigate the very large time behavior of the critical curves for $C(t)$ and $D(t)$. They are plotted in Figure \ref{figCDlog}, using a logarithm timescale. Interestingly, for very large $t\sim e^6$, both the functions go to zero after a short period of oscillation. Once again, similar behavior is expected for correlation functions in the spherical or quartic models, which goes to zero after a “plateau” with macroscopic time for $T<T_c$ (see Appendix \ref{appendix2} and references \cite{DeDominicisbook, Cugliandolo2,Castellani1}). We expect to lack the plateau but not the power law behavior because they reach the transition from high temperatures.

\paragraph{on shell, 4-point function.} From the construction given in the previous section, it follows that a typical Feynman graph contributing to the $4$-point function can be obtained from an off shell vacuum tree by deleting two loops on two different leaves. To obtain a contribution for $\Gamma_{\bar{\varphi}\phi\phi\phi}^{(4)}$ -- the 4-points effective vertex function, we have to cancel one propagator of type $G_{\bar{\varphi}\phi}$ and a second of type $G_{\phi\phi}$. A typical Feynman graph has therefore the structure of a chain whose links are made of loops of length two:
\begin{align}
\nonumber \vcenter{\hbox{\includegraphics[scale=0.8]{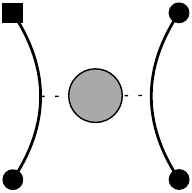}}}\,&\equiv\,\vcenter{\hbox{\includegraphics[scale=0.8]{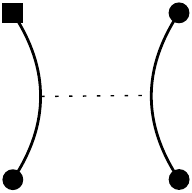}}}\,+\,\vcenter{\hbox{\includegraphics[scale=0.8]{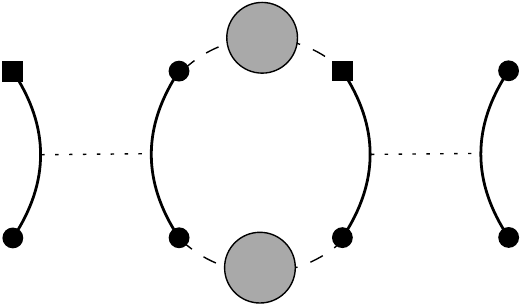}}}\\\nonumber
&+\,\vcenter{\hbox{\includegraphics[scale=0.8]{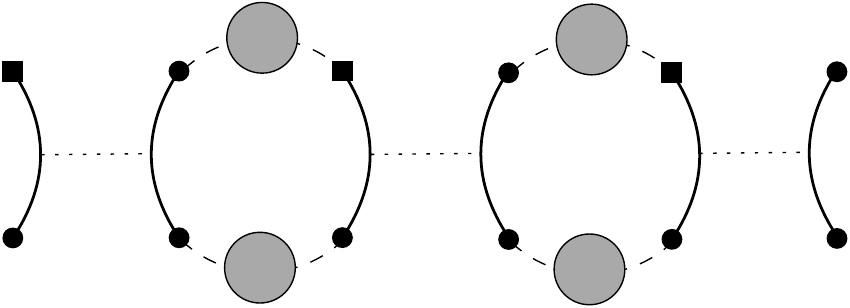}}}\\
&+\, \vcenter{\hbox{\includegraphics[scale=0.8]{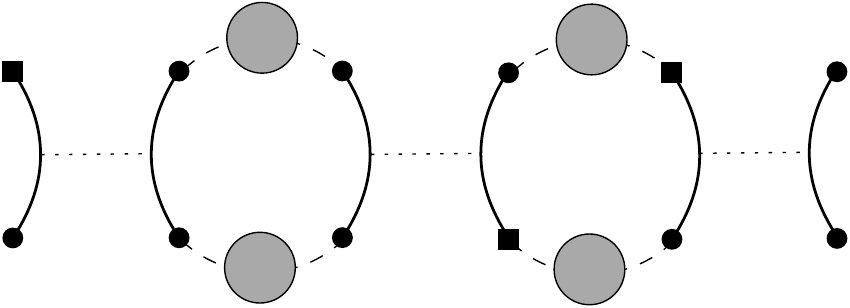}}}\,+\,\cdots\label{sum4pointsbis}
\end{align}
As discussed in \cite{Canet_2011,lahoche2022stochastic,Lahoche:2021tyc}, $\Gamma_{\phi\phi\phi\phi}^{(4)}$ (and in fact any component that does not involve the response field) has to vanish on the shell, this requires in particular that:
\begin{equation}
\vcenter{\hbox{\includegraphics[scale=0.8]{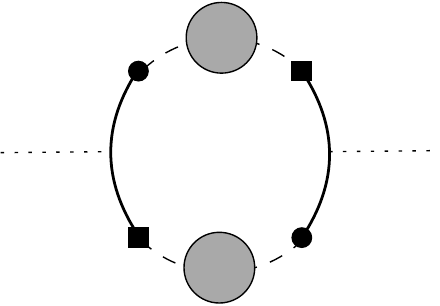}}}\,\equiv \, 0\,.
\end{equation}
Hence, on the shell, the expansion \eqref{sum4pointsbis} involves only one kind of loop, made on one $G_{\bar{\varphi}\phi}$ and one $G_{\phi\phi}$. The geometric progression can be resumed, and a straightforward calculation leads to the effective $4$-points function:
\begin{equation}
\kappa_{2,\text{eff}}=\frac{\kappa_2}{1+4\kappa_2 \mathcal{A}_{2}}\,,\label{4pointsclosed}
\end{equation}
with:
\begin{equation}
\mathcal{A}_{2}:= \int \frac{d\omega}{2\pi} \int \mu(\lambda) d\lambda \, G_{\bar{\varphi}\phi}(\lambda, \omega) G_{\phi\phi}(\lambda,\omega)\,.
\end{equation}
Once again, this relation has to be valid on shell. \\

\section{Functional renormalization group for a stochastic process}\label{FRG}
In this section, we give the basics of the Wetterich-Morris FRG formalism for a stochastic system \cite{Canet_2007,Canet_2011,Prokopec_2018,wilkins2021functional,wilkins2021functional2,castellana2013renormalization,Balog_2020,Tarjus_2020,Balog_2018,canet2022functional}. We consider a coarse-graining in both frequency and eigenvalue, choosing a regulator that respects the causality and the time-reversal symmetry of the action \eqref{actionbare}.
\medskip

To be more formal, let us define $\Phi=(\phi,\bar{\varphi})$ the doublet with $2N$ component built from the pair $\phi$ and $\bar{\varphi}$. In the same way, we define the source doublet $J=(j,\tilde{j}\,)$. The components of these fields are $\Phi_{\alpha,\lambda}$ and $J_{\alpha,\lambda}$, for $\alpha=1,2$ and $\lambda=\llbracket 0,N \rrbracket$, with for instance:
\begin{equation}
J_{1,\lambda}\equiv j_\lambda\,,\qquad J_{2,\lambda}=\tilde{j}_\lambda\,.
\end{equation}
We modify the classical action \eqref{classicalaction} as $S\to S_k:= S+\Delta {S}_k$, adding to it the regulator $\Delta S_k$ defined as in \cite{Delamotte1}:
\begin{align}
\Delta{S}_k:=\int dt dt^\prime \sum_{\lambda} r_k(p)\Big(i \bar{\varphi}_\lambda(t) \rho_k^{(1)}(t-t^\prime)  \phi_\lambda(t^\prime)+\frac{1}{2}\bar{\varphi}_\lambda(t)\rho_k^{(2)}(t-t^\prime)\bar{\varphi}_\lambda(t^\prime)
\Big)\,. \label{regulatordef}
\end{align}
We denote as $Z_k[J]$ the corresponding generating functional. We furthermore introduce the classical field $\Xi$ with components $\Xi_{\alpha,\lambda}$ such that:
\begin{equation}
\Xi_{\alpha,\lambda}=\frac{\partial }{\partial J_{\alpha,\lambda}} \ln Z_k[J]\,,
\end{equation}
and we define $\Gamma_k[\Xi]$, the effective average action as:
\begin{equation}
\Gamma_k[\Xi]+\frac{1}{2} \int dt dt^\prime\sum_{\lambda,\alpha,\beta} \Xi_{\alpha,\lambda}(t) (R_k)_{\alpha \beta}(\lambda,t-t^\prime)\, \Xi_{\beta,\lambda}(t^\prime)=-W_k[J]+\int dt\sum_{\lambda,\alpha}  \Xi_{\alpha,\lambda} J_{\alpha,\lambda} \,,
\end{equation}
where $W_k[J]:=\ln Z_k[J]$. The matrix $(R_k)_{\alpha \beta}$ is called \textit{regulator}, and has to be chosen carefully, in agreement with some physical conditions, especially in regard to causality and time-reversal symmetry (which holds the ergodic phase). Furthermore, it has to be designed such that $\Gamma_k[\Xi]$ interpolates smoothly between classical action $\Gamma_{k=\Lambda}[\Xi]\to S$ for some UV cut-off $\Lambda$ and the effective action $\Gamma_{k=0}[\Xi]=\Gamma$. The fundamental equation describing the RG flow is the first-order differential equation:
\begin{equation}
\Gamma_k^\prime[\Xi]= \frac{1}{2}\int dtdt^\prime \sum_{\lambda, \alpha,\beta} ({R}_k^\prime)_{\alpha,\beta}(\lambda,t-t^\prime) G_{k,\alpha \beta}(\lambda, t-t^\prime)\,,\label{Wetterich}
\end{equation}
where $G_{k\alpha \beta}$, the entry $\alpha\beta$ of the propagator matrix $\mathbf{G}_k$, explicitly:
\begin{equation}
G_{k\alpha \beta}(\lambda,t-t^\prime):= \langle \Xi_{\alpha,\lambda}(t) \Xi_{\beta,\lambda}(t^\prime) \rangle \equiv (\Gamma_k^{(2)}+R_k)^{-1}_{k,\alpha \beta}(\lambda,t-t^\prime) \,,
\end{equation}
$\Gamma_k^{(2)}$ denoting the second derivative's matrix of $\Gamma_k$ with respect to classical fields. Finally, the “prime” is defined as
\begin{equation}
X^\prime:= \frac{d X}{ds}\equiv k \frac{dX}{dk}\,.
\end{equation} 
We assume that:
\begin{equation}
({R}_k)_{\alpha,\beta}(\lambda,t-t^\prime)= \rho_{k,\alpha,\beta}(t-t^\prime) r_k(p)\,,
\end{equation}
with $p$ defined by equation \eqref{equationp}.
Furthermore, the scale function defined in \eqref{regulatordef} may help to derive the elements $ \rho_{k,\alpha,\beta}(t-t^\prime)$. It can be shows (see \cite{Duclut_2017}) that time reversal symmetry given by the field transformation \eqref{fieldtransform} imposes the relation:
\begin{equation}
\boxed{
\rho_k^{(1)}(t)-\rho_k^{(1)}(-t)+\dot{\rho}_k^{(2)}(-t)-\dot{\rho}_k^{(2)}(t)=0\,,}\label{relationTRUE}
\end{equation}
Causality furthermore imposes: $\rho_k^{(1)}(t) \propto \theta(t)$. There are many functions that match all these requirements. In this paper we follow the references \cite{Duclut_2017} and we choose:
\begin{equation}
\rho_k^{(1)}(\omega)=\alpha \frac{Z_k k}{k-i Z^{-1}_k Y_k\beta\omega}\,, \label{rho1}
\end{equation}
which corresponds to the Fourier transform of $\rho_k^{(1)}(t)$.
From \eqref{relationTRUE} we get:
\begin{equation}
\rho_k^{(2)}(\omega)=-\alpha \frac{Y_k k \beta}{k^2+Z^{-2}_k Y_k^2\beta^2\omega^2}\,.
\end{equation}
The role played by parameters $Z_k$ and $Y_k$ will be clarified in the next subsections. The parameters $\alpha$ and $\beta$ allow us to investigate the dependency of the results on the choice of the regulator and to address the reliability issue of our results from the \textit{minimal sensitivity principle} (MSP). In particular, for $\beta=0$, we have no coarse-graining in frequency. 
\medskip

\paragraph{Regulator along eigenvalues.} The regulator $r_k(p)$ on the eigenvalues must be chosen with more care. We impose that, in the large $N$ limit, a sum like:
\begin{equation}
S[f]:=\frac{1}{N}\sum_p \, r_k(p) f(p) \,,
\end{equation}
converges on the interval $ \chi_\mathcal{D}:=[0,\max(\mathcal{D})-\min(\mathcal{D})]$:
\begin{equation}
S[f] \to \int_{\chi_\mathcal{D}} dp\,\mu(p) \bar{r}_k(p) f(p)\,,
\end{equation}
where $\bar{r}_k(p)$ is a continuous function, which vanishes outside $\chi_\mathcal{D}$. We assume the following scaling relations:
\begin{equation}
p=k x\,, \qquad \sigma=k \bar{\sigma}\,, \label{scaling1}
\end{equation}
and: $\lambda=k \bar{\lambda}$ such that both for Wigner and Wishart matrices:
\begin{equation}
\mu(\lambda)d \lambda \equiv \bar{\mu}(\bar{\lambda})d \bar{\lambda} \,,
\end{equation}
where the distribution $\bar{\mu}$ expresses only in terms of $\bar{\lambda}$ and $\bar{\sigma}$. Explicitly, the regulator $r_k(p)$ can be constructed from any ‘‘good'' regulator $g_k$ considered in the literature, as follows. Let $h(x)$ be a smooth enough interpolation between $\chi_\mathcal{D}$ and $[0, \infty[$. We are aiming to define $r_k$ as the composition function $\bar{r}_k\propto g_k\circ h$. In this paper, we mainly focus on the Litim's regulator \cite{Litim_2001}, and we define $\bar{r}_k(p)=k \tau(x)$, with $p:=kx$ and:
\begin{equation}
\boxed{\tau(x)=\frac{x}{{h(x)}} \left(1-{h(x)}\right)\theta(1-{h(x)})\,.}\label{eqtau}
\end{equation}
Many explicit choices of functions $h$ should be considered. In this paper we will especially consider the function:
\begin{equation}
h(x)=\frac{x}{1-\frac{x}{\vert\bar{\chi}_\mathcal{D} \vert}}\,\label{eqh}
\end{equation}
but other functions can be considered, for instance:
\begin{equation}
h(x)=\frac{2 \vert\bar{\chi}_\mathcal{D} \vert }{\pi} \tan \left(\frac{\pi x}{2 \vert\bar{\chi}_\mathcal{D} \vert} \right)\,,
\end{equation}
where in both cases: $\vert\bar{\chi}_\mathcal{D} \vert:=\max(\mathcal{\bar{D}})-\min(\mathcal{\bar{D}})$. In both cases, for $x \ll 1$, $h(x) \sim x$. With the first regulator, the upper boundary fixed by $1-h(x)=0$ is:
\begin{equation}
x_{\text{max}}=\frac{1}{1+\frac{1}{ \vert\bar{\chi}_\mathcal{D} \vert}}\,.
\end{equation}
For the second choice:
\begin{equation}
x_{\text{max}}=\frac{2 \vert\bar{\chi}_\mathcal{D} \vert}{\pi} \arctan \left(\frac{\pi}{2 \vert\bar{\chi}_\mathcal{D} \vert} \right)\,.
\end{equation}
Hence:
\begin{enumerate}
\item For $k\to \infty$, $x_{\text{max}}\to \vert\bar{\chi}_\mathcal{D} \vert$ in both cases, as required. 

\item For $k\to 0$, we have $x_{\text{max}}\to 1$ for the two functions. 
\end{enumerate}

\noindent
Finally, let us note that the scaling \eqref{scaling1} implies the trivial flow equation:
\begin{equation}
\boxed{\frac{d\bar{\sigma}}{ds} =-\bar{\sigma}\,.} \label{eqflowsigma}
\end{equation}
\medskip

The flow equation \eqref{Wetterich} is exact but hard to solve in practice, and approximations are required to understand the nonperturbative behavior of the RG. Approximations generally look like truncation of the infinite-dimensional functional space where the flow for $\Gamma_k$ takes place, and we consider only projection along the subspace spanned by the parameters of the truncation. In this paper, we mainly focus on the so-called \textit{local potential approximation} (LPA), and we assume the following ansatz for $\Gamma_k$:
\begin{equation}
\Gamma_k[\Xi]=\int dt \sum_\lambda\, \left( Y_k \frac{\bar{\varpi}^2_\lambda}{2}+i \bar{\varpi}_\lambda \left(Y_k \dot{M}_\lambda+Z_k(\lambda-\min (\mathcal{D})) M_\lambda+U^\prime_k(M^2) M_\lambda\right) \right)\,,\label{ansatztruncation}
\end{equation}
where $\Xi=:(\varpi,M)$ and the function $U^\prime_k(M)$ is assumed to be an $O(N)$-invariant. The potential $U_k(M)$ characterizes the equilibrium distribution, $P_{\text{eq}} \sim e^{-2\Gamma_{k,\text{eq}}}$, with:
\begin{equation}
\Gamma_{k,\text{eq}}[M]:=\frac{1}{2}\left[\sum_{\lambda} Z_\lambda (\lambda-\min (\mathcal{D})) M_\lambda^2 +N U_k(M^2)\right]\,,
\end{equation}
and we have again: $M^2:=\sum_\lambda M_\lambda^2$. Note that the truncation \eqref{ansatztruncation} ensures that time reversal symmetry \eqref{fieldtransform} holds. In particular, the truncation imposes that the response field $\varpi$ does not propagate:
\begin{equation}
\boxed{G_{k,\bar{\varpi}\bar{\varpi}}=0\,,}\label{conditionproparesponse}
\end{equation}
where $G_{k,\bar{\varpi}\bar{\varpi}}$ is the ‘‘$\bar{\varpi}\bar{\varpi}$'' of the propagator $\mathbf{G}_k$. This result can be established nonperturbatively for the partition function \eqref{expressionZ}, see Appendix \ref{appendix3} and reference \cite{Aron_2010}.

\section{Vertex and effective vertex expansion}\label{vertexexp}
\subsection{Vertex expansion}
\begin{figure}
\begin{center}
\includegraphics[scale=1]{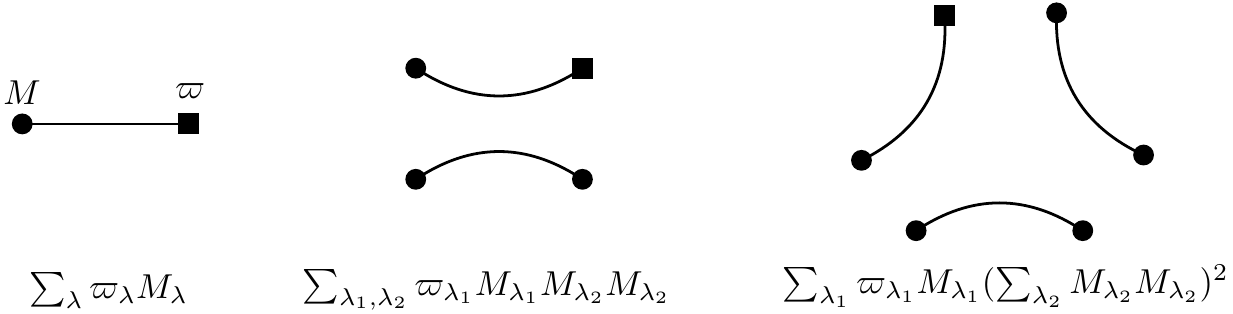}
\end{center}
\caption{Graphical rules for interaction vertices.}\label{figdiag}
\end{figure}

\begin{figure}
\begin{center}
\includegraphics[scale=0.8]{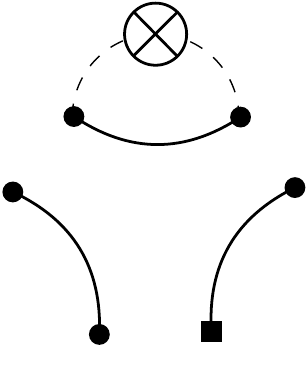}\qquad\qquad \includegraphics[scale=0.8]{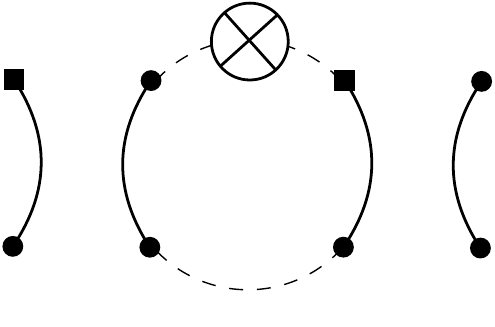}
\end{center}
\caption{Relevant diagrams for the flow equation for $\mu_2$.}\label{relevantdiag}
\end{figure}

The vertex expansion assumes that $U^\prime_k(M^2)$ looks like a series in power of $M^2$ with finite radius of convergence, namely:
\begin{equation}
U^\prime_k(M^2)=\mu_1(k) M^2+\frac{\mu_2(k)}{N} (M^2)^2+\frac{\mu_3(k)}{N^2} (M^2)^3+\cdots\,,\label{expansiontruncation}
\end{equation}
and the projection of the RG flow corresponds to be $\beta$-functions for each coupling $\mu_n$. For the computation, we introduce the graphical notation summarized in Figure \ref{figdiag}.
The vertex has two kinds of nodes, black circles and black squares, corresponding respectively to fields $M$ and $\varpi$, the solid edges materializing the sums over $\lambda$. Using this graphical convention, the flow equation for $\Gamma^{(2)}_k$ (the component ‘‘$\varpi M$'' of the matrix of second derivatives $\Gamma_k^{(2)}$) reads:
\begin{equation}
\frac{d}{ds}\Gamma_{k,\varpi M}^{(2)}(\lambda,\omega)\delta_{\lambda\lambda^\prime}=-\frac{1}{2} \, \left(\, \vcenter{\hbox{\includegraphics[scale=0.8]{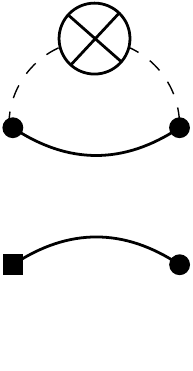}}} \quad + \quad \vcenter{\hbox{\includegraphics[scale=0.8]{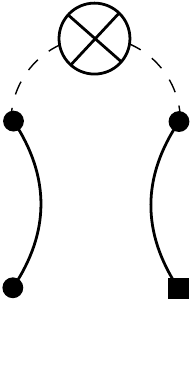}}}\,\right)\,,\label{flow2points}
\end{equation}
where dotted edges materialize the effective propagator and the crossed disc materializes the regulator. Accordingly, with the large $N$ expansion discussed in section \ref{largeN}, only the first term is relevant in the large $N$ limit because it creates one closed face, whereas the second contribution involves only an open cycle. Thus, the first contribution scales as $N^0$ and the second one as $1/N$. Furthermore, because the relevant contribution does not depend on the external momenta $p:=\lambda-\min(\mathcal{D})$, the derivative with respect to it vanishes. This has to be true also for the derivative with respect to the external frequency $\omega$ since we have no dependency on it on the left-hand side, hence:
\begin{equation}
\frac{d Z_k}{d k}=0\,,\qquad \frac{d Y_k}{d k}=0\,.
\end{equation}
There are two contributions to the effective loops, explicitly:
\begin{equation}
\vcenter{\hbox{\includegraphics[scale=1]{vertex1.pdf}}}\quad=\quad\vcenter{\hbox{\includegraphics[scale=1]{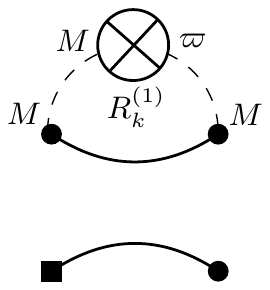}}}\,+\,\vcenter{\hbox{\includegraphics[scale=1]{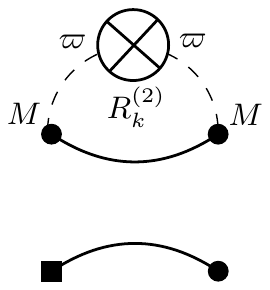}}}\,.
\end{equation}
We introduce the dimensionless variables $x:=p/k$ and $y=Z^{-1}_k Y_k \omega/k\equiv \omega/k$, as well as the dimensionless couplings $\bar{\mu}_n$ as:
\begin{equation}
\mu_n=: Z_k^n k^n \bar{\mu}_n\,,
\end{equation}
such that the flow equation for $\bar{\mu}_1$ reads:
\begin{equation}
\boxed{\beta_{\mu_1}:=\frac{d}{ds}\bar{\mu}_1=-\bar{\mu}_1-4\bar{\mu}_2 (\, \bar{I}_2+2\bar{J}_2)\,,}\label{equationmu1}
\end{equation}
dimensionless integrals $\bar{I}_2:=k I_2$ and $\bar{J}_2:=k J_2$ being defined as:
\begin{equation}
I_2=-\frac{1}{k}\int \frac{dy}{2\pi} \int dx\, \bar{\mu}(x) \frac{ \Delta_2(x,y)}{\vert -i\, y+x+\bar{\mu}_1+\bar{\rho}^{(1)}(y) \tau(x)\vert^2}\,,\label{I2}
\end{equation}
and
\begin{equation}
J_2=\frac{1}{k} \int \frac{dy}{2\pi} \int dx\, \bar{\mu}(x) \Delta_1(x,y)g_{1}(x,-y) \frac{1+\bar{\rho}^{(2)}(y) \tau(x)}{\vert -i \, y+x+\bar{\mu}_1+\bar{\rho}^{(1)}(y) \tau(x)\vert^2}\,,\label{J2}
\end{equation}
with:
\begin{equation}
g_{1}(x,y)=\frac{1}{i y+x+\bar{\mu}_1+\bar{\rho}^{(1)}(-y) \tau(x)} \,,
\end{equation}
\begin{equation}
\bar{\rho}^{(1)}(y)=\frac{\alpha}{1-i \beta y}\,, \quad 
\bar{\rho}^{(2)}(y)=-\frac{\alpha\beta}{1+\beta^2 y^2}\,.
\end{equation}
Furthermore, $\Delta_1(x,y)$ and $\Delta_2(x,y)$ are given by:
\begin{equation}
\Delta_1(x,y)=\tau(x)\bar{\rho}^{(1)}(y)- \tau(x)y\partial_y\bar{\rho}^{(1)}(y)-\bar{\rho}^{(1)}(y) x\partial_x\tau(x)\,,\label{delta1}
\end{equation}
and:
\begin{equation}
\Delta_2(x,y)=-\tau(x)y\partial_y\bar{\rho}^{(2)}(y)-\bar{\rho}^{(2)}(y) x\partial_x\tau(x)\,.\label{delta2}
\end{equation}
Flow equations for higher couplings can be derived with the same strategy, relevant diagrams involved in the flow equations for $\mu_2$ in Figure \ref{relevantdiag}, and we get:
\begin{equation}
\boxed{\beta_{\mu_2}:=\frac{d\bar{\mu}_2}{ds}=-2\bar{\mu}_2-3 \bar{\mu}_3 \, ( \bar{I}_2+2\bar{J}_2)-8 (\bar{\mu}_2)^2\,(\bar{I}_3+3\bar{J}_3)\,,}\label{equationmu2}
\end{equation}
where:
\begin{equation}
\bar{I}_3=\int \frac{dy}{2\pi} \int dx\, \bar{\mu}(x) \Delta_2(x,y)\,g_{1}^2(x,y)g_{1}(x,-y)\,,
\end{equation}
\begin{equation}
\bar{J}_3= -\int \frac{dy}{2\pi} \int dx\, \bar{\mu}(x) \Delta_1(x,y)\vert g_{1}(x,-y)\vert^2 \frac{1+\bar{\rho}^{(2)}(y) \tau(x)}{\vert -i \, y+x+\bar{\mu}_1+\bar{\rho}^{(1)}(y) \tau(x)\vert^2}\,.
\end{equation}

\begin{figure}
\begin{center}
\includegraphics[scale=0.36]{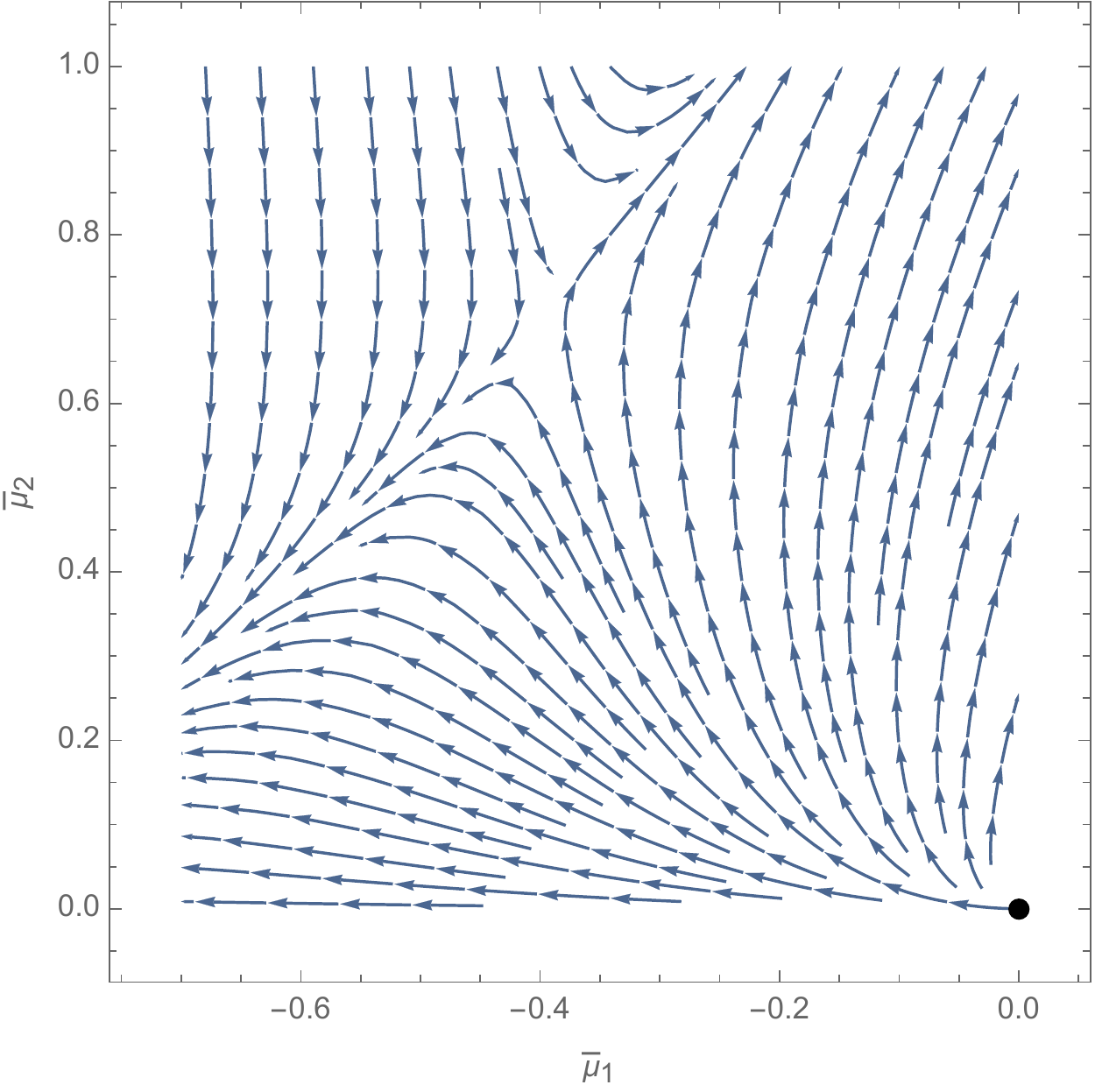}\quad \includegraphics[scale=0.36]{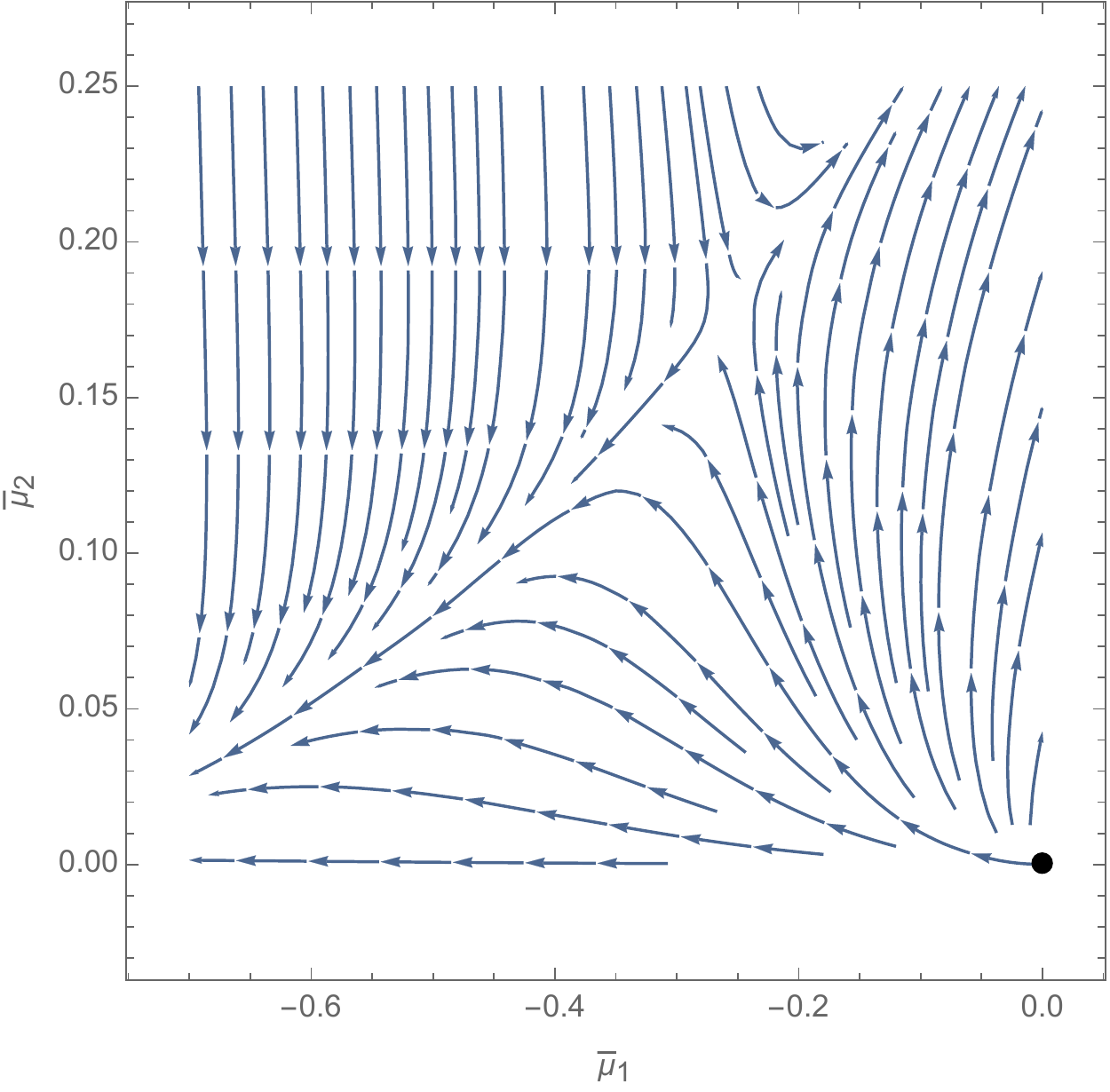}\quad \includegraphics[scale=0.36]{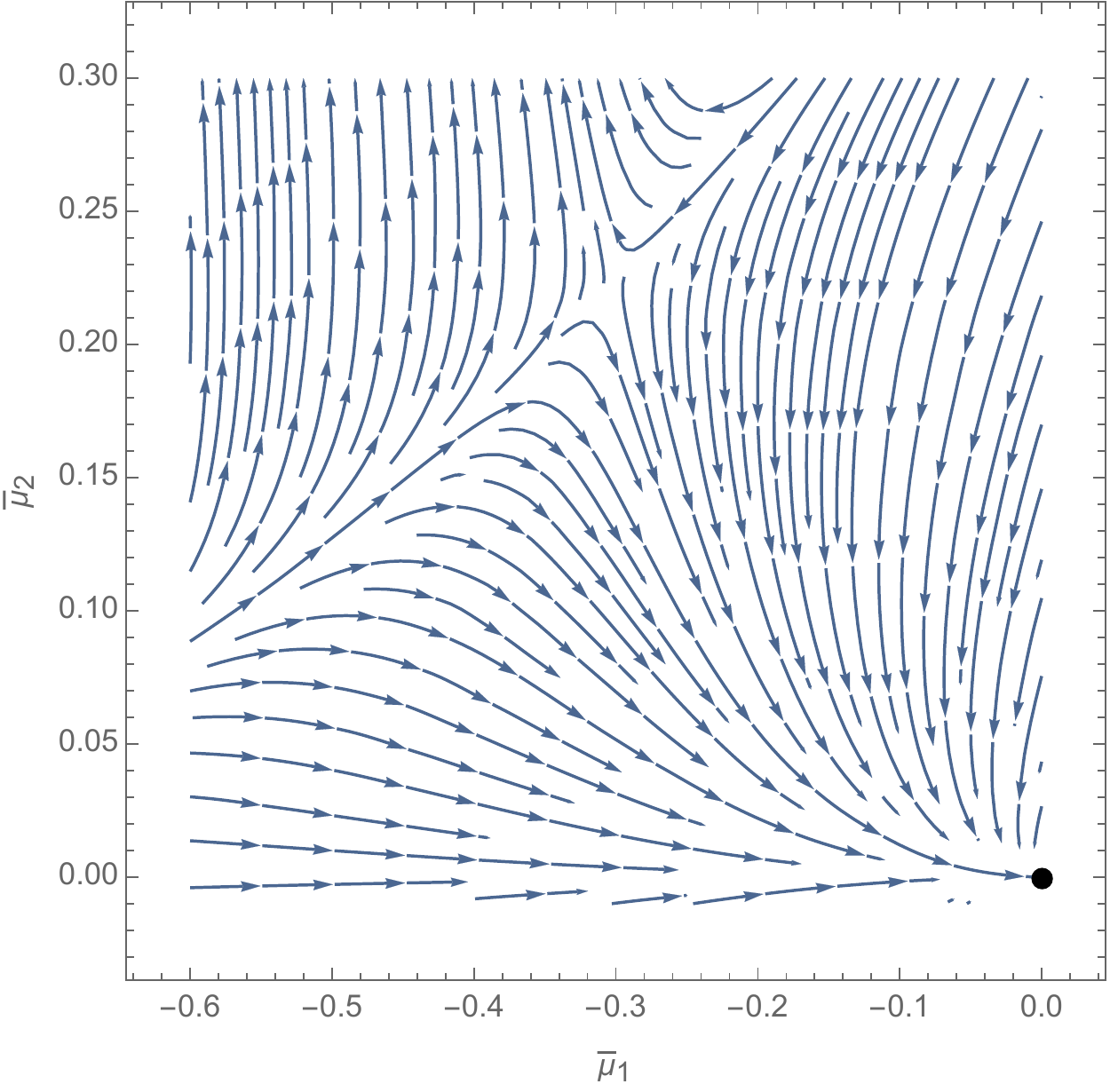}
\end{center}
\caption{Flow diagrams for fixed $\bar{\sigma}$. On the left for Wigner distribution and on the middle for MP with $c=4$ and on the right for $c=1$.}\label{figdiagramflow}
\end{figure}

\begin{figure}
\begin{center}
\includegraphics[scale=0.3]{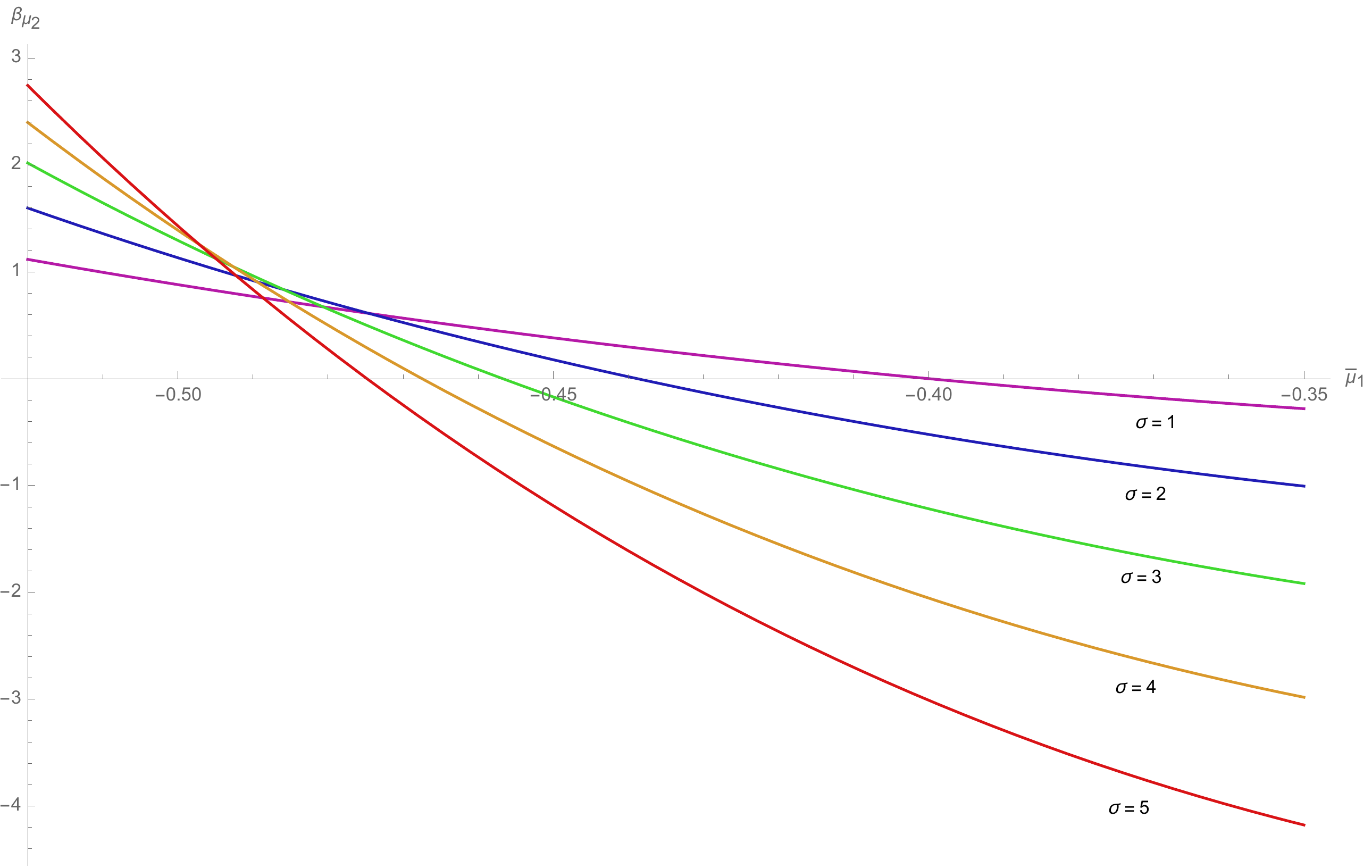}
\end{center}
\caption{The value of the mass $\mu_1$ at the fixed point for different values of the parameter $\sigma$. The figure shows that the spacing between two fixed point solution decreases as $\sigma$ increases.}\label{figdependencesigma}
\end{figure}

The resulting equations can be investigated numerically, and we mainly focus on the Wigner distribution. Figure \ref{figdiagramflow} shows the behavior of the RG flow keeping $\bar{\sigma}$ fixed, for Wigner and MP distributions and for a quartic truncation. Note that in the absence of a global fixed point, we have no fix structure to address the issue of the convergence of the vertex expansion in that case. We however provide an argument in the second section. The figures show the existence of a fixed point of Wilson-Fisher type, reminiscent of a second order phase transition. We have however to keep in mind that $\bar{\sigma}$ has its own flow equation, and change for each scale. Hence, the Figure \ref{figdiagramflow} provides an instantaneous view of the RG stream, and the position of the fixed point changes for each $s$. The dependence of this fixed point on $\bar{\sigma}$ is shows on Figure \ref{figdependencesigma} for the Wigner distribution, where the value for $\mu_1$ at the fixed point is given by the zero of the function $\beta_{\mu_2}$ where we replaced the solution $\mu_2(\mu_1)$ given by the condition $\beta_{\mu_1}=0$. Note that numerically, we find the same kind of diagram as on Figure \ref{figdiagramflow} for the MP distribution with $c=1$, suggesting a phase transition scenario as well in that case. However, from remark \ref{remarkdim}, for $c=1$, the large scale distributions for momenta corresponds to a one dimensional field theory, for which no continuous phase transition is expected. This observation illustrates the failure of the vertex expansion, and the effective potential formalism discussed in the next section shows that symmetry is restored for MP law with $c=1$, as expected.
\medskip

\begin{figure}
\begin{center}
\includegraphics[scale=0.55]{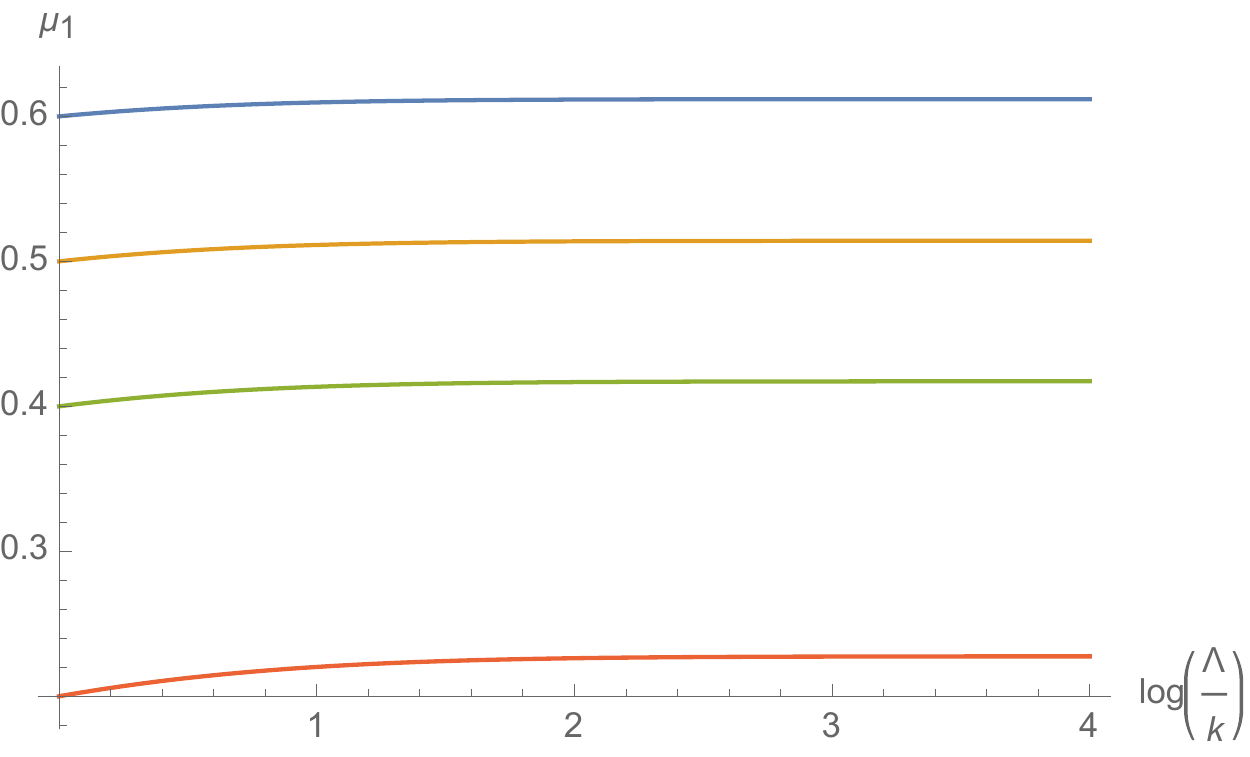}\quad\includegraphics[scale=0.55]{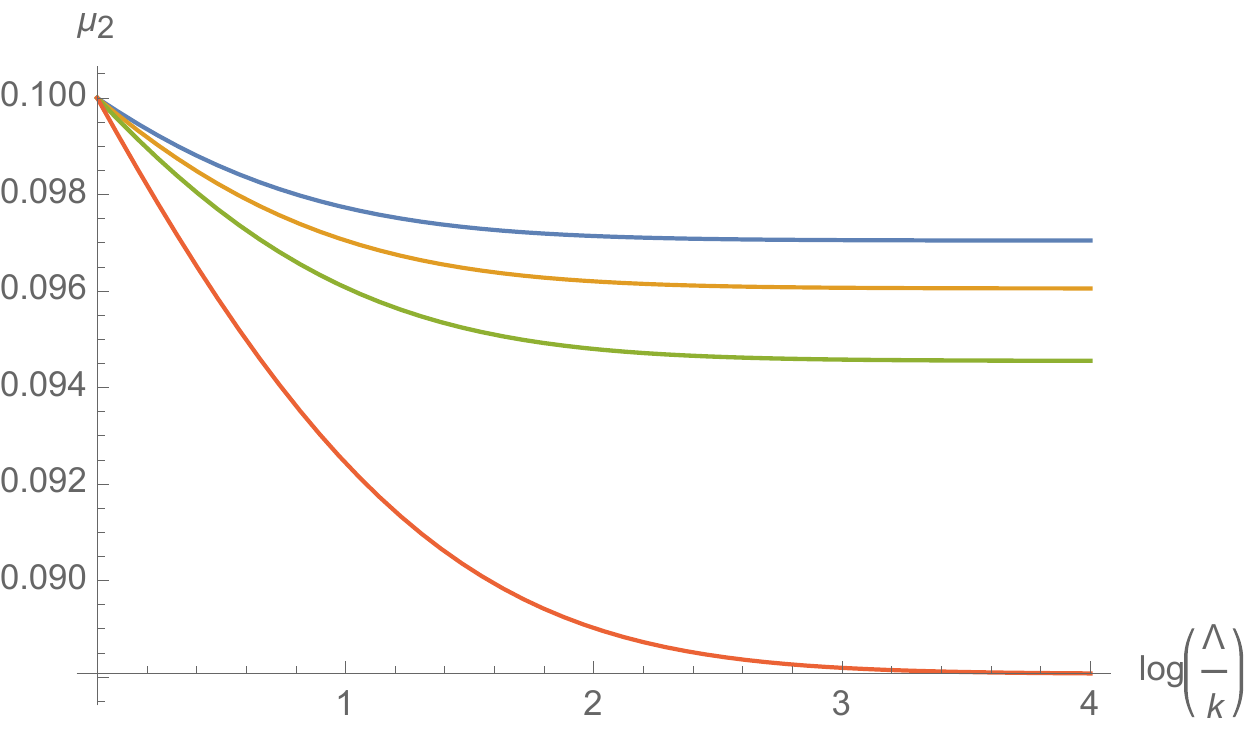}
\end{center}
\caption{Evolution of mass and quartic coupling for several positive initial conditions for the mass.}\label{figdiagramflow2}
\end{figure}

\begin{figure}
\begin{center}
\includegraphics[scale=0.55]{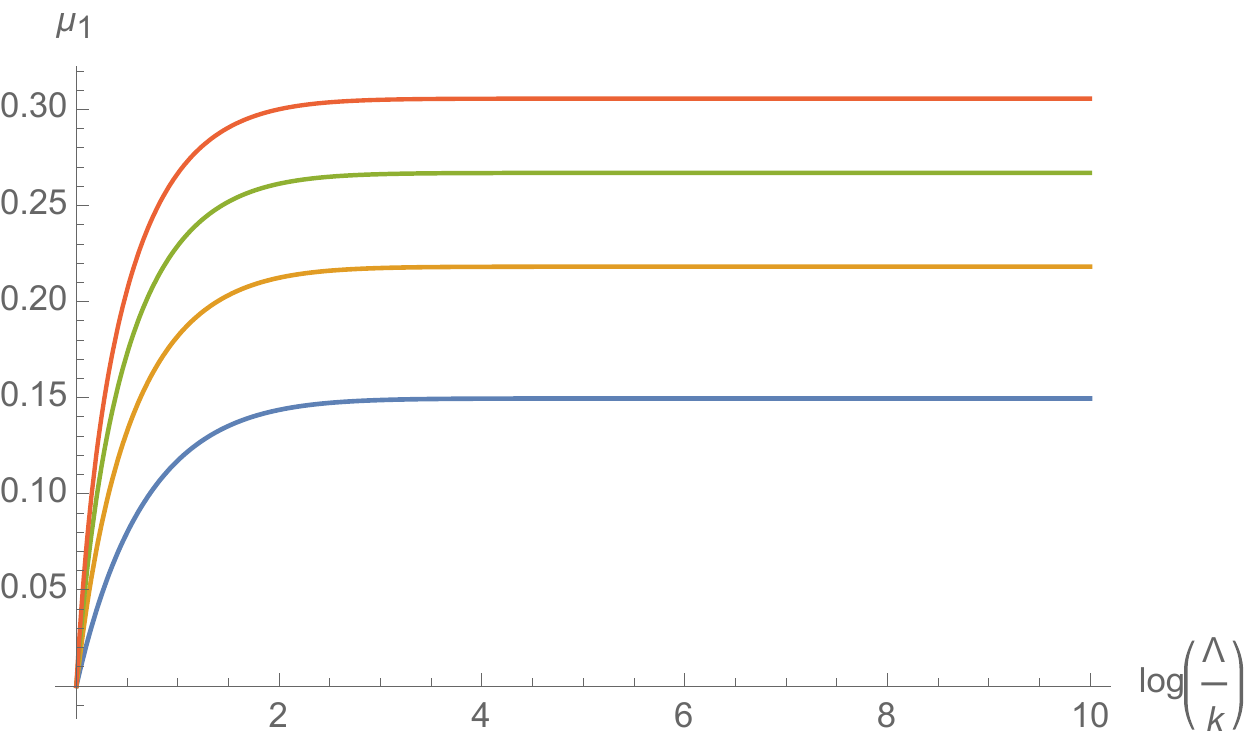}\quad\includegraphics[scale=0.55]{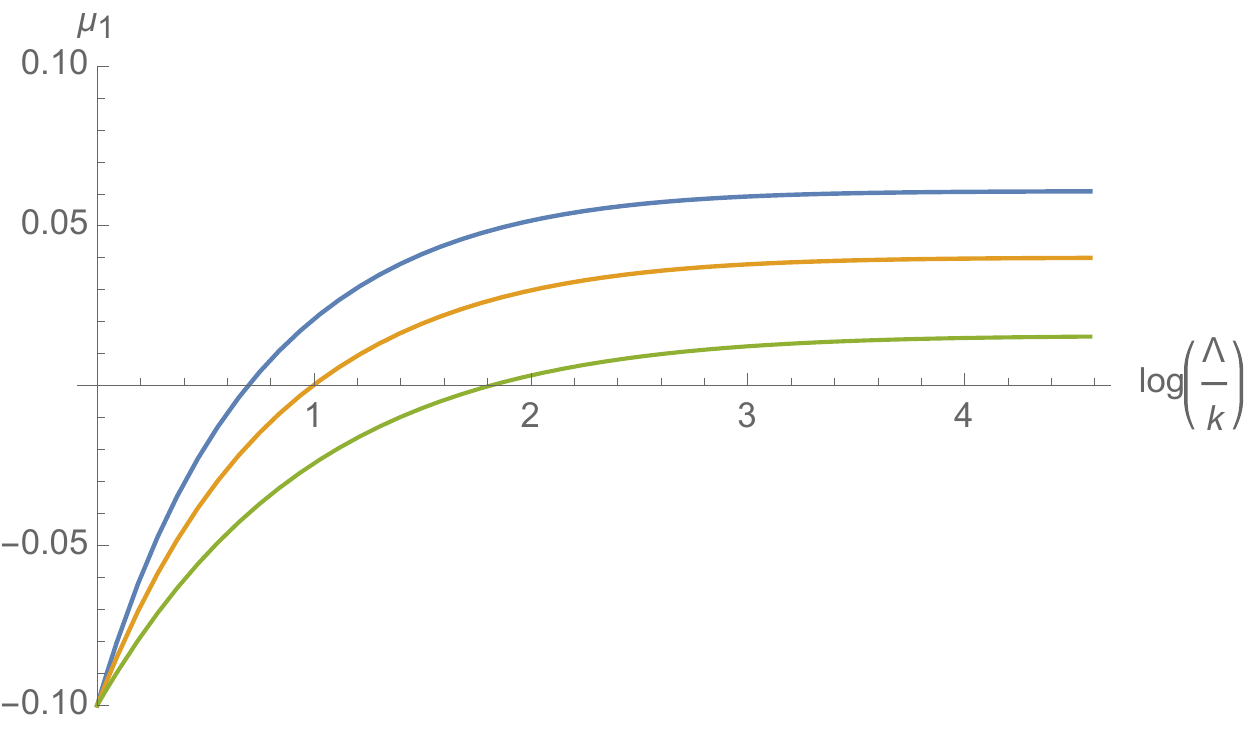}
\end{center}
\caption{On left, the evolution of $\mu_1$ for different initial conditions for $\mu_2$, taking $\mu_1=0$ as initial condition. On the right, evolution of $\mu_1$ for a negative initial conditions and different values of the quartic coupling.}\label{figdiagramflow3}
\end{figure}

\begin{figure}
\begin{center}
\includegraphics[scale=0.55]{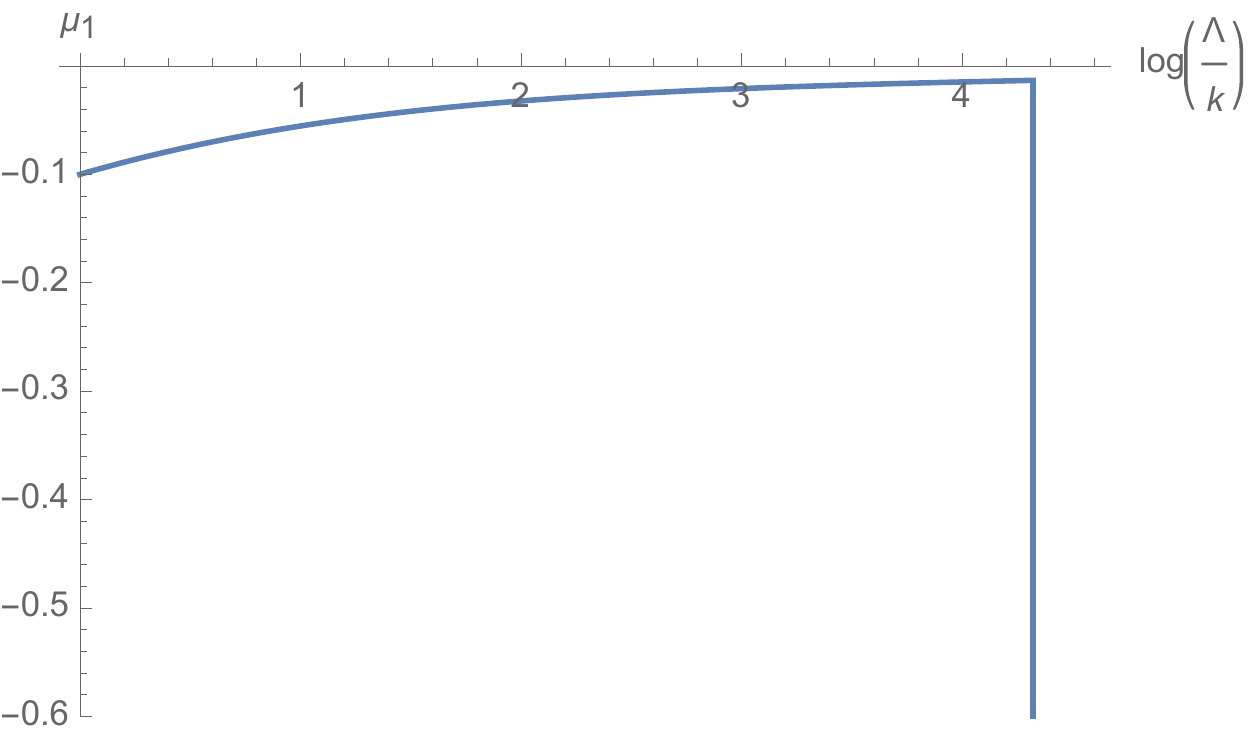}\quad\includegraphics[scale=0.55]{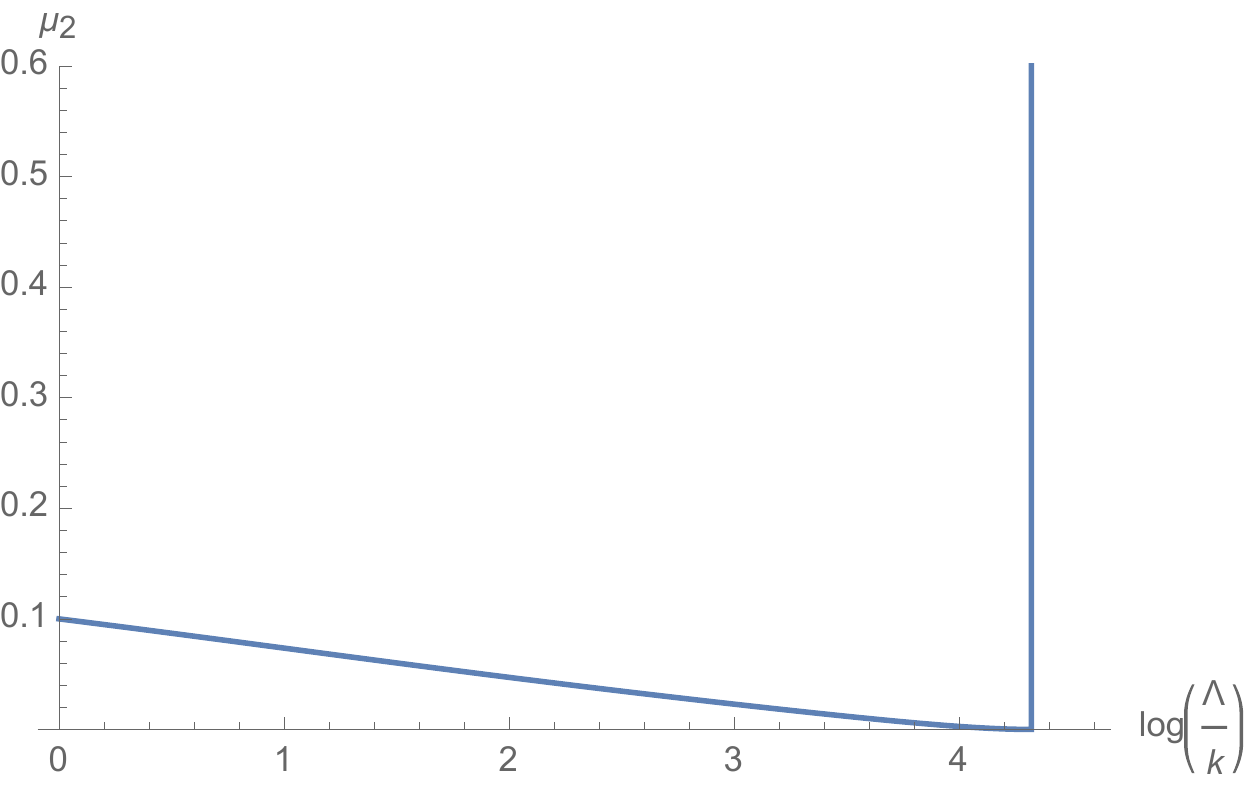}
\end{center}
\caption{Finite time singularity for mass and coupling below the critical tenperature.}\label{figdiagramflow4}
\end{figure}

The global behavior of the flow, taking into account the dependency of $\bar{\sigma}$ on the scale $s$ can be investigated as well, and the result are show on Figures \ref{figdiagramflow2}, \ref{figdiagramflow3} and \ref{figdiagramflow4} for the Wigner law and for a quartic truncation. On Figure \ref{figdiagramflow2} one can show the evolution of $\mu_1$ and $\mu_2$ in the equilibrium phase, for a positive initial value for $\mu_1$. The results show a typical convergent behavior toward a purely scaling regime, where mass and couplings are essentially constants. Figure \ref{figdiagramflow3} on the left shows the behavior of $\mu_1$ for different initial conditions for $\mu_2$, taking $\mu_1=0$ initially, showing that the end value for $\mu_1$ is all the greater as the initial value of $\mu_2$ is large. On the right, the Figure shows that the convergence phenomena holds for negatives initial conditions for $\mu_1$, as long as $\mu_2$ is large enough. But as the end value equal zero, equilibrium is dynamically broken, and the RG flow becomes singular, as Figure \ref{figdiagramflow4} shows. These observations agree with the conclusions of analytical results of Appendix \ref{appendix2} and qualitative discussion of the section \ref{Closed} We will recover in fact the same phenomena for all the approaches discussed in this paper. Regarding higher truncation, their conclusions are essentially the same qualitatively, but as recalled before, it is difficult to investigate the convergence of the vertex expansion in the absence of a global invariant structure as a nonperturbative fixed point.

\subsection{Effective vertex expansion} 

For 1D systems, the vertex expansion could be of limited interest because all the couplings $\{\mu_n\}$ are relevant regarding the canonical dimension. This failure of the vertex expansion for low dimensional field theories has been pointed out by many authors \cite{synatschke2009flow,lahoche2020revisited,eichhorn2014towards,defenu2015truncation,keitel2012zero}. Our aim in this section is to investigate the reliability of the vertex expansion exploiting the structure of leading order Schwinger-Dyson equations, a method called effective vertex expansion (EVE) in a recent series of papers in \cite{lahoche2018nonperturbative,lahoche2021no} in the context of quantum gravity. In this section, we focus on the quartic theory, even if the method can be extended for higher interactions \cite{lahoche2019ward}. We consider two different approximation schemes. In the first one, we exploit the relations between sixtic and quartic interactions to close the hierarchy. In the second one, we use the closed equation \eqref{closedSigma1} for the $2$-point, as considered in section \ref{largeN}. The reader might find it surprising or confusing that we consider two different approaches when the second one already offers a priori the complete solution. But we have seen in the section \ref{largeN} that the closed equation has no solution below the critical temperature. 
\medskip

\begin{figure}
\begin{center}
\includegraphics[scale=0.8]{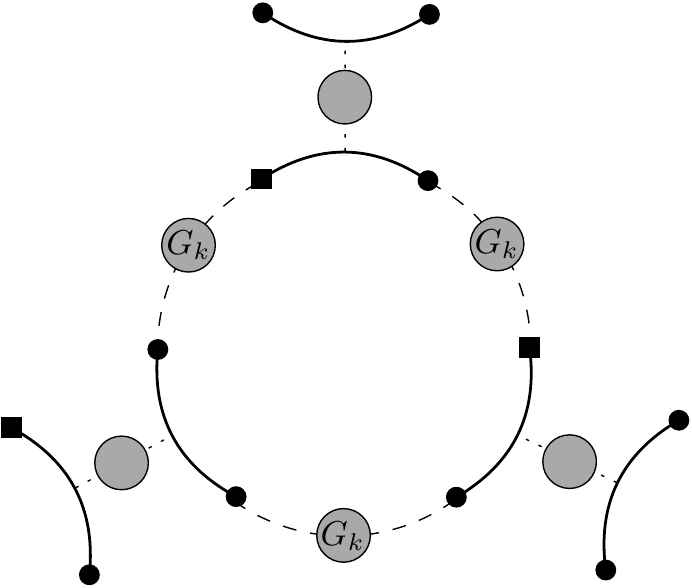}
\end{center}
\caption{The effective vertex,  $\Gamma_k^{(6)}$ expressed in terms of $\Gamma^{(4)}_k$ and $G_k$.}\label{contributionphi64}
\end{figure}

\paragraph{Closing hierarchy around local $4$-point function.} The leading order resumed series for $\Gamma_k^{(6)}$ takes the form pictured on Figure \ref{contributionphi64} where we materialized the effective $4$-point vertex accordingly with \eqref{sum4pointsbis} -- details about the derivation could be found in Appendix \ref{EVEAdendeum}. We furthermore pictured the effective $2$-point function as dotted edges, with a gray disc materializing the formal sum of the perturbation series as well. The effective $6$-point coupling express therefore as follows:
\begin{equation}
\bar{\mu}_3(k)=3\bar{\mu}_2^3 \bar{K}_3\,,
\end{equation}
where the loop integral $\bar{K}_3$ involves exact propagators and approximations are required to compute it. Because leading order contributions come from the mass shift in the large N limit (see equation \eqref{closedSigma1}), the simpler approximation replaces the effective propagator by their truncated expressions given by vertex expansion. We thus define:
\begin{equation}
\boxed{\bar{K}_3\approx - \int \frac{dy}{2\pi} \int \bar{\mu}(x) dx \vert g_{1}(x,y)\vert^2 \frac{1+\bar{\rho}^{(2)}(y) \tau(x)}{\vert -i \, y+x+\bar{\mu}_1+\bar{\rho}^{(1)}(y) \tau(x)\vert^2} \,.}
\end{equation}
Hence, the flow equation for $\bar{\mu}_2$ (equation \eqref{equationmu2}) becomes:
\begin{align}
\nonumber\frac{d\bar{\mu}_2}{ds}&=-2\bar{\mu}_2-8 (\bar{\mu}_2)^2\,(\bar{I}_3+3\bar{J}_3)\\
&+9\bar{\mu}_2^3 \left[ \int \frac{dy}{2\pi} \int \bar{\mu}(x) dx \vert g_{1}(x,y)\vert^2 \frac{1+\bar{\rho}^{(2)}(y) \tau(x)}{\vert -i \, y+x+\bar{\mu}_1+\bar{\rho}^{(1)}(y) \tau(x)\vert^2} \right] \, ( \bar{I}_2+2\bar{J}_2)\,.\label{equationmu22}
\end{align}
 Numerical investigations lead essentially to the same conclusions as for the quartic truncation discussed in the previous subsection. Hence, we mainly focus on the numerical results on the second approach, closing the hierarchy around $2$-point function, which provides a preliminary to our 2PI investigations of section \ref{section2PI}. However, the EVE shows explicitly that vertex expansion provides reliable results and converges and converges rapidly, if we disregard the bad conclusions about the case $c=1$ for MP law, for which EVE seems to predict false results as well regarding the existence of a continuous phase transition. We will return to this aspect in the next section.

\begin{figure}
\begin{center}
\includegraphics[scale=0.6]{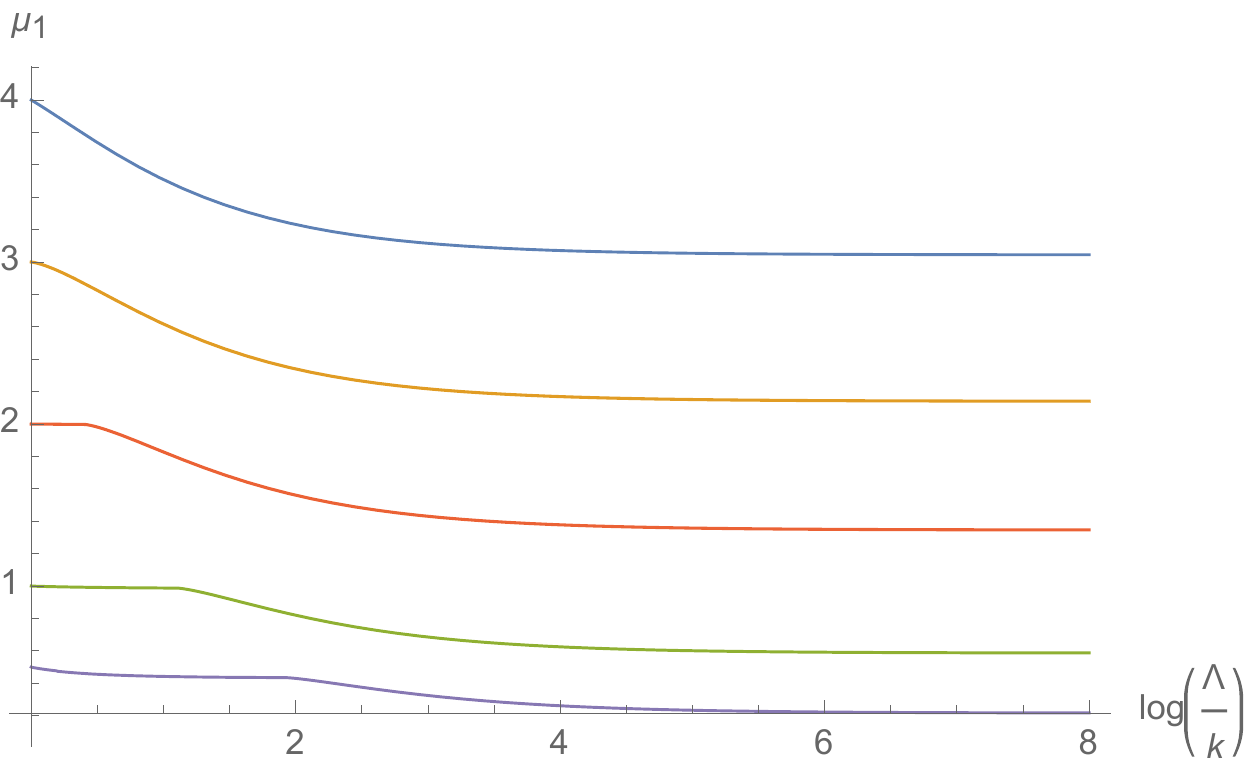}\,\includegraphics[scale=0.62]{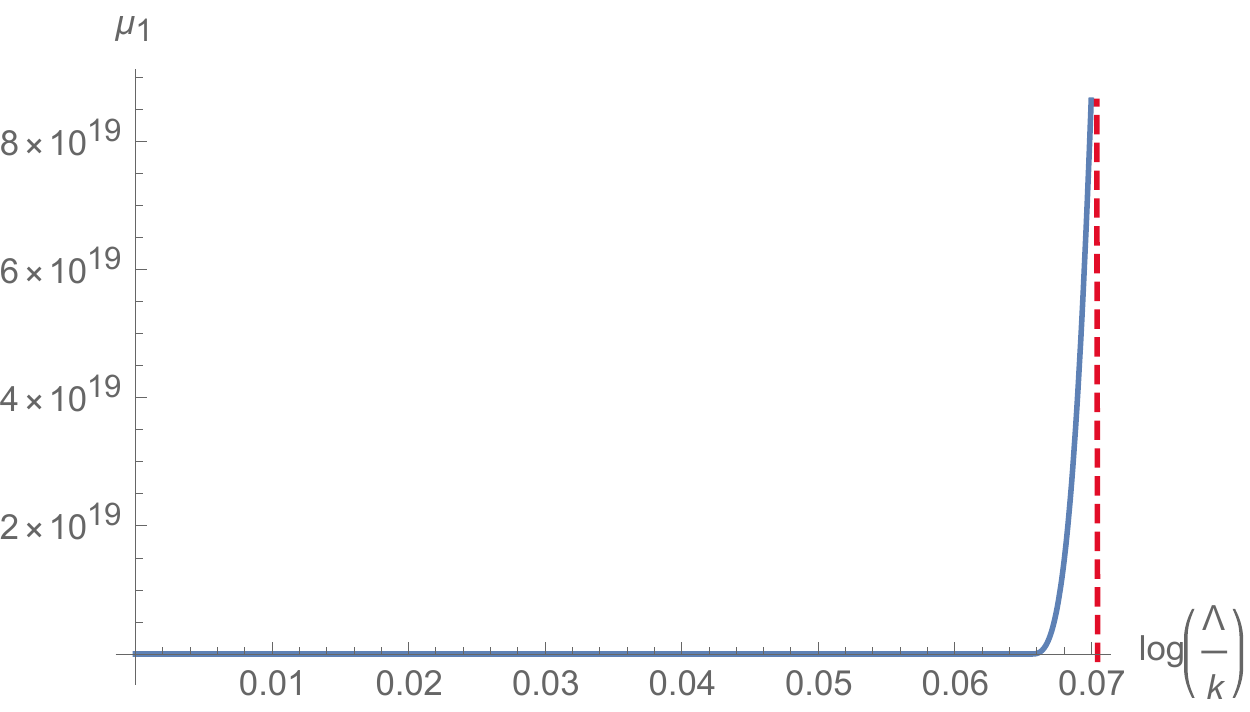}
\end{center}
\caption{Evolution of $\mu_1(s)$ in the IR regime for Wigner distribution. On left for $\bar{\mu}_1 > \bar{\mu}_1^{(c)}$. On right for $\bar{\mu}_1 < \bar{\mu}_1^{(c)}$. }\label{figflow}
\end{figure}

\paragraph{Closing hierarchy around local 2-point function.} Let us denote as $\mathbf{\Sigma}_k$ the self-energy, the index $k$ referring to the presence of the regulator in the classical action. From section \ref{largeN}, the component $\Sigma_{k,\bar{\varphi},\phi}$ is fixed by a self-consistent equation (see \eqref{closedzero}):
\begin{equation}
\Sigma_{k,\bar{\varphi} \phi}=-2\kappa_2 \, \int dp\, \mu(p)  \int \frac{d\omega}{2\pi}\frac{1+r_k(p)\rho^{(2)}_k(\omega)}{\vert -i \omega+p+\kappa_1-\Sigma_{k,\bar{\varphi} \phi}+r_k(p)\rho_k^{(1)}(\omega)\vert^2}\,.\label{zeroorderclosed}
\end{equation}
We define $R_k^{(1)}(p,\omega):=r_k(p)\rho_k^{(1)}(\omega)$, $R_k^{(2)}(p,\omega):=r_k(p)\rho_k^{(2)}(\omega)$ and we introduced the effective mass:
\begin{equation}
\mu_1(k):=\kappa_1-\Sigma_{k,\bar{\varphi} \phi}\,,
\end{equation}
such that, computing the derivative of the closed equation with respect to $k$ leads to:
\begin{equation}
\frac{d \mu_1}{ds}=-4\kappa_2\, \frac{\mathcal{I}_1(k)-\frac{1}{2}\mathcal{I}_2(k)}{1+4\kappa_2\mathcal{I}_3(k,\Lambda)}\label{dmu}
\end{equation}
where:
\begin{equation}
\mathcal{I}_1(k):=\int dp\, \mu(p)  \int \frac{d\omega}{2\pi}\, \frac{dR_k^{(1)}}{ds}(p,\omega)\vert g_{1}(p,\omega)\vert^4(\mu_1+R_k^{(1)}(p,\omega))(1+R_k^{(2)}(p,\omega)),
\end{equation}
\begin{equation}
\mathcal{I}_2(k):=\int dp\, \mu(p)  \int \frac{d\omega}{2\pi}\, \frac{dR_k^{(2)}}{ds}(p,\omega)\vert g_{1}(p,\omega)\vert^2,
\end{equation}
\begin{equation}
\mathcal{I}_3(k,\Lambda):=\int dp\, \mu(p)  \int_{-\Lambda}^\Lambda \frac{d\omega}{2\pi}\, \vert g_{1}(p,\omega)\vert^4(\mu_1+R_k^{(1)}(p,\omega))(1+R_k^{(2)}(p,\omega))\,,
\end{equation}
and we introduced some UV cut-off $\Lambda$ and we assume $\Lambda \gg k$. In that limit, we expect that $\Lambda$ contributions should be discarded, and the leading scaling relation with respect to $k$ should be simple enough. It is not hard to find the global scaling for each integral:
\begin{equation}
\mathcal{I}_1(k)=:\frac{1}{k} \bar{\mathcal{I}}_1(k)\,,\quad \mathcal{I}_2(k)=:\frac{1}{k} \bar{\mathcal{I}}_2(k) \,,\quad \mathcal{I}_3(k,\infty)=:\frac{1}{k^2} \bar{\mathcal{I}}_3(k)\,,
\end{equation}
where overlined quantities are dimensionless. Furthermore, the bare coupling $\kappa_2$ has dimension $2$, and we can define the dimensionless coupling at scale $\Lambda$ as: $\kappa_2=:\bar{\kappa}_2 \Lambda^2$. We have to distinguish two cases:
\begin{enumerate}
    \item If $\bar{\kappa}_2$ is such that $\bar{\kappa}_2 \Lambda^2=\mathcal{O}(1)$, for $k$ small enough, the left-hand side of \eqref{dmu} is essentially independent of bare coupling.
    \item If $\bar{\kappa}_2 \Lambda^2=\mathcal{O}(k^2)$, $\bar{\kappa}_2$ remains an adjustable quantity in the equation. 
\end{enumerate}
Because $\kappa_2 \propto T$ -- the temperature, it follows that the first item corresponds to a large scale state with \textit{finite temperature} as, $k \to 0$ whereas the second one has \textit{zero temperature} in the same limit. We expect the second item assumption out from our equilibrium treatment, valid for $T$ larger than the critical temperature (expected to be non-zero), and we only retain the first case. In that way, the flow equation for $\bar{\mu}_1:=\mu_1/k$ is, for $k$ small enough:
\begin{equation}
\frac{d \bar{\mu}_1}{ds}\approx -\bar{\mu}_1-\frac{\bar{\mathcal{I}}_1(k)-\frac{1}{2}\bar{\mathcal{I}}_2(k)}{\bar{\mathcal{I}}_3(k)}\label{flowclosed}
\end{equation}
\begin{remark}
Equation \eqref{4pointsclosed} implies that the effective quartic coupling is essentially independent as well of the initial bare coupling $\kappa_2$ for $k$ small enough.
\end{remark}
The equation \eqref{flowclosed} 
can be investigated numerically, but we have to remember that it is not valid in the large, $k$ i.e., in the deep UV regime. In that way, the initial condition $\bar{\mu}_1(s_0)$ is for $e^{s_0}\ll 1$, and blind integrated out UV effects. The results for the Wigner distribution are summarized in Figure \ref{figflow}. For $\bar{\mu}_1(s_0)$ large enough, $\mu_1(s)$ decreases exponentially toward a finite value. As $\bar{\mu}_1(s_0)$ decreases, a plateau arises at a finite timescale, but ultimately the coupling goes to a finite value $\mu_1(-\infty)$ for $-s_0$ large enough. But it exists in a finite value $\bar{\mu}_1
^{(c)}$ such that for $\bar{\mu}_1(s_0)<\bar{\mu}_1
^{(c)}$ the flow becomes singular at a finite timescale. The asymptotic value $\mu_1(-\infty)$ decreases as $\bar{\mu}_1(s_0)$ decreases, and the critical value $\bar{\mu}_1
^{(c)}$ is such that $\mu_1(-\infty)=0$, in agreement with our analytical statement of the section \ref{Closed}, where we showed that $\Sigma_+=0$ at the transition. 
\bigskip

\section{Effective local potential in the broken phase}\label{LPAformalism}
In this section, we outperform the previous vertex expansions and consider a truncation scheme called the effective potential method, which considers the shape of the effective potential at equilibrium as global data. We thus assume again the validity of the truncation \eqref{ansatztruncation}, that defines the so-called Local Potential Approximation (LPA) \cite{Delamotte1} where wave functions normalization $Z_k$ and $Y_k$ are assumed to be pure numbers, independent of the classical field configuration $\Xi$. More precisely, the approximation corresponding to the truncation scheme \eqref{ansatztruncation}, including the field strength $Y_k$ and $Z_k$ is called LPA$^\prime$ in the literature. But in Appendix \ref{appendix4}, we show that flow equations for $Y_k$ and $Z_k$ vanish in the large $N$ limit, and LPA$^\prime$ reduces to LPA. 
\medskip

\subsection{Field configurations,  stability, and flow equation}
In this section, we assume, $M^2 \sim N$ i.e., we work in the broken phase of the effective spherical model. The shape of $U_k(M^2)$ can be fixed by a specific expansion around some \textit{non-zero} (running) vacuum $\kappa(k)$, for instance as:
\begin{equation}
U_k(M^2)= \frac{\mu_2(k)}{2}\left(\frac{M^2}{N}-\kappa\right)^2+\frac{\mu_3(k)}{3}\left(\frac{M^2}{N}-\kappa\right)^3+\cdots\label{explicitUk}
\end{equation}
but this assumption is not required for nowadays powerful numerical methods, and arbitrary shapes for $U_k(M^2)$ can be considered. We furthermore assume that $M_\lambda$ is a uniform field in time, i.e., that $M_\lambda$ does not depend on time. However, we have to be careful regarding the $\lambda$ dependency. Indeed, if we assume that the field $M_\lambda$ describes the large timescale regime of the system, that is, for $k \ll 1$, such that the typical frequencies are small i.e., time derivatives are negligible. In this regime, $\lambda$ must be small as well, and one expects that only the component $M_{-2\sigma}$ contributes significantly. Such an approximation however blinds the non-local structure of the effective potential due to the $O(N)$ invariance. In this paper, we will consider essentially two different approximation schemes for uniform field solutions, which we call respectively \textit{scheme 1}, \textit{scheme 2}. 

\begin{itemize}
    \item In the \textit{scheme 1}, we assume to project the flow along \textit{a uniform} field configuration in the eigenspace, namely:
\begin{equation}
\boxed{M_\lambda=\sqrt{\chi}\,.} \label{scheme1}
\end{equation}
\item In the \textit{scheme 2}, that we call \textit{staggered} field configuration, we project the flow along:
\begin{equation}
\boxed{M_\lambda^2 = \chi \frac{\delta (\lambda+2\sigma)}{\mu(2\sigma)}\,.}\label{sheme2}
\end{equation}
\end{itemize}
Note that both of them assume that:
\begin{equation}
\frac{1}{N}\sum_{\lambda=1}^N M_\lambda^2\to  \int_{\mathcal{D}} d\lambda\,\mu(\lambda)  M_\lambda^2 = \chi\,,
\end{equation}
It is easy to check that for the equilibrium dynamics, $\bar{\varpi}_\lambda \equiv \langle \bar{\varphi}_\lambda \rangle$ must vanish at large scale, for $k\sim 0$ (see Appendix \ref{appendix3}). In that limit, the quantum equation of motion reads:
\begin{equation}
0\equiv \frac{\delta \Gamma_k}{\delta \bar{\varpi}_\lambda}\approx Y_k  \bar{\varpi}_\lambda+ i \mathcal{U}_k^\prime(\chi) M_\lambda\,,\label{eq1}
\end{equation}
\begin{equation}
0\equiv \frac{\delta \Gamma_k}{\delta M_\lambda}\approx i \sum_{\lambda^\prime}\bar{\varpi}_{\lambda^\prime} \left(\mathcal{U}_k^\prime(\chi)\delta_{\lambda{\lambda^\prime}}+2\frac{M_\lambda M_{\lambda^\prime}}{N}\mathcal{U}_k^{\prime\prime}(\chi) \right)\,,\label{eq2}
\end{equation}
where $\mathcal{U}_k(\chi):=U_k(M^2=N\chi)$. 
The first equation \eqref{eq1} shows that, for $M_\lambda \neq 0$, $\bar{\varpi}_\lambda \propto \mathcal{U}_k^\prime(\chi)$. Hence, the equilibrium constraint $\bar{\varpi}_\lambda=0$ must hold only for $\mathcal{U}_k^\prime(\chi)=0$, that solves again the second equation \eqref{eq2}. Hence, equilibrium requires the existence of a stable vacuum. 
\medskip

The flow equation for the effective potential $U_k^\prime$ can be derived from the Wetterich equation \eqref{Wetterich}, imposing scheme 1 or scheme 2 on both sides of the equation. Formally, it reads:
\begin{equation}
N^2\mathcal{V}\frac{d}{ds}U_k^\prime[\frac{\chi}{N}]=-i\int dt\sum_\lambda\frac{1}{M_\lambda(t)}\, \frac{d}{ds} \frac{\partial \Gamma_k}{\partial \bar{\varpi}_\lambda(t)} \equiv -\frac{1}{2}\,\vcenter{\hbox{\includegraphics[scale=1]{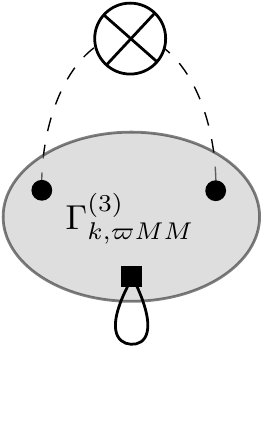}}}\,,\label{floweqpotential1}
\end{equation}
where the gray disc materializes the $3$-point vertex $\Gamma_k^{(3)}$, the solid self-loop materializes the sum and time integral over the variable of $\varpi_\lambda(t)$,  and we regularized time integrals as $\int dt =: \mathcal{V}$. To compute the left-hand side, we used the Wetterich equation, namely:
\begin{equation}
\int dt \sum_\lambda \frac{\partial}{\partial \bar{\varpi}_\lambda(t)} \frac{d}{ds} \Gamma_k=\frac{1}{2}\int dtdt^\prime dt^{\prime\prime} \sum_{\lambda, \alpha,\beta} ({R}_k^\prime)_{\alpha,\beta}(\lambda,t-t^\prime)  \frac{\partial}{\partial \bar{\varpi}_\lambda(t^{\prime\prime})}G_{k,\alpha \beta}(\lambda, t-t^\prime)\,.
\end{equation}
The $3$-points vertex can be computed exactly assuming that $M_\lambda$ does not depends on time, and we get straightforwardly in terms of the potential $\mathcal{U}_k$:
\begin{align}
\nonumber\frac{\delta^3\Gamma_k}{\delta \bar{\varpi}_\lambda(t) \delta M_{\lambda_1}(t_1)\delta M_{\lambda_2}(t_2)} =&\frac{1}{N}\Big(2\mathcal{U}_k^{\prime\prime}\Big[\delta_{\lambda \lambda_1} M_{\lambda_2}+\delta_{\lambda \lambda_2} M_{\lambda_1}+\delta_{\lambda_1 \lambda_2} M_{\lambda}     \Big]\\
&+4\frac{M_{\lambda}M_{\lambda_1}M_{\lambda_2} }{N}\,\mathcal{U}_k^{\prime\prime\prime} \Big)\delta(t-t_1)\delta(t-t_2)\,.\label{decomp3pts2}
\end{align}
The computation of the flow equation requires moreover the knowledge of the propagator. For the two approximation schemes considered in this paper, $\Gamma_k^{(2)}=A_k+B_k$, with:
\begin{equation}
(A_k)_{\lambda_1,\lambda_2}=u \delta_{\lambda_1\lambda_2}\,,\qquad (B_k)_{\lambda_1,\lambda_2}= v_1 \delta_{\lambda_1\lambda_2}+\frac{v_2}{N} e_{\lambda_1}e_{\lambda_2}\,,\label{ABnews}
\end{equation}
where $e_\lambda:=M_\lambda/\sqrt{\chi}$ and the $u$, $v_1$ and $v_2$ can be computed from the truncation \eqref{ansatztruncation}:
\begin{equation}
u=Y_k+R_k^{(2)}(\lambda,\omega)\,,\quad v_1=Y_k \omega + i Z_k \lambda +i\mathcal{U}^\prime_k+iR_k^{(1)}(\lambda,\omega)\,,\quad v_2=2i\chi \mathcal{U}^{\prime\prime}_k\,. 
\end{equation}
The inverse of the matrix $B_k$ can be straightforwardly computed by perturbation theory\footnote{It organizes explicitly as \begin{equation}
(B_k)_{\lambda_1\lambda_2}^{-1}=\frac{1}{v_1}\delta_{\lambda_1\lambda_2}-\frac{v_2}{v_1^2}\frac{e_{\lambda_1}e_{\lambda_2}}{N}+\frac{v_2^2}{v_1^3}\frac{e_{\lambda_1}e_{\lambda_2}}{N}-\cdots\,.
\end{equation}}, assuming $v_2/v_1$ small enough and using $\sum_\lambda e_\lambda^2=N$. We get:
\begin{equation}
(B_k)_{\lambda_1\lambda_2}^{-1}=\frac{1}{v_1}\delta_{\lambda_1\lambda_2}-\frac{v_2}{v_1(v_1+v_2)} \frac{e_{\lambda_1}e_{\lambda_2}}{N}\,.
\end{equation}
The inverse matrix $\textbf{G}_k$ is written as:
\begin{equation}
\textbf{G}_k= \begin{pmatrix}
0&B^{-1}(-\omega)\\
B^{-1}(\omega)& -B^{-1}(\omega) A B^{-1}(-\omega)
\end{pmatrix}\,.
\end{equation}
We introduce the following graphical convention:
\begin{equation}
\textbf{G}_k=\vcenter{\hbox{\includegraphics[scale=1]{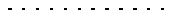}}}\,+\,\vcenter{\hbox{\includegraphics[scale=1]{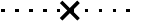}}}\,\equiv \,  \vcenter{\hbox{\includegraphics[scale=1]{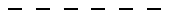}}}\,,
\end{equation}
the dotted edge materializing the connected part of the propagator ($\propto \delta_{\lambda \lambda^\prime}$) and the dotted edge, with a cross, corresponds to the disconnected part ($\propto e_\lambda e_{\lambda^\prime}$). Hence, the right-hand side of the flow equation \eqref{floweqpotential1} becomes:
\begin{align}
\vcenter{\hbox{\includegraphics[scale=0.8]{vertex3pts.pdf}}}\,=\,\vcenter{\hbox{\includegraphics[scale=0.8]{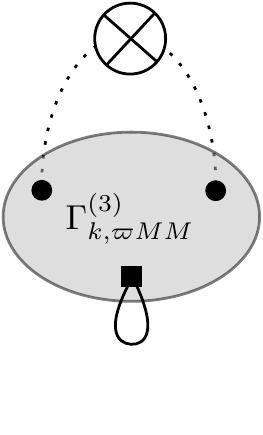}}}\,+2\times \,\vcenter{\hbox{\includegraphics[scale=0.8]{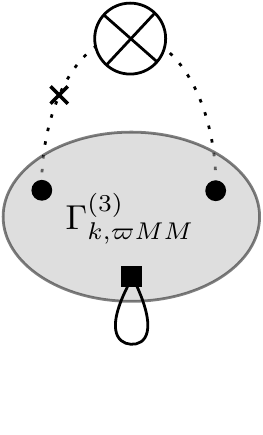}}}\,+\,\vcenter{\hbox{\includegraphics[scale=0.8]{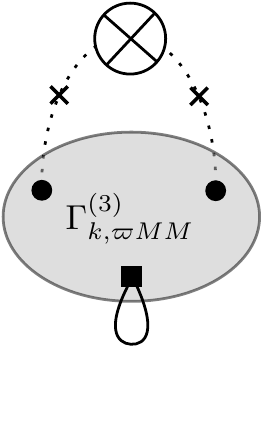}}}\label{decompositionconnexe}
\end{align}
In the large $N$ limit, only a few contractions will survive, depending on the approximation scheme that we consider. Let us consider the connected contribution, for instance. In the expansion \eqref{decomp3pts2}, only the contributions that maximize the number of faces (i.e., closed cycles) will be relevant in the large $N$ limit. For the first three terms in \eqref{decomp3pts2}, only the third one will generate a closed cycle, and the corresponding contribution in \eqref{decompositionconnexe} scale globally as $N$. The three last ones in contrast scale as $N^0$ and should be discarded. 

\subsection{RG for staggered classical field}
\begin{figure}
\begin{center}
\includegraphics[scale=0.55]{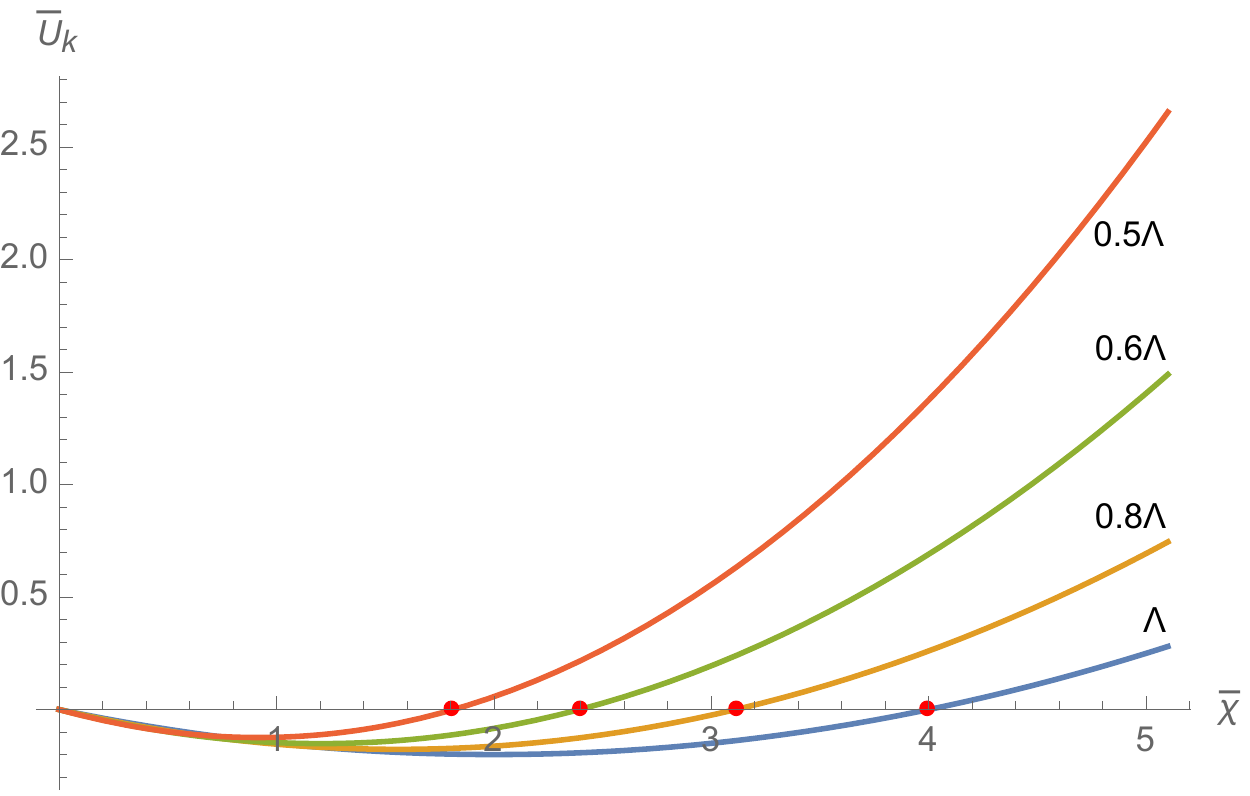}\qquad \includegraphics[scale=0.55]{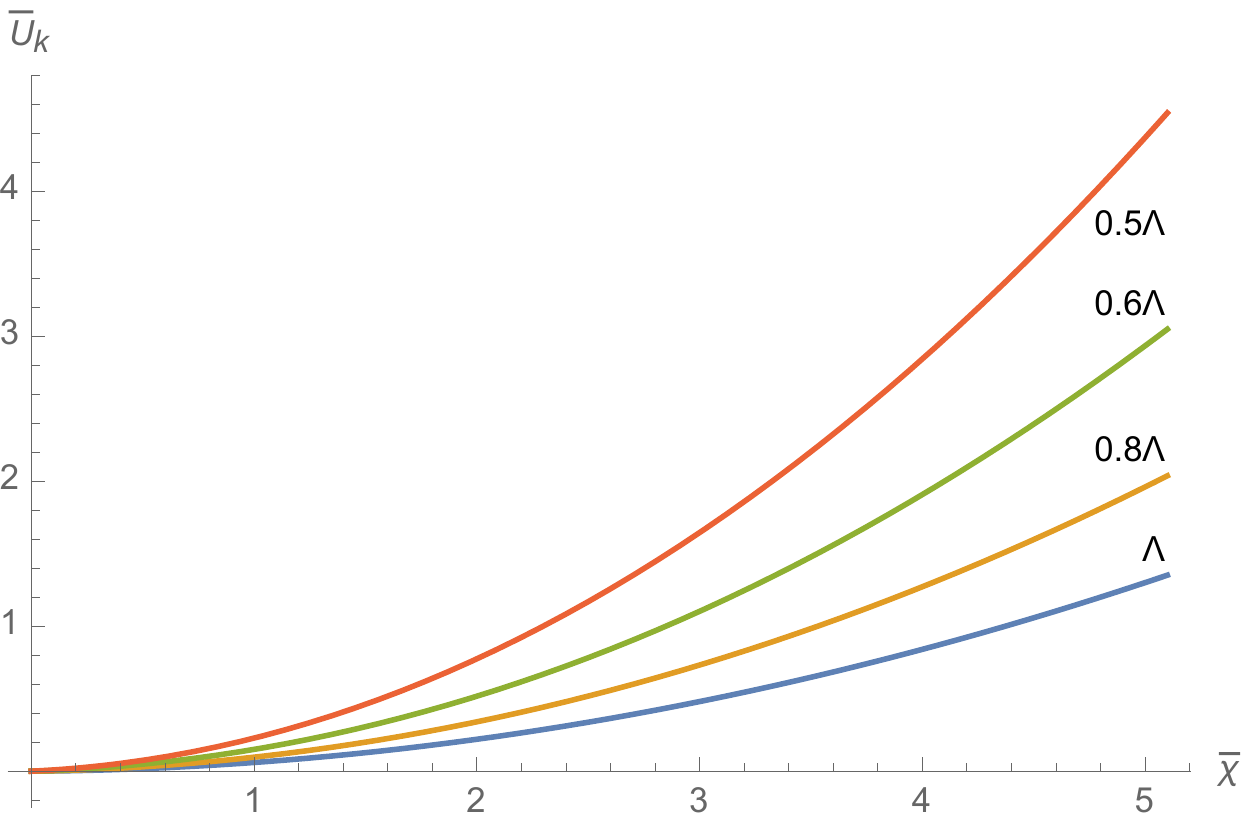}
\end{center}
\caption{Evolution of the quartic potential for staggering initial condition and Wigner distribution. On the right for a positive initial mass. On the left for a negative initial mass. }\label{figflowstaggered}
\end{figure}

\begin{figure}
\begin{center}
\includegraphics[scale=0.55]{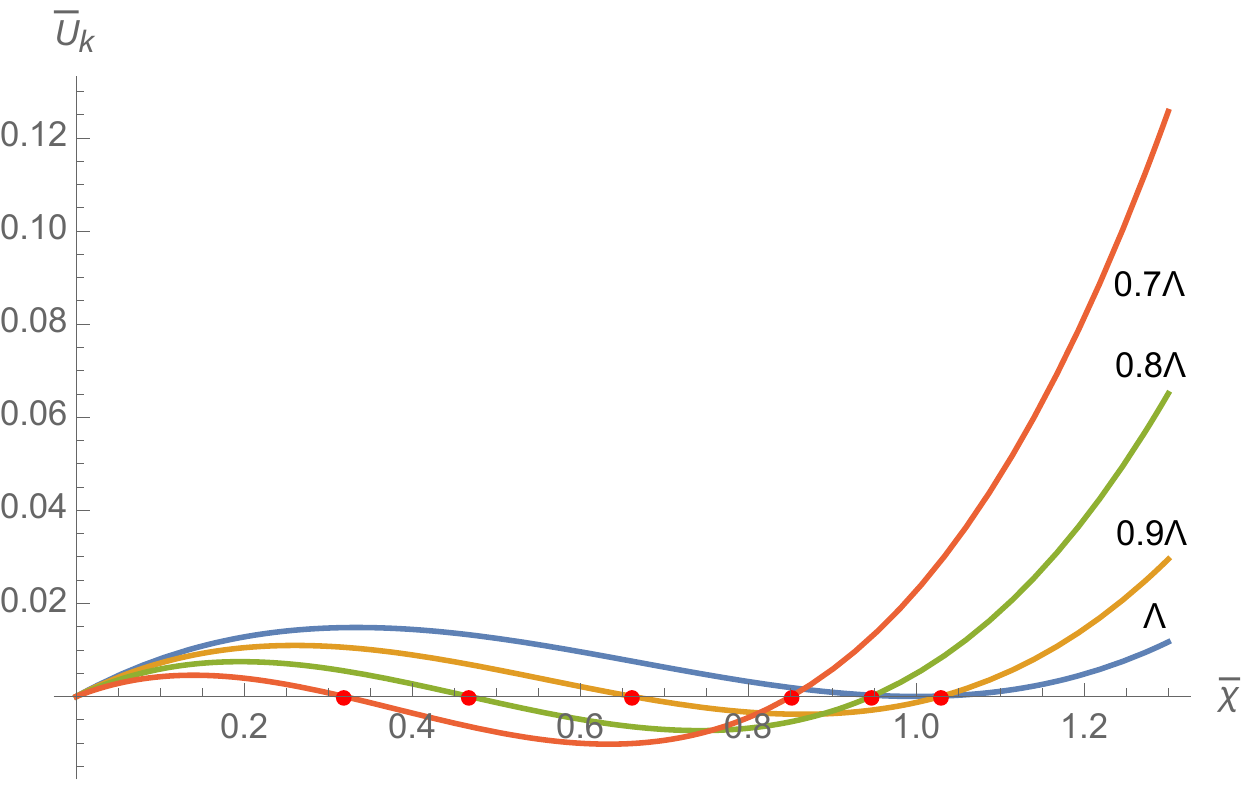}\qquad \includegraphics[scale=0.55]{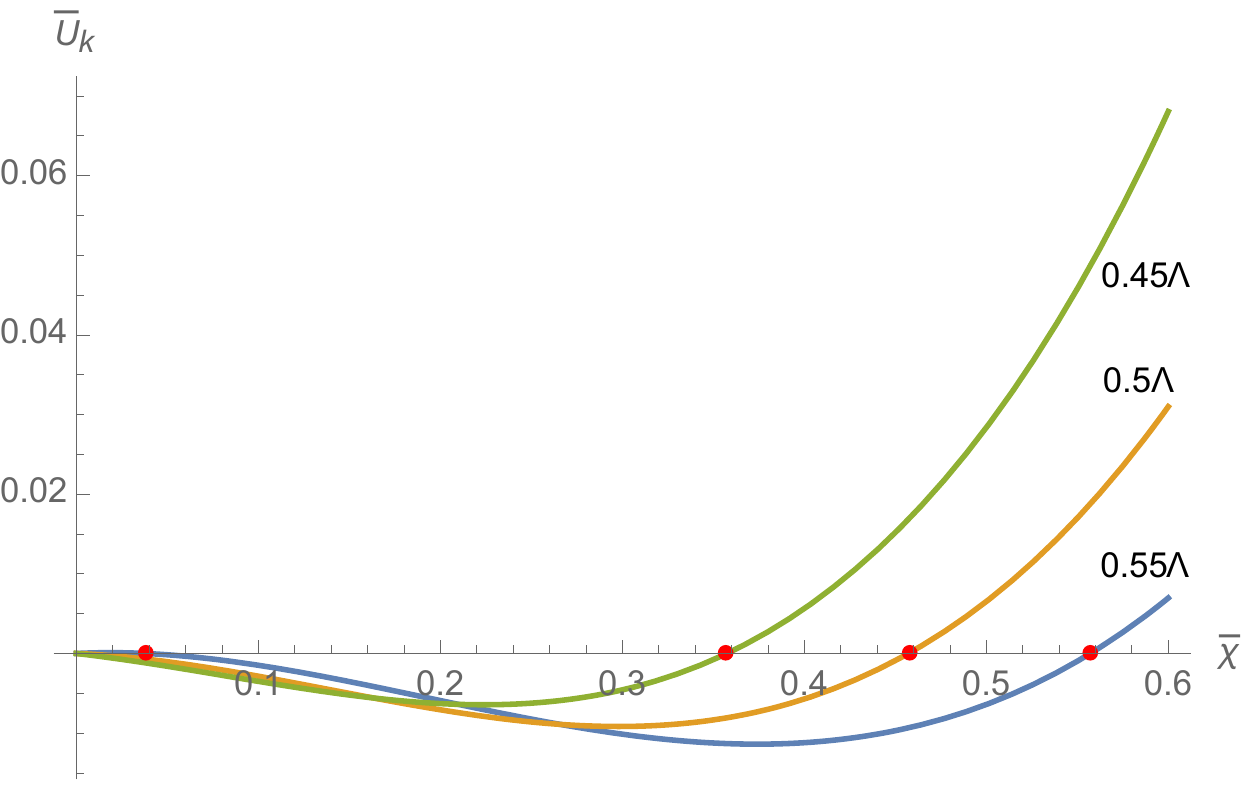} \\
\includegraphics[scale=0.55]{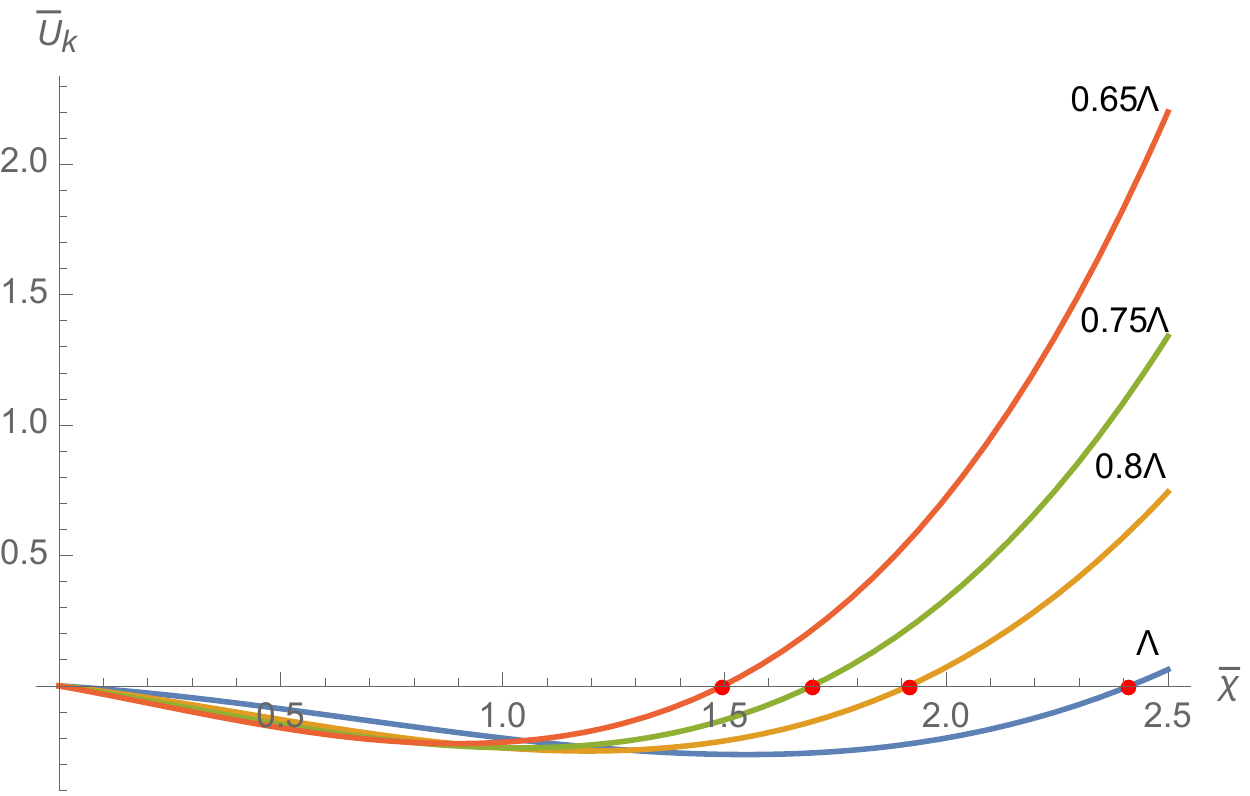}\qquad \includegraphics[scale=0.55]{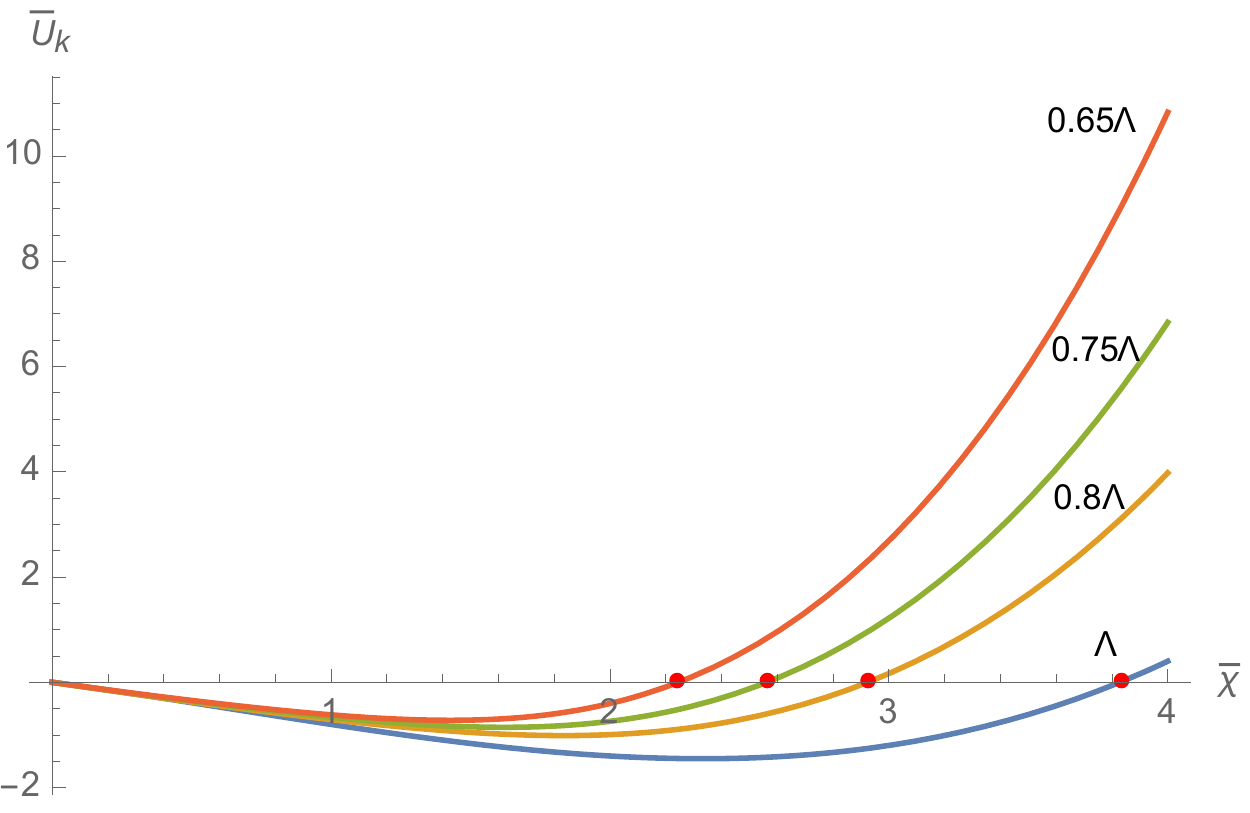}
\end{center}
\caption{Evolution of the sixtic potential for staggering initial condition and Wigner distribution. On the top for a positive initial mass. On the left for a negative initial mass. In both cases, the quartic coupling has a negative value.}\label{figflowstaggered2}
\end{figure}

Because it is simpler, we first consider the staggering classical field configuration. In that way, the vector $e_\lambda$ reads:
\begin{equation}
e_\lambda \propto \sqrt{N}\,\delta_{-2\sigma,\lambda}\,,
\end{equation}
leading to
\begin{equation}
(A_k)_{\lambda_1,\lambda_2}=u \delta_{\lambda_1\lambda_2}\,,\qquad (B_k)_{\lambda_1,\lambda_2}= v_1 \delta_{\lambda_1\lambda_2}+v_2 \delta_{-2\sigma\lambda_1}\delta_{-2\sigma,\lambda_2}\,.\label{ABnewsbis}
\end{equation}
There are both diagonal matrices, and the inverse of $B$ computes straightforwardly:
\begin{equation}
(B_k)^{-1}_{\lambda_1\lambda_2}=\frac{\delta_{\lambda_1\lambda_2}}{v_1+v_2 \delta_{-2\sigma,\lambda_1}\delta_{-2\sigma,\lambda_2}}\,.
\end{equation}
Because the propagator is diagonal, only the connected contribution in \eqref{connectedcomponentflow} has to be retained. Furthermore, because we are aiming to write the flow equation for dimensionless quantities and keeping $\bar{\chi}$ fixed, we define the dimensionless derivatives $\bar{\mathcal{U}}_k^{(n)}$ as:
\begin{equation}
\mathcal{U}_k^{(n)}= k^{n} \bar{\mathcal{U}}_k^{(n)}
\end{equation}
and:
\begin{equation}
\frac{d}{ds}\mathcal{U}_k^\prime(\chi)= k \left[\frac{d}{ds}\bar{\mathcal{U}}_k^\prime(\bar{\chi})+\left(\bar{\mathcal{U}}_k^\prime(\bar{\chi})+\bar{\chi}{\mathcal{U}}_k^{\prime\prime}(\bar{\chi})\right) \right]\,,
\end{equation}
the dimensionless field $\bar{\chi}$ being:
\begin{equation}
\bar{\chi}:= k \chi\,.
\end{equation}
Note that it is easy to check that in the large $N$ limit, the LPA enforces that flow equations for $Z_k$ and $Y_k$ vanish identically (see Appendix \ref{appendix4}). Hence, defining:
\begin{equation}
\bar{I}_2^\prime=\frac{1}{k}I_2^\prime\,,\qquad \bar{J}_2^\prime=\frac{1}{k}J_2^\prime\,,
\end{equation}
where $I_2^\prime$ and $J_2^\prime$ are obtained from $I_2$ and $J_2$ given by equations \eqref{I2} and \eqref{J2}, replacing the dimensionless mass $\bar{\mu}_1$ by $\bar{\mu}_1^\prime$ defined as:
\begin{equation}
\bar{\mu}_1^\prime=\frac{1}{k}\, \mathcal{U}_k^\prime(\chi)\,.
\end{equation}
Hence with these definitions, the flow equation reduces to:
\begin{equation}
\boxed{\frac{d}{ds}\bar{\mathcal{U}}_k^\prime(\bar{\chi})=-\bar{\mathcal{U}}_k^\prime(\bar{\chi})-\bar{\chi}{\mathcal{U}}_k^{\prime\prime}(\bar{\chi})-\bar{\mathcal{U}}_k^{\prime\prime}(\bar{\chi}) (\bar{I}_2^{\prime}+2\bar{J}_2^{\prime})}\,.
\end{equation}
As for the vertex expansion, the previous flow equation can be investigated numerically. Figure \ref{figflowstaggered} and \ref{figflowstaggered2} shows the behavior of the potential for quartic and sixtic initial conditions for $U_{\Lambda}$, for some arbitrary UV scale $\Lambda$. Let us describes each of them separately. Note that we focus on the Wigner distribution to begin, and extend the discussion for MP law at the end of the section. 
\medskip

\paragraph{Quartic potential with Wigner distribution.} If the initial potential is in the high-temperature phase in the UV, it remains in the high-temperature phase in the IR, and the symmetry is unbroken (Figure \ref{figflowstaggered} on left). What happens for a potential in the broken phase at scale $\Lambda$ depends essentially on the value of the variance $\sigma$. On the right of Figure \ref{figflowstaggered} we show the behavior of the potential for $\sigma=1$ using a logarithmic scale for the flow. One can show that the spacing between two dimensionless vacuum decreases toward IR scales. If it decreases faster than $k$, the dimensional vacuum $\rho(k)$ goes to $0$, as soon as $k\to 0$, meaning that the symmetry is restored. If it goes to zero for a finite value of $k=k_0$, the theory is above the critical point and has finite correlation time $\tau:=k_0^{-1}$. As $k_0\to 0$, the correlation time diverges, and the theory reaches the critical region, where $\rho(k)/k$ reaches a finite value as $k\to 0$. Finally, in the broken regime, for $\sigma$ strong enough, the vacuum $\rho(k)$ remains finite in the deep IR. The critical value $\sigma_c$ depends on the values for $\kappa_1$ and $\kappa_2$ in the initial potential, indeed, accordingly to the analytical insights of Appendix \ref{appendix2}. 
\medskip

\begin{figure}
\begin{center}
\includegraphics[scale=0.55]{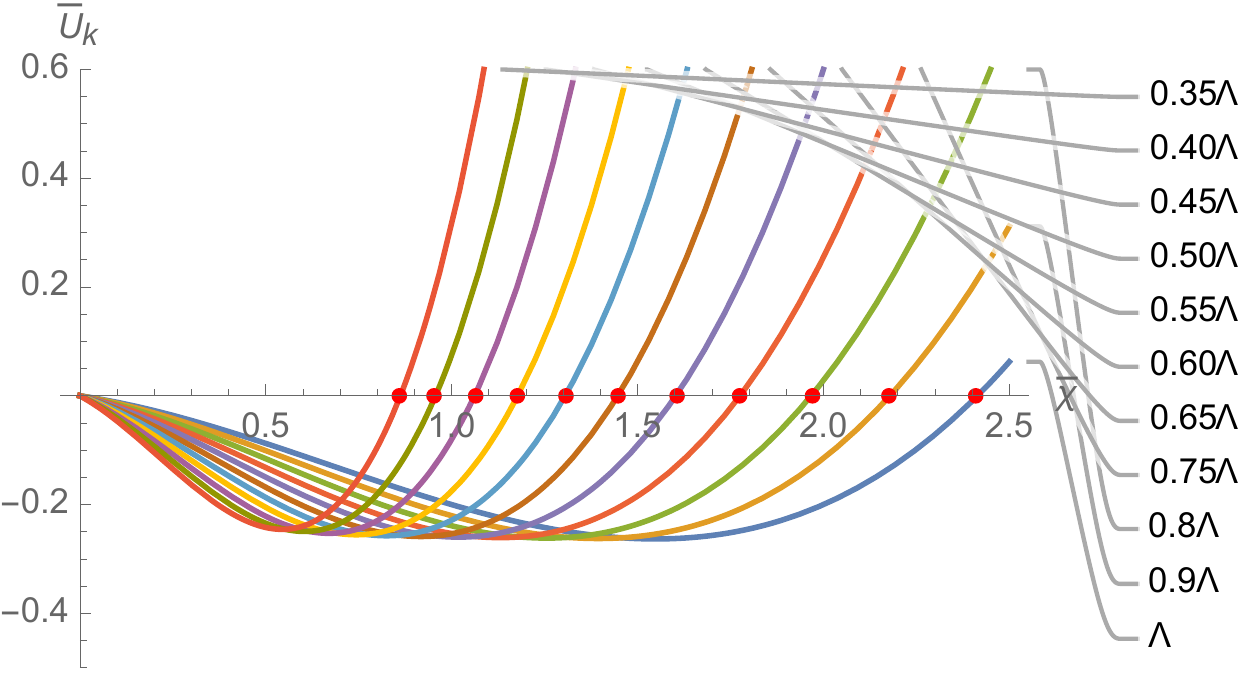}\qquad \includegraphics[scale=0.55]{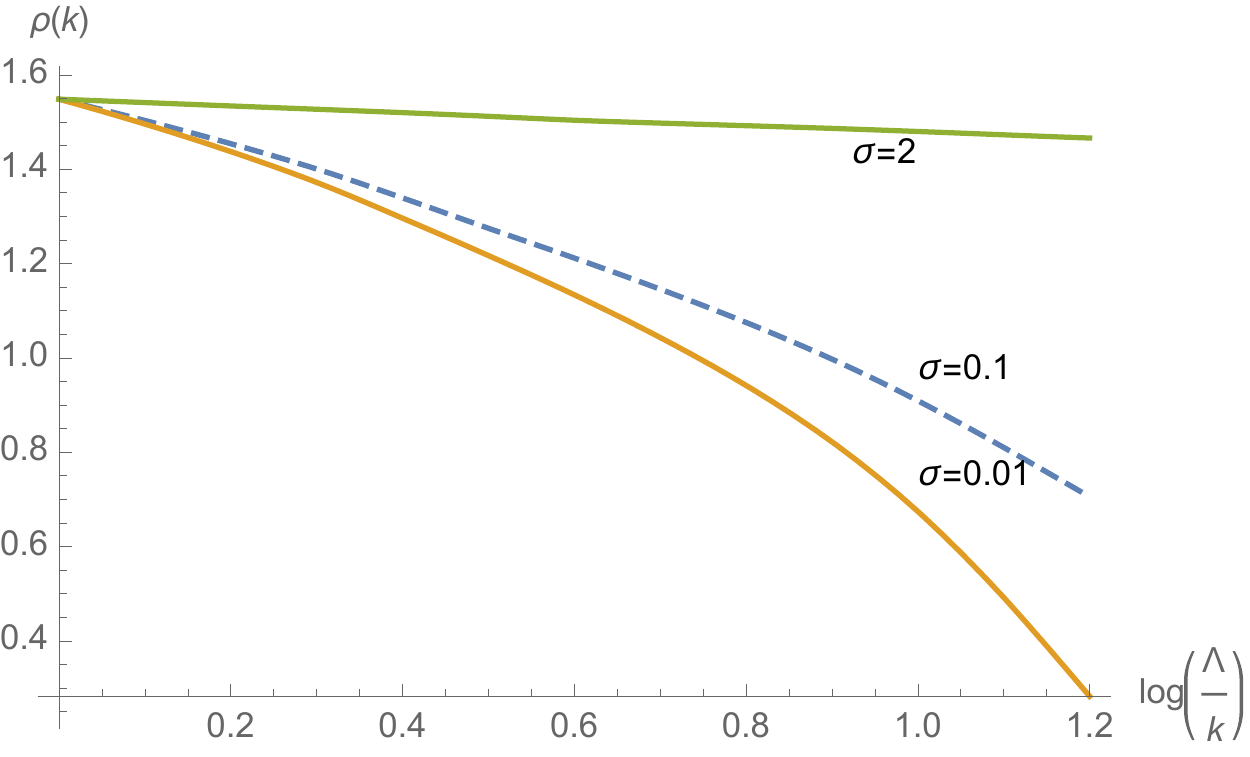}
\end{center}
\caption{On the right, evolution of the sixtic potential in the ‘‘deep IR’’ for $\sigma=2$ with a negative initial mass. On the left, the evolution of the nonzero vacuum $\rho(k)$ for large and small $\sigma=0.01, 0.1$ (above the critical surface) and $\sigma=2$ (below the critical surface).}\label{figflowstaggered4}
\end{figure}

\paragraph{Sixtic potential with Wigner distribution.} Assuming the initial quartic coupling $\kappa_2$ is negative, what the IR potential becomes depends on the sign of the mass $\kappa_1$. Typical evolution in both cases is pictured in Figure \ref{figflowstaggered2}. For positive mass, the shape of the potential exhibits a phase coexistence at scale $\Lambda$. This coexistence survives up to a finite frequency scale $k_1$ corresponding to the timescale $\tau_1:=(k_1)^{-1}$, above which the solution $\rho(k)=0$ becomes unstable. For $\kappa_1$ negative, however, the solution $\rho(k)=0$ is unstable at the beginning, and the stable (non-zero) vacuum moves toward the origin as $k$ decreases. Figure \ref{figflowstaggered4} shows the evolution of the vacuum $\rho(k)$. As for the quartic cases, for large time scales, the vacuum field is a macroscopic occupation number for the component $M_0$, corresponding to the smallest eigenvalue. 

\begin{figure}
\begin{center}
\includegraphics[scale=0.55]{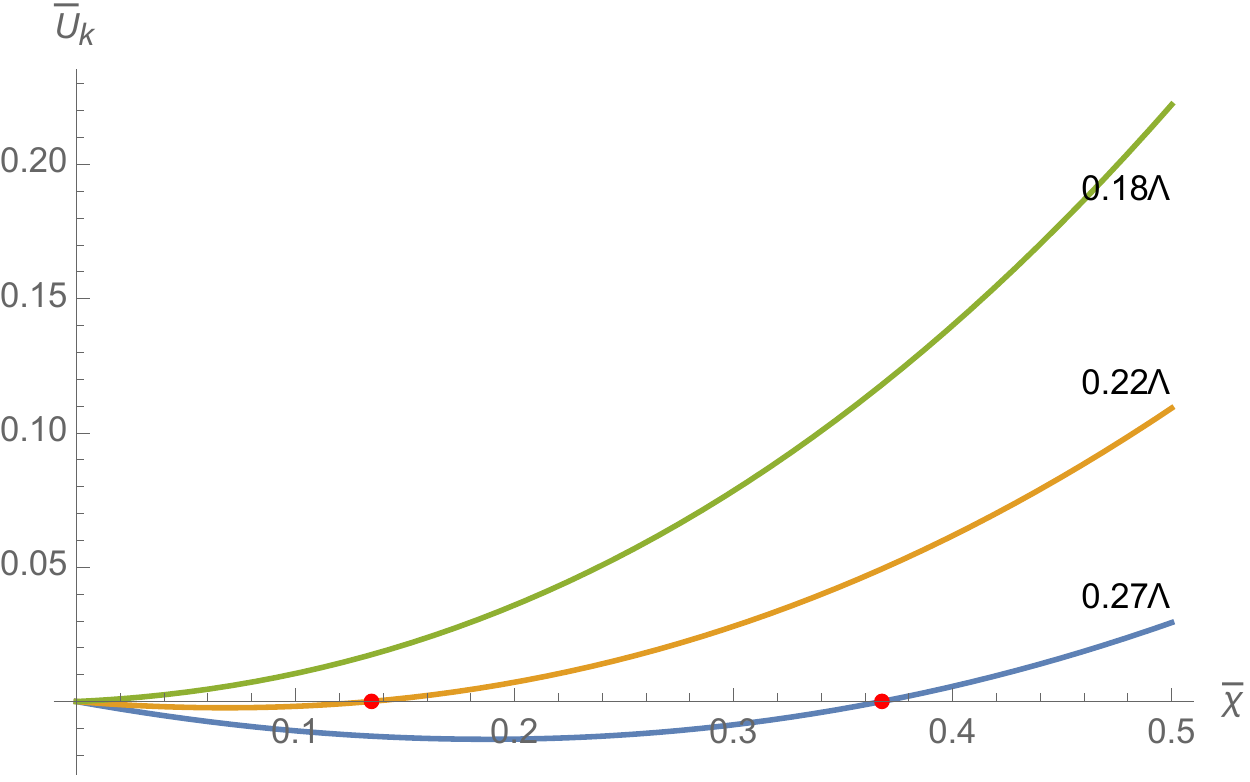}\qquad \includegraphics[scale=0.55]{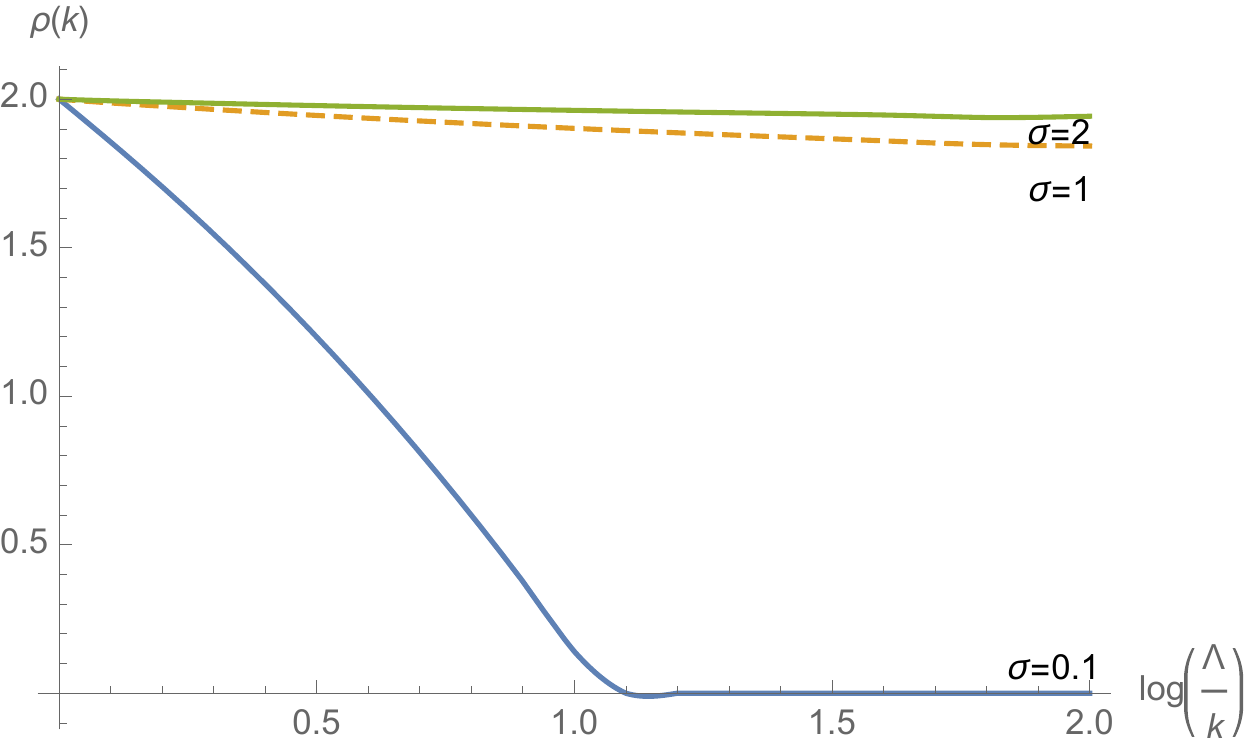}
\end{center}
\caption{On the right, evolution of the quartic potential in the ‘‘deep IR’’ for $\sigma=0.1$ with a negative initial mass. On the left, the evolution of the vacuum $\rho(k)$ for large and small $\sigma=0.1, 1$ and $2$ (above and below the critical surface).}\label{figflowstaggered3}
\end{figure}

\begin{figure}
\begin{center}
\includegraphics[scale=0.55]{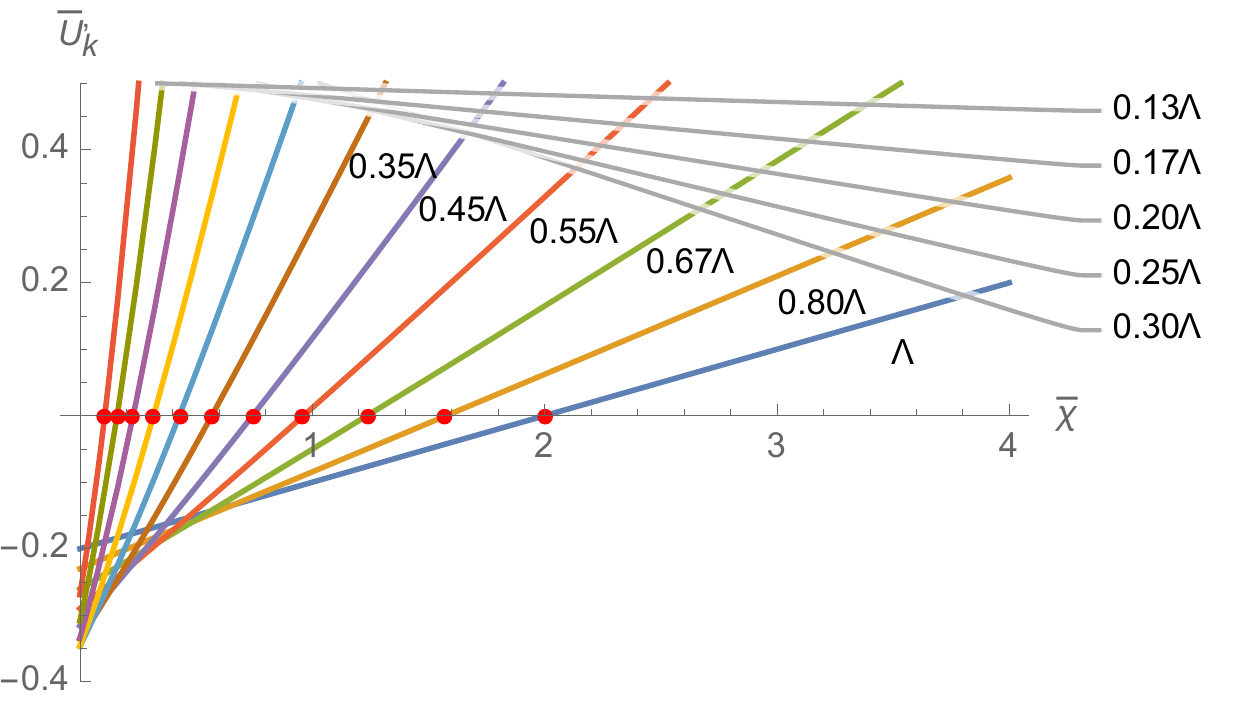}\qquad \includegraphics[scale=0.55]{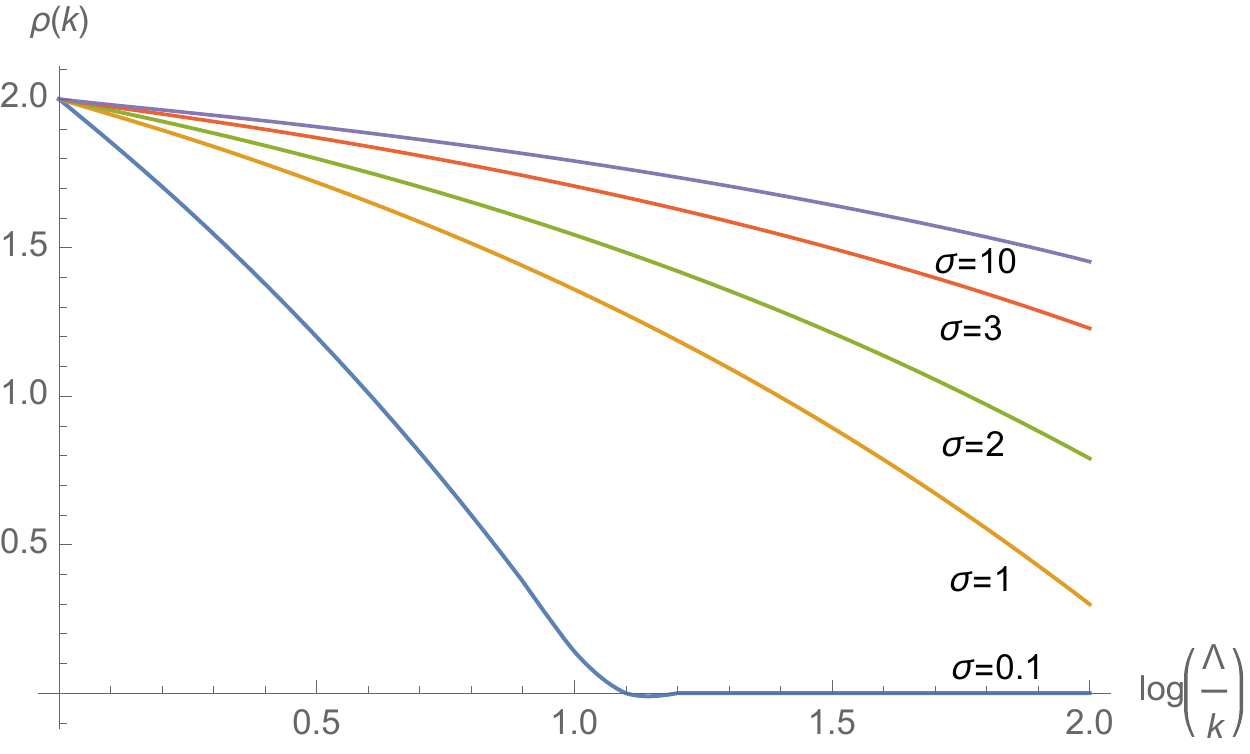}
\end{center}
\caption{On the right: Evolution of the derivative of the quartic potential $\bar{U}_k^\prime$ in the ‘‘deep IR’’ for $\sigma=2$ with a negative initial mass for the $c=1$ MP distribution. On the left: The evolution of the nonzero vacuum $\rho(k)$ for $\sigma=0.1, 1, 2, 3$ and $10$.}\label{figflowstaggered5}
\end{figure}

\begin{figure}
\begin{center}
\includegraphics[scale=0.55]{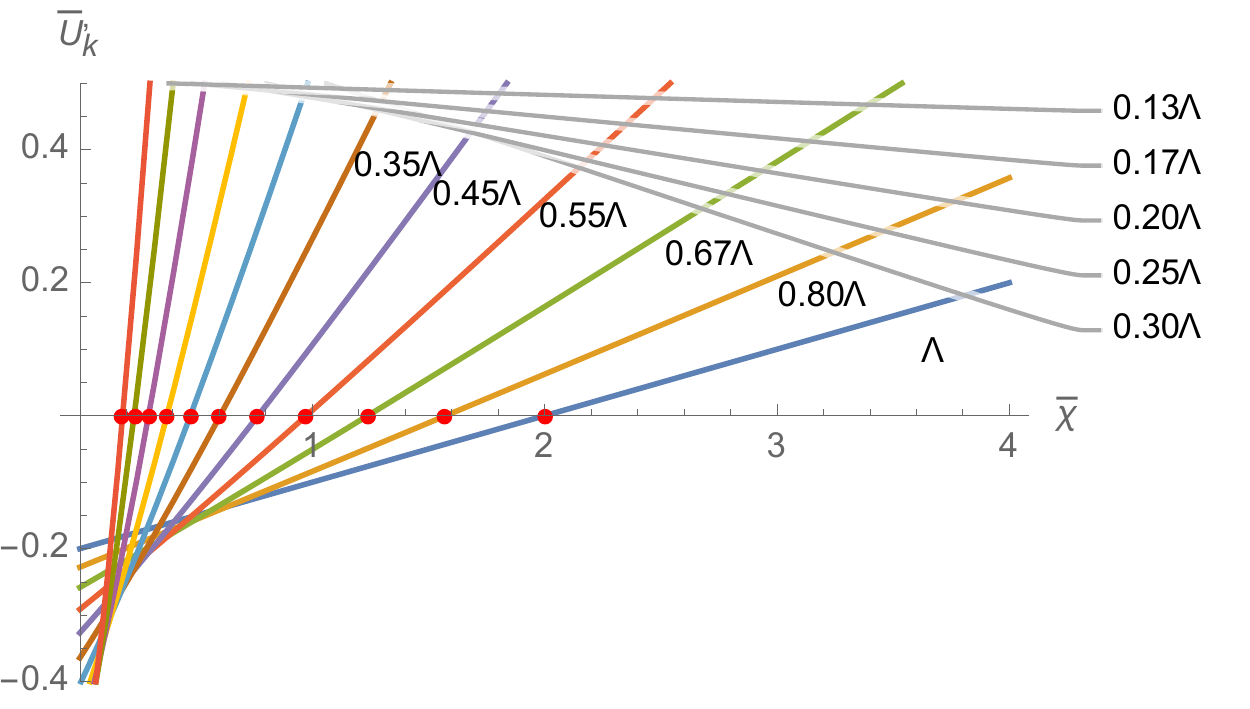}\qquad \includegraphics[scale=0.55]{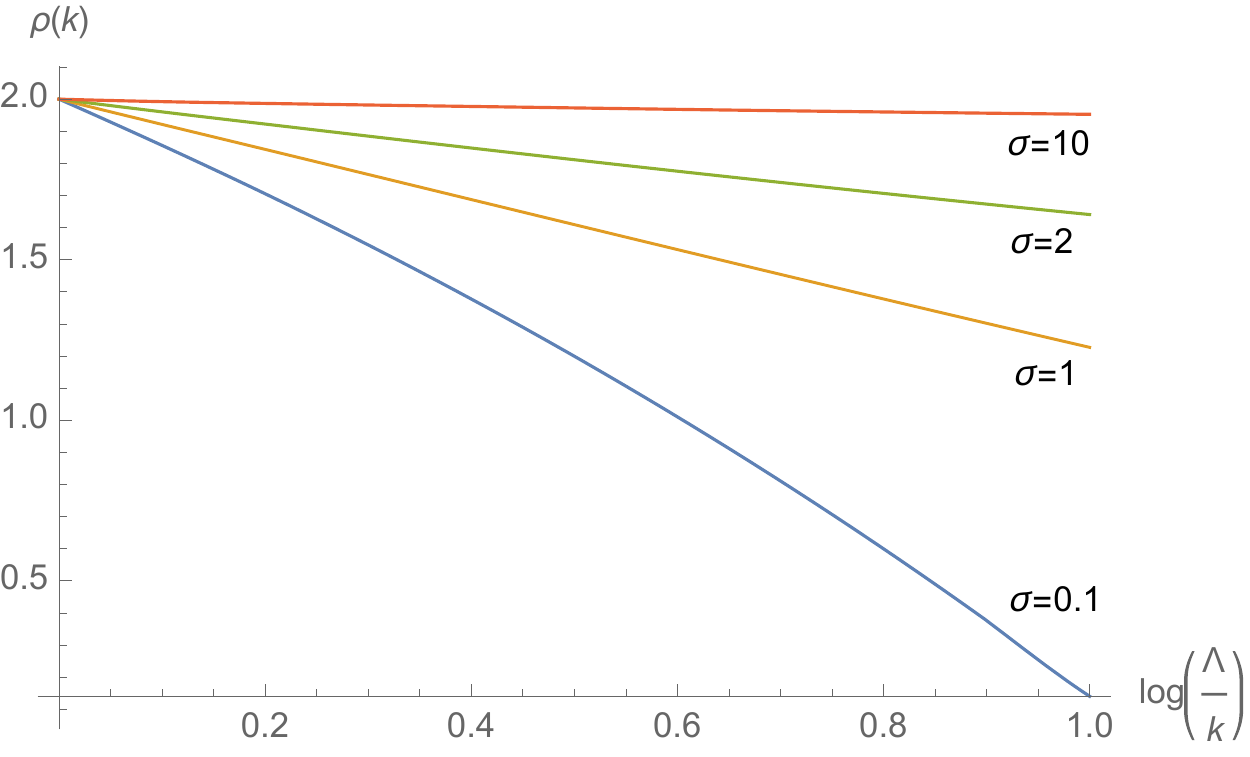}
\end{center}
\caption{On the right: Evolution of the derivative of the quartic potential $\bar{U}_k^\prime$ in the ‘‘deep IR’’ for $\sigma=2$ with a negative initial mass for the $c=0.25$ MP distribution. On the left: The evolution of the nonzero vacuum $\rho(k)$ for $\sigma=0.1, 1, 2$ and $10$.}\label{figflowstaggered6}
\end{figure}

\paragraph{Marchenko-Pastur distribution.} For MP distribution, the large-scale behavior of the RG flow depends on the parameter $c$ defined in \eqref{mp}. For $c=1$ i.e. $X$ in \eqref{MPMAT} is a square matrix, the MP distribution behaves as $1/\sqrt{p}$ for small $p$. As discussed in remark \eqref{remarkdim}, this behavior is reminiscent of the momenta distribution for a one dimensional field theory. We know from the Mermin-Wagner theorem \cite{mermin1966absence,coleman1973there} that there is no phase transition with spontaneous symmetry breaking in dimension smaller than $2$. However, the analytic arguments given in Appendix \ref{appendix2} show that the critical temperature does not vanish. Nerveless, as we show in Figure \ref{figflowstaggered5}, even if the macroscopic vacuum $\rho(k)$ can survive for large timescale, breaking the $O(N)$ symmetry, it reaches zero ultimately for timescale large enough, in agreement with Mermin-Wagner theorem. Finally, for $c<1$, the momentum distribution behaves as $\sqrt{p}$, and the results are similar to what we obtained for the Wigner distribution. Figure \ref{figflowstaggered6} summarizes the results for $c=0.25$ MP law.

\subsection{RG for uniform classical field}
In this section, we investigate the first approximation scheme, namely the uniform field configuration. We voluntarily separated the numerical study from the theoretical derivation because of the volume of the numerical insights. 

\subsubsection{Derivation of the flow equations}
Let us consider the uniform classical field $M_\lambda = \sqrt{\chi}$. The connected contribution of the decomposition \eqref{decompositionconnexe} reads:
\begin{equation}
\vcenter{\hbox{\includegraphics[scale=0.8]{vertex3ptscon.pdf}}}= 4\pi \delta(0) N \mathcal{U}_k^{\prime\prime}(\bar{\chi}) \sqrt{\chi}(I_2^\prime+2J_2^\prime)\,,\label{connectedcomponentflow}
\end{equation}
Let us address the contribution of the disconnected pieces. It is easy to check that only the last term in the expansion \eqref{decomp3pts2} have to be retained in the large $N$ limit, and:
\begin{equation}
2\times \vcenter{\hbox{\includegraphics[scale=0.8]{vertex3ptscon2.pdf}}}\,+\,  \vcenter{\hbox{\includegraphics[scale=0.8]{vertex3ptscon3.pdf}}}=\,  8\pi \delta(0) N (\chi)^{3/2} \mathcal{U}_k^{\prime\prime\prime} (I_2^{\prime\prime}+2J_2^{\prime\prime})\,,
\end{equation}
the integrals $I^{\prime\prime}_k$ and $J^{\prime\prime}_k$ being defined as:
\begin{align}
I_2^{\prime\prime}=-\frac{1}{Z_k^2k^2}\int \frac{dy}{2\pi} \int \bar{\mu}(x) dx  \Delta_2(x,y) \vert g_1^\prime(x,y)\vert^2 \bigg[2\Xi(x,y) +\vert \Xi(x,y) \vert^2\bigg]
\end{align}
and:
\begin{align}
\nonumber J_2^{\prime\prime}=&\frac{1}{Z_k^2k^2} \int \frac{dy}{2\pi} \int \bar{\mu}(x) dx \Delta_1(x,y)\Big[2g_{1}^\prime(x,-y) \Xi(x,-y)g_{2}^\prime(x,y)\\
&+2g_{1}^\prime(x,-y)g_{2}^\prime(x,y)\Upsilon[x,y]+g_{1}^\prime(x,-y)g_{2}^\prime(x,y)\Upsilon[x,y]\Xi(x,-y)\Big]\, 
\end{align}
where:
\begin{equation}
\Upsilon[x,y]:=\Xi(x,y)+\Xi(x,-y)+\vert \Xi(x,y)\vert^2\,,
\end{equation}
and:
\begin{equation}
\Xi(x,y):=-\dfrac{2 \bar{\chi}}{i y+x+\bar{\mu}^\prime_1+\bar{\rho}^{(1)}(-y) \tau(x)}\,.
\end{equation}
Finally, the flow equation for the effective potential reads:
\begin{equation}
\boxed{
\frac{d}{ds}\mathcal{U}_k^\prime(\chi)=-\mathcal{U}_k^{\prime\prime}(\chi) (I_2^\prime+2J_2^\prime)-2 \chi \mathcal{U}_k^{\prime\prime\prime} (I_2^{\prime\prime}+2J_2^{\prime\prime})\,.}
\end{equation}
As in the previous section, we define the dimensionless integrals as:
\begin{equation}
\bar{I}_2^{\prime\prime}=\frac{1}{ k^2}I_2^{\prime\prime}\,,\qquad \bar{J}_2^{\prime\prime}=\frac{1}{ k^2}J_2^{\prime\prime}\,,
\end{equation}
and the flow equation for the dimensionless potential, keeping the dimensionless variable $\bar{\chi}$ fixed is:
\begin{equation}
\boxed{\frac{d}{ds}\bar{\mathcal{U}}_k^\prime(\bar{\chi})=-\bar{\mathcal{U}}_k^\prime(\bar{\chi})-\bar{\chi}\bar{\mathcal{U}}_k^{\prime\prime}(\bar{\chi})-\bar{\mathcal{U}}_k^{\prime\prime}(\bar{\chi}) (\bar{I}_2^\prime+2\bar{J}_2^\prime)-2 \bar{\chi} \bar{\mathcal{U}}_k^{\prime\prime\prime} (\bar{I}_2^{\prime\prime}+2\bar{J}_2^{\prime\prime})\,.}\label{flowU}
\end{equation}

\subsubsection{Numerical investigations and reliability}
The equation \eqref{flowU} can be investigated numerically, without assumption on the power field expansion of $\bar{\mathcal{U}}_k^\prime(\bar{\chi})$. This is especially relevant where local interactions have increasing divergence degrees and where crude truncation are illegals because of their bad convergence properties (as the vertex expansion showed). For this reason, our numerical investigations follow the potential evolution on its own globally. Because we are essentially aiming to investigate the reliability of the RG flow, we mainly focus on the case of a quartic potential, for which we have analytic insights. 
\medskip

\begin{figure}
\begin{center}
\includegraphics[scale=0.15]{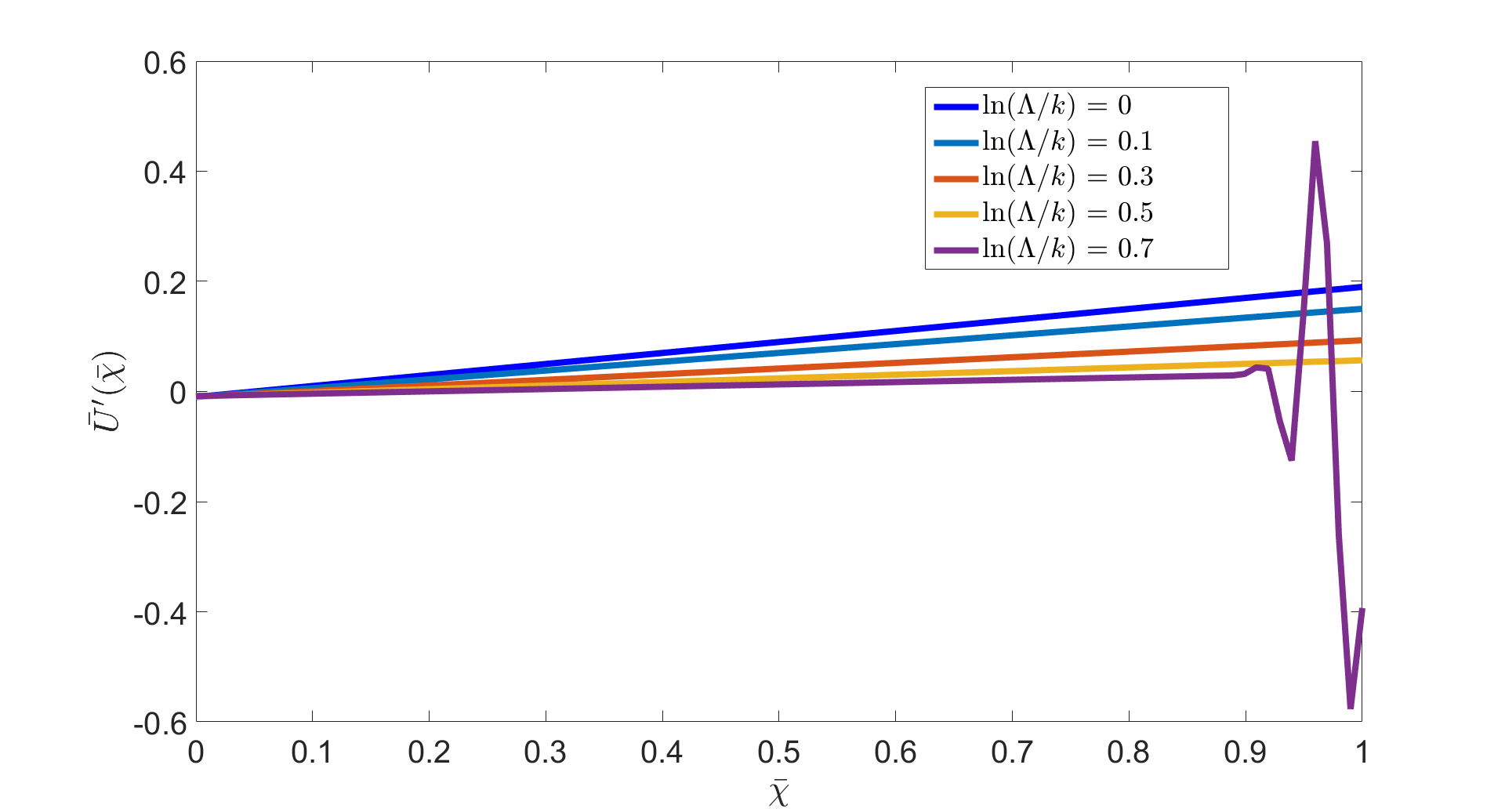}\,\includegraphics[scale=0.15]{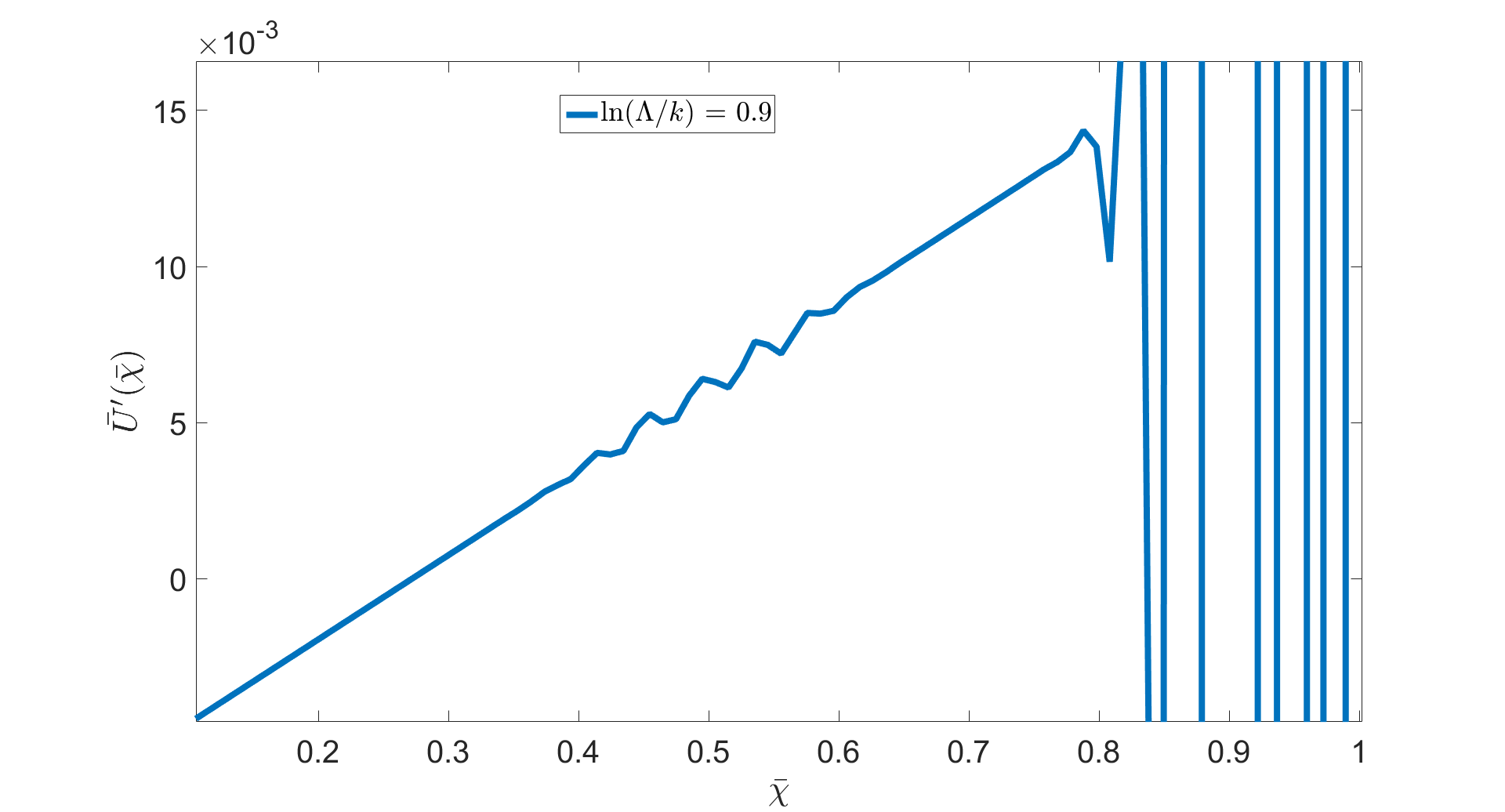}\\
\includegraphics[scale=0.15]{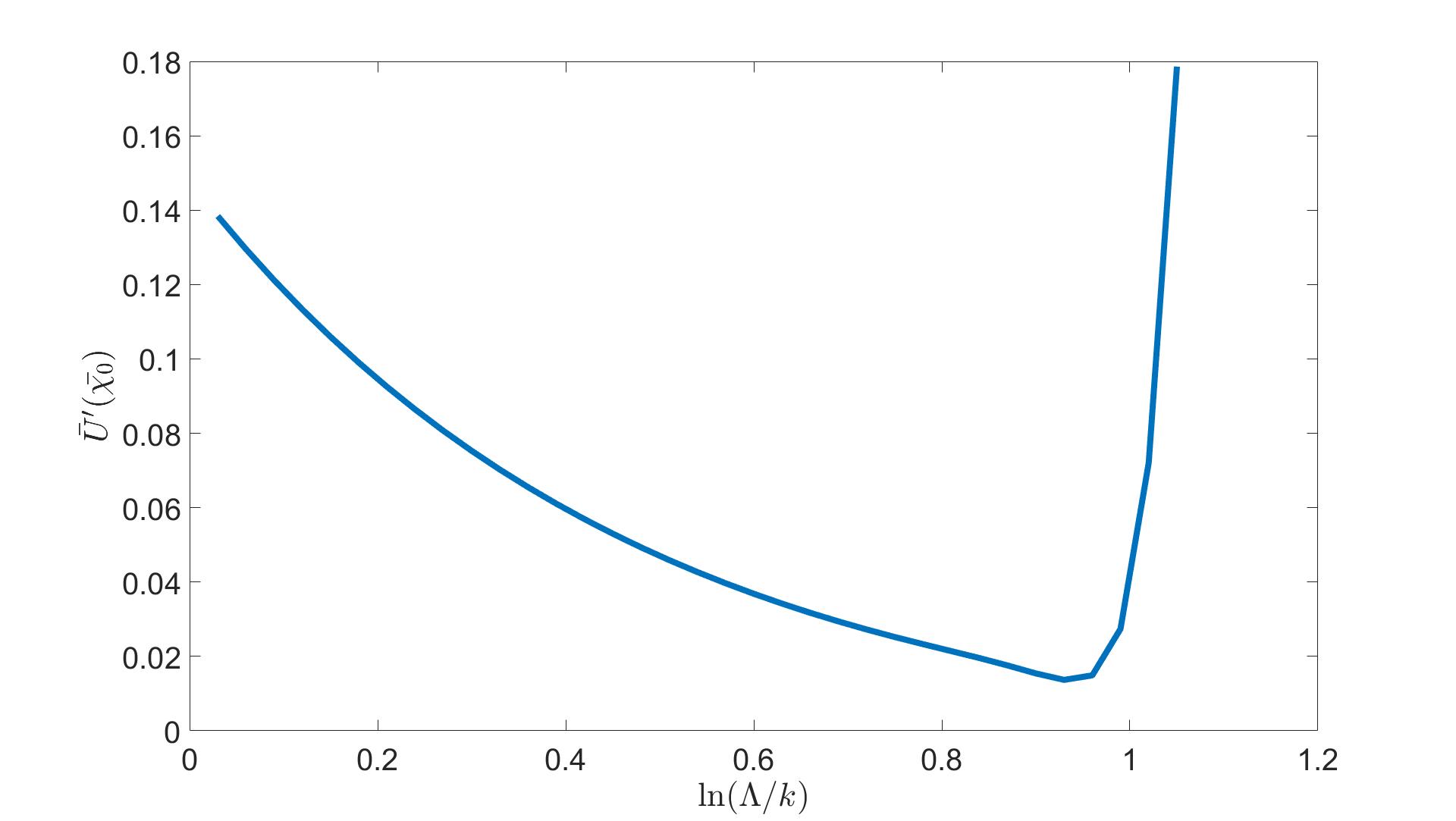}\,\includegraphics[scale=0.15]{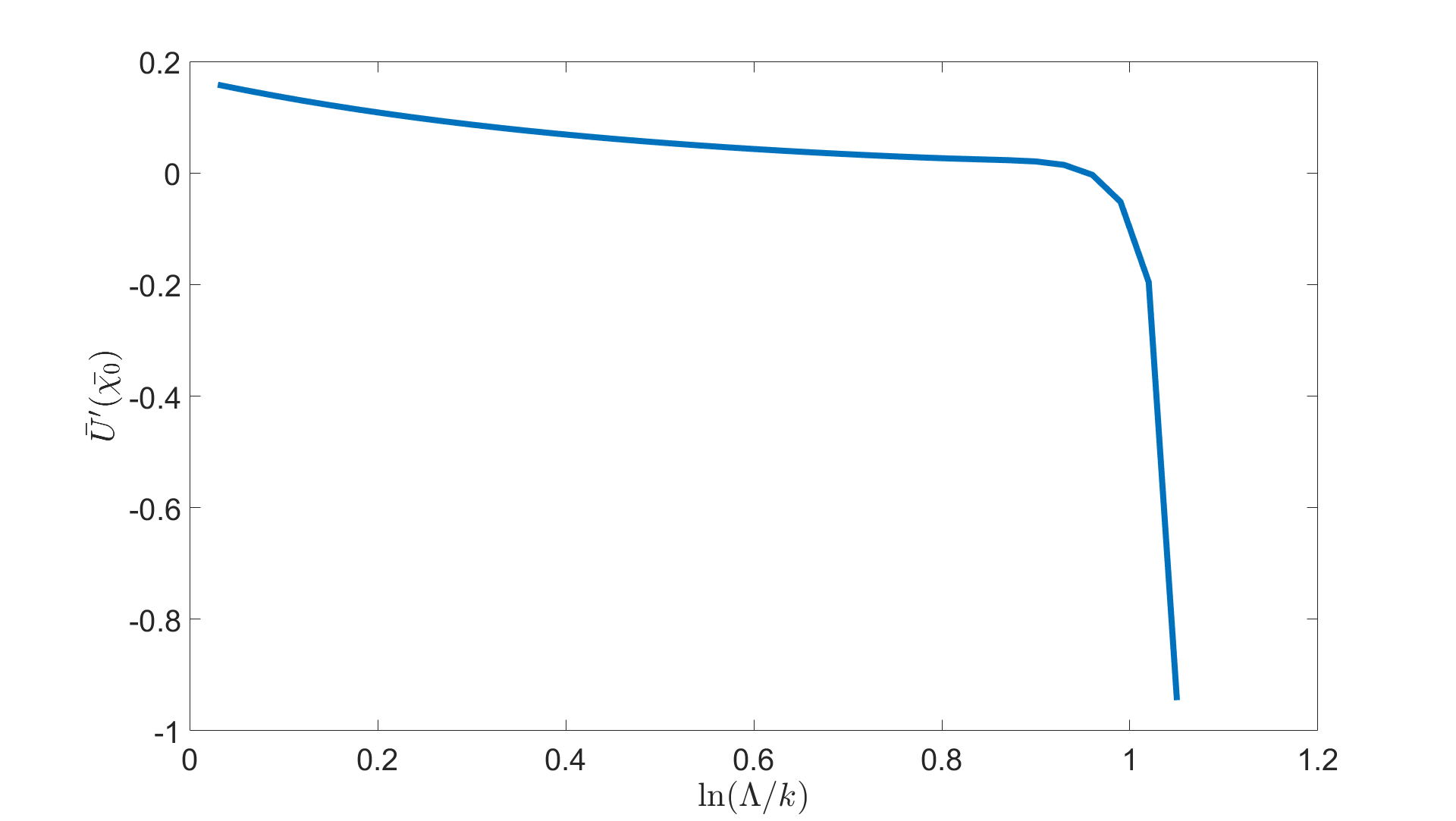}\\
\includegraphics[scale=0.15]{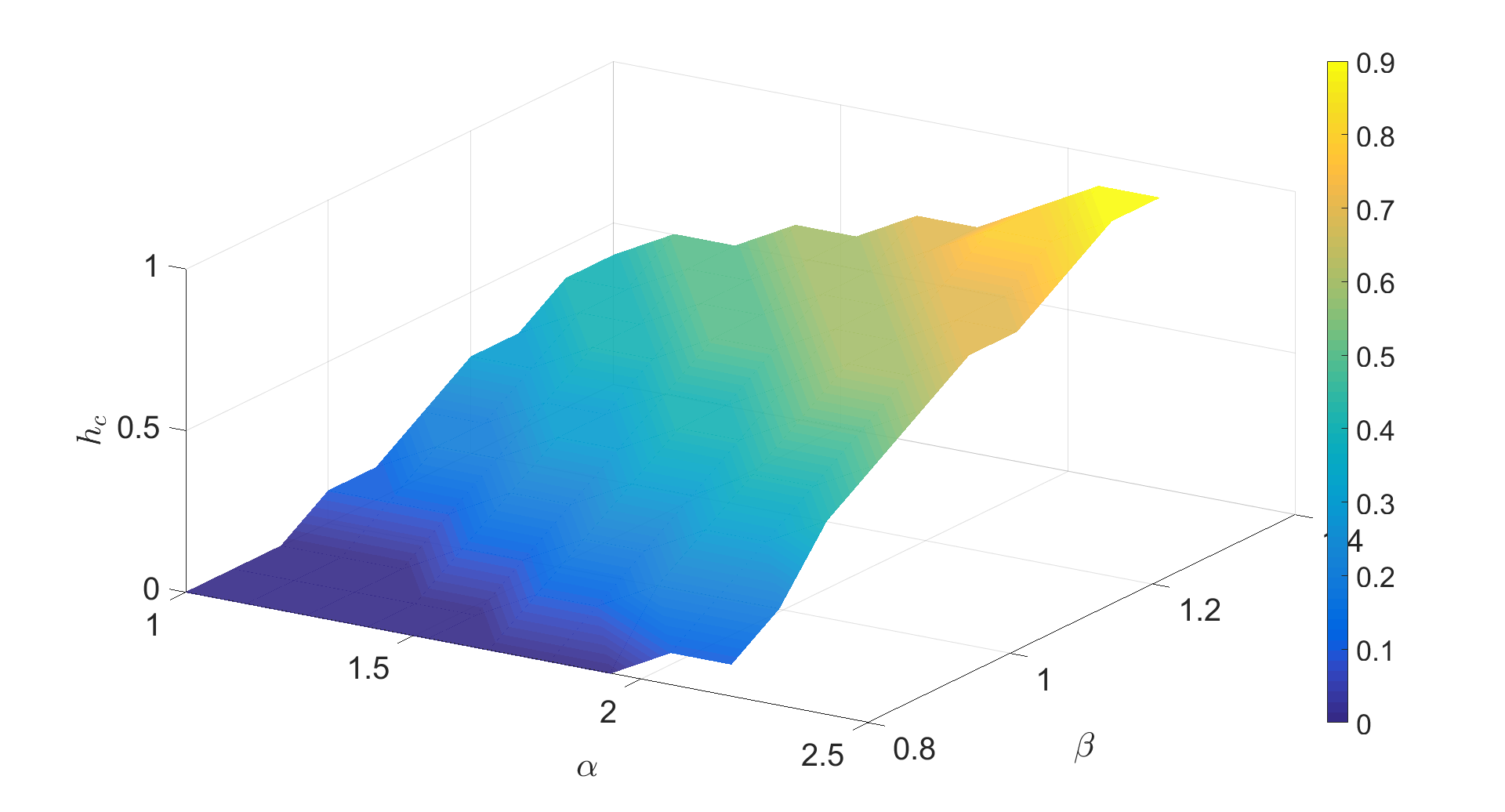}\,\includegraphics[scale=0.15]{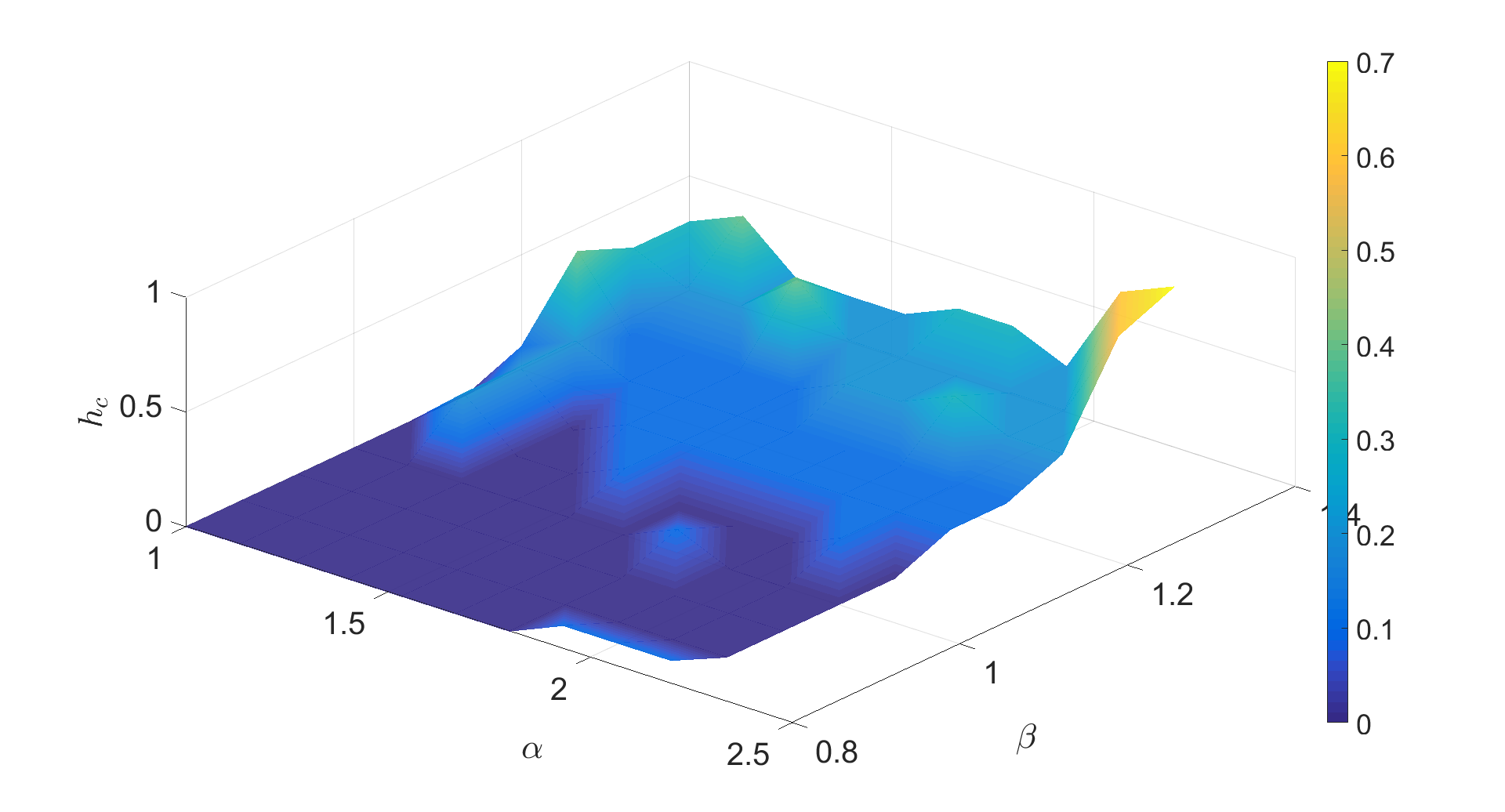}
\end{center}
\caption{Summary of numerical investigation for Wigner distribution with quartic potential. On the top, one can see the evolution of the derivative of the potential $\bar{\mathcal{U}}_k^\prime(\bar{\chi})$. In the middle, the evolution of a specific value $\bar{\mathcal{U}}_k^\prime(\bar{\chi}_0)$. On the bottom, we can show the dependency of $h_c$ on the parameters $\alpha$ and $\beta$ defining the regulator family. On the left for $\tau(y)=1$ and on the right for $\tau(y) \neq 1$.}\label{figevolWig1}
\end{figure}

Let us begin with the Wigner distribution, and consider a quartic potential initially in the broken phase in the UV regime i.e., having a negative mass:
\begin{equation}
U_\Lambda^{\prime}(\chi)=h_0+h_1 \chi\,,
\end{equation}
with $h_0<0$, for some UV scale $\Lambda$. The typical evolution for $\bar{\mathcal{U}}_k^\prime(\bar{\chi})$ is shown on the top of Figure \ref{figevolWig1}. If we remain close to the UV regime, the shape of the potential does not change significantly. However, the flow becomes singular after some steps, and very large variations of arbitrary size appear, with numerous zeros. The diagrams in the middle of Figure \ref{figevolWig1} illustrate the behavior of $\bar{\mathcal{U}}_k^\prime(\bar{\chi})$ for a given value of $\bar{\chi}$, and show that a sharp singularity occurs at finite scale $k\approx \Lambda/\exp(0.9)$. Analytic computations as given in Appendix \ref{appendix1}-\ref{appendix2} show that the system indeed never reaches equilibrium accordingly with an exponential low, and has rather a power law decay. Hence, under the critical temperature, the system is not in equilibrium dynamics, and our assumptions leading to the effective field theory \eqref{expressionZ} break down. One expects that these finite scale singularities for the uniform field configurations are a consequence of this breakdown of the equilibrium dynamics assumption. Indeed, this statement agrees with the analytical result that the system remains in an equilibrium dynamics for staggered initial conditions (see Appendix \ref{appendix1}), explaining why no singularities have been observed in the previous section for the staggered classical field configuration. 
\medskip

The analytic critical temperature $T_c$, given by \eqref{criticaltempWing} is defined for $h_0<0$, i.e $T_c \sim -h_0/h_1$. Nerveless, it happens that for some choices of regulator $R_k$, singularities happens also for $h_0>0$, meaning that RG flow can become singular even if the system is in the equilibrium dynamics,   accordingly with the analytical results. This suggests a reliability criterion based on the ability of the RG flow to agree with the analytical statement that the system relax toward equilibrium for $h_0>0$. We thus define as $h_c$ the critical values for $h_0$, such that the equilibrium assumption breaks down and singularities occur for $h_0<h_c$. It is difficult to investigate the dependency of $h_c$ on the regulator, which leaves in a functional space of infinite dimension. Plots on the bottom of the Figures \ref{figevolWig1} show the dependency of $h_c$ on the parameters $\alpha$ and $\beta$ defined by equation \eqref{rho1}. On the left, we imposed $\tau(y)=1$ i.e., with no coarse graining in eigenvalues is performed, while on the left $\tau(y)$ is given by equations \eqref{eqtau}-\eqref{eqh}. In both case, we observe the existence of a region where $h_c$ vanishes and which is larger in the case where we perform the coarse-graining both in frequencies and eigenvalues. Furthermore, the variations of $h_c$ outside this region seem to be smaller in the last case as in the case where $\tau(y)=1$. This shows that, not only a reliable RG can be constructed, accordingly with the criterion considered before, but this illustrates the superiority of a coarse graining in both eigenvalues and frequencies. Note that other diagrams like Figure \ref{figevolWig1} are obtained for $\tau(y)\neq 1$. 

\begin{figure}
\begin{center}
\includegraphics[scale=0.15]{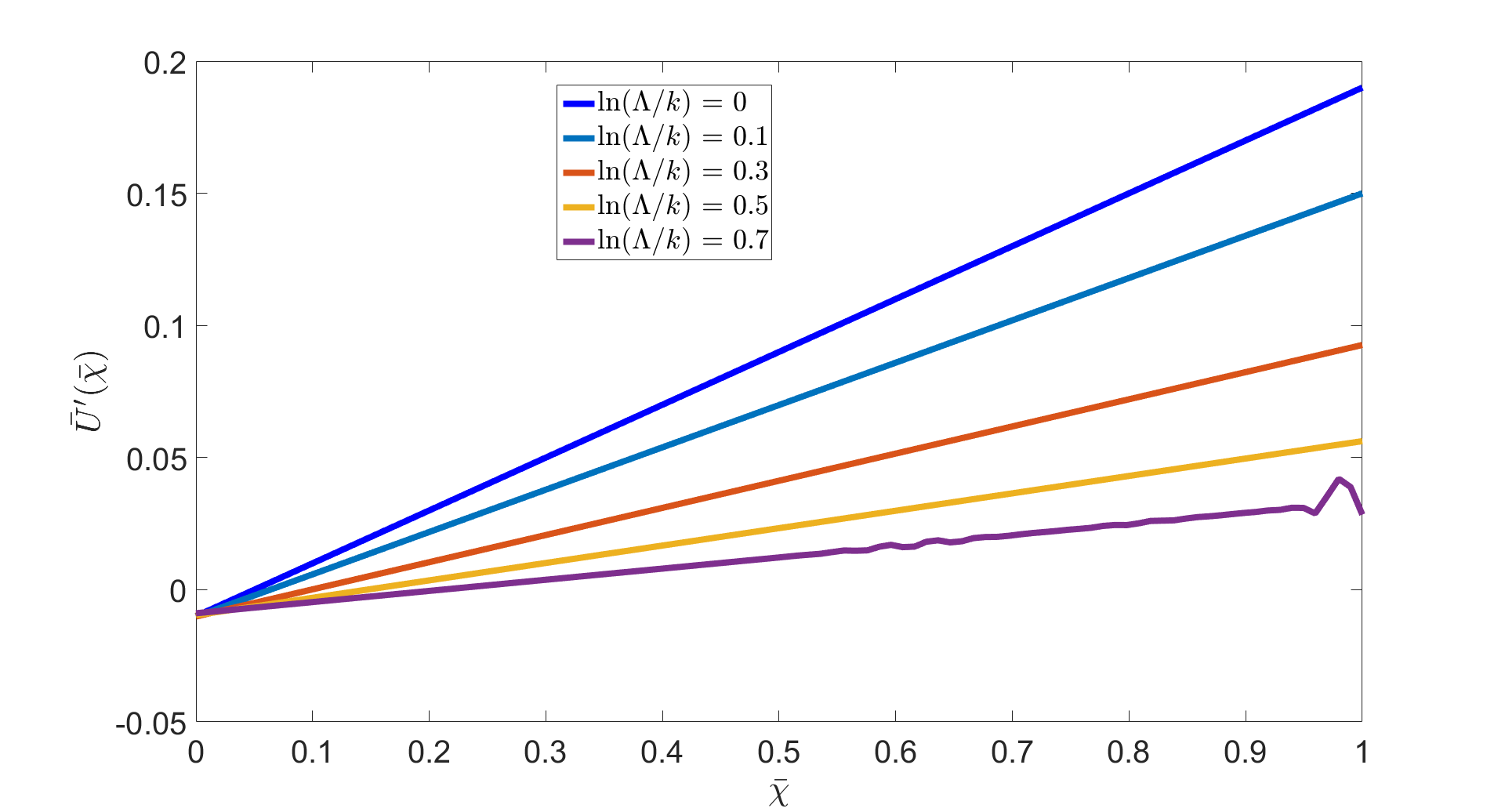}\,\includegraphics[scale=0.15]{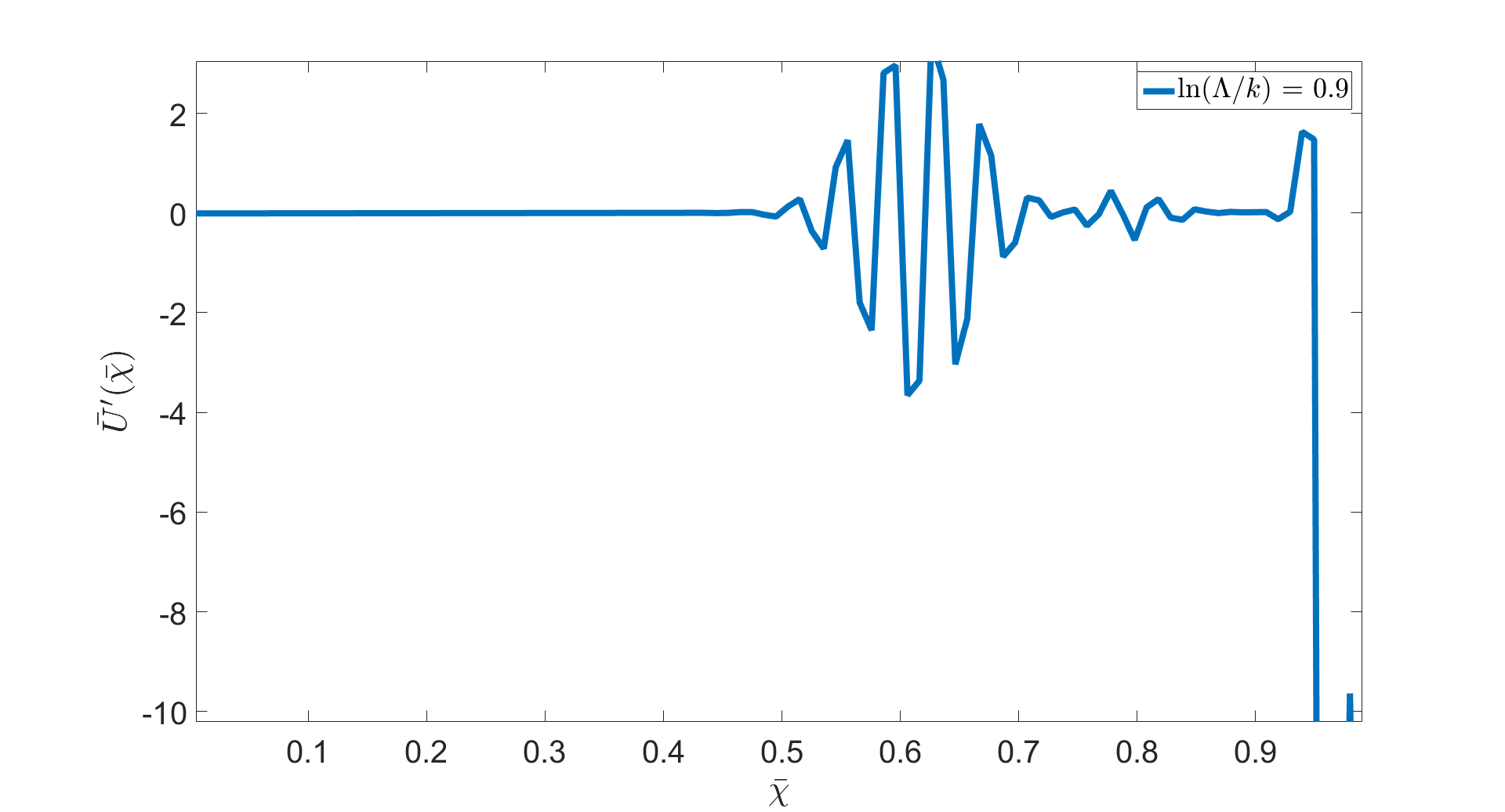}\\
\includegraphics[scale=0.15]{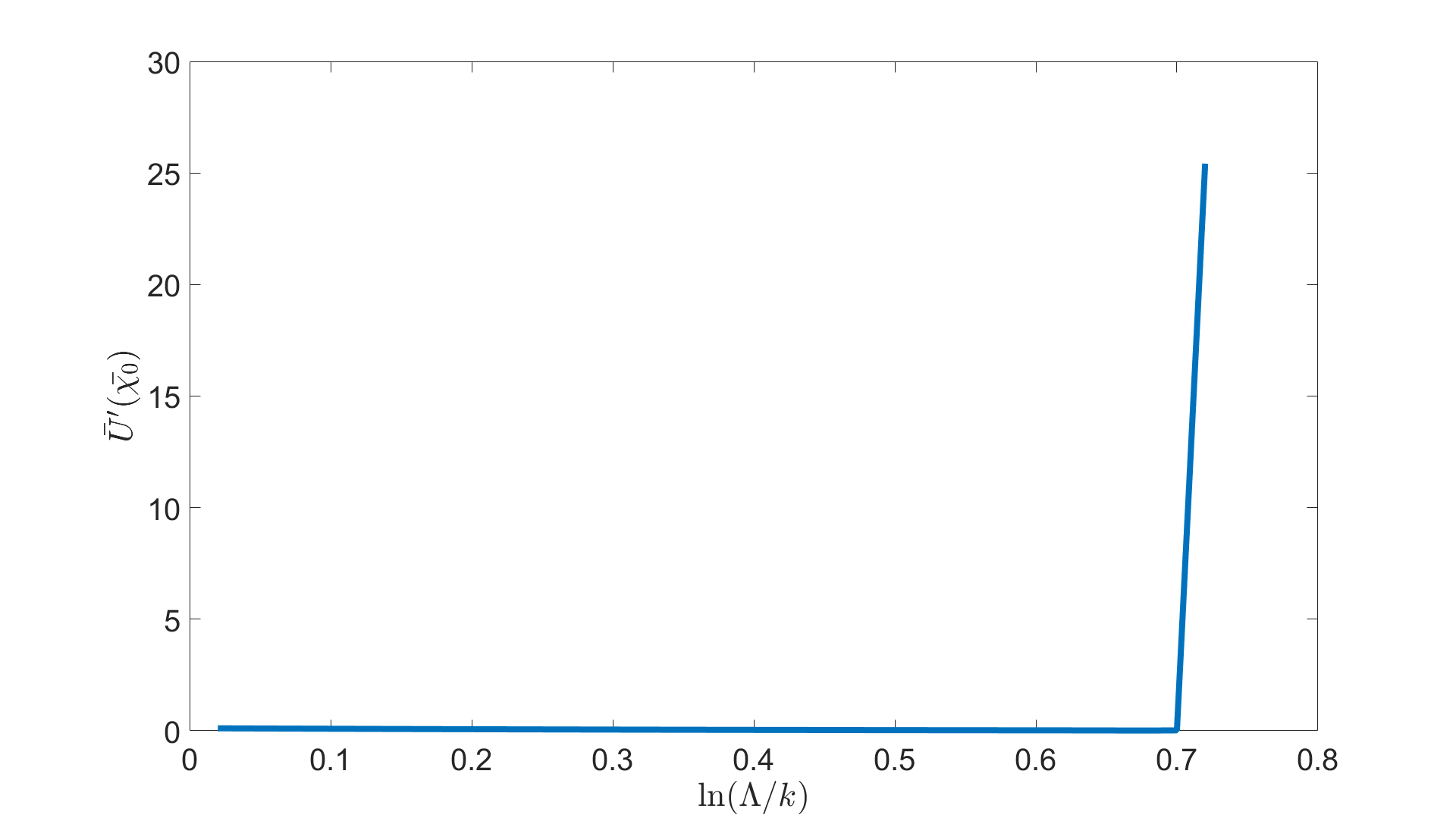}\,\includegraphics[scale=0.15]{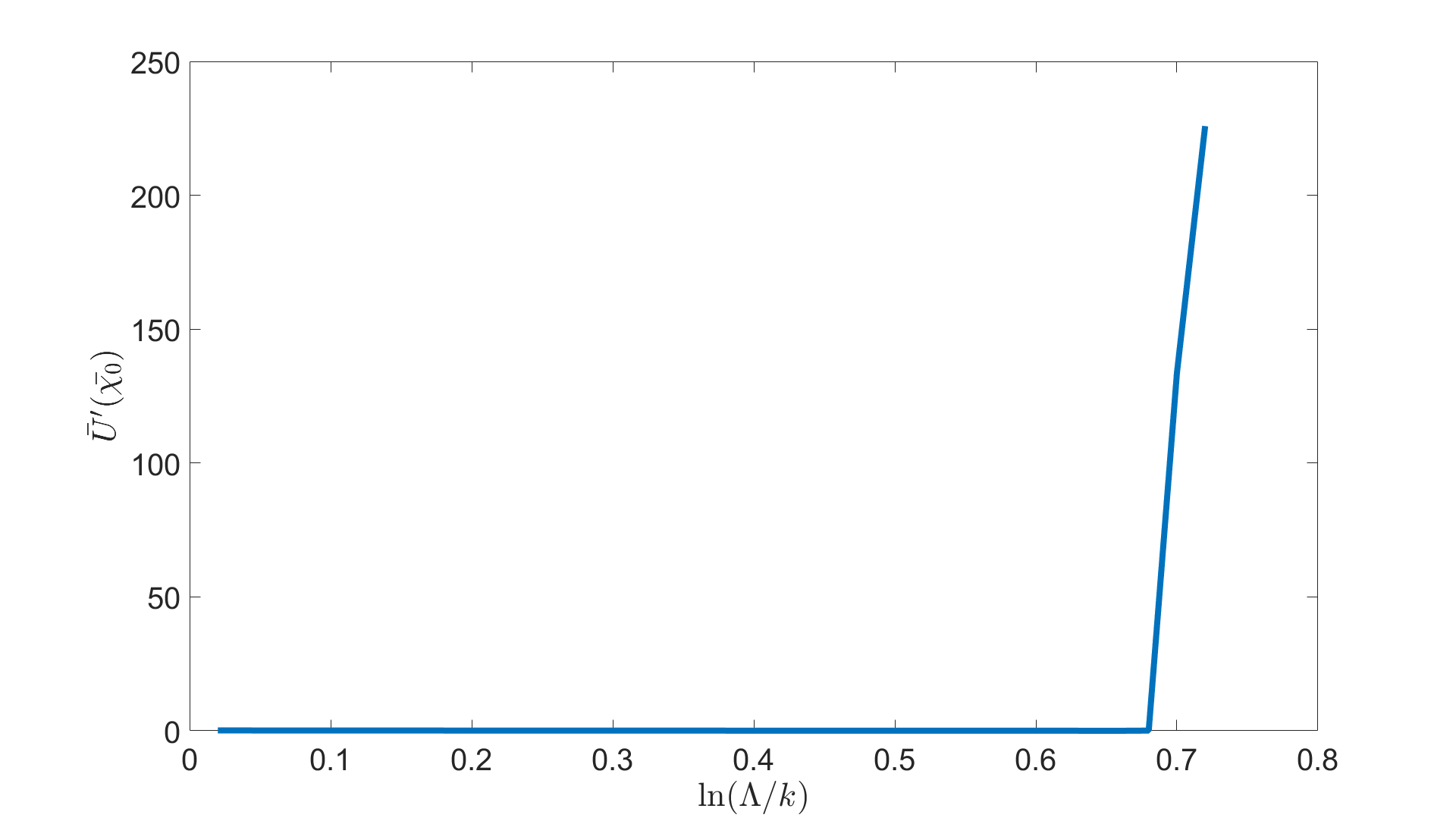}\\
\includegraphics[scale=0.15]{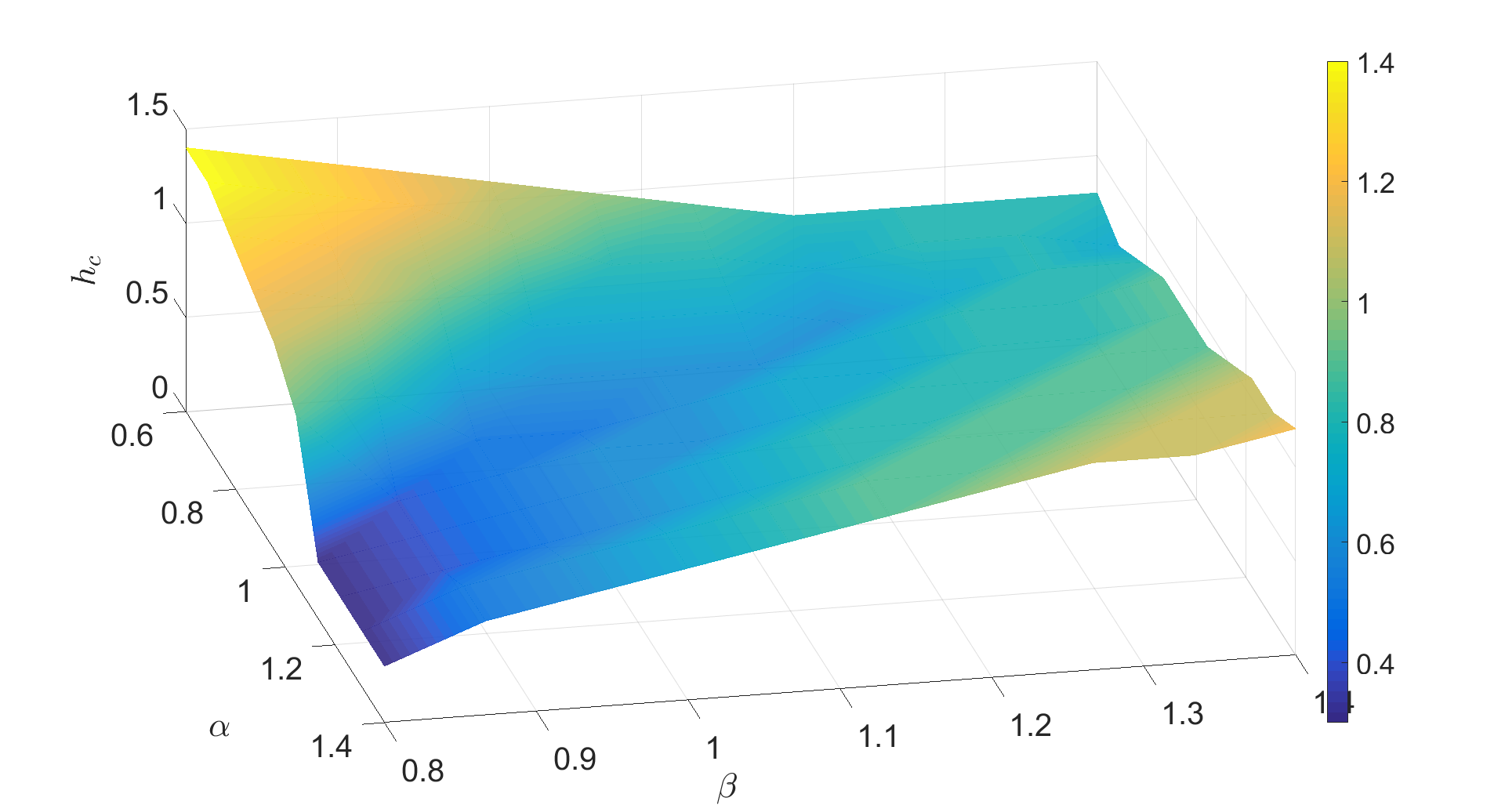}\,\includegraphics[scale=0.15]{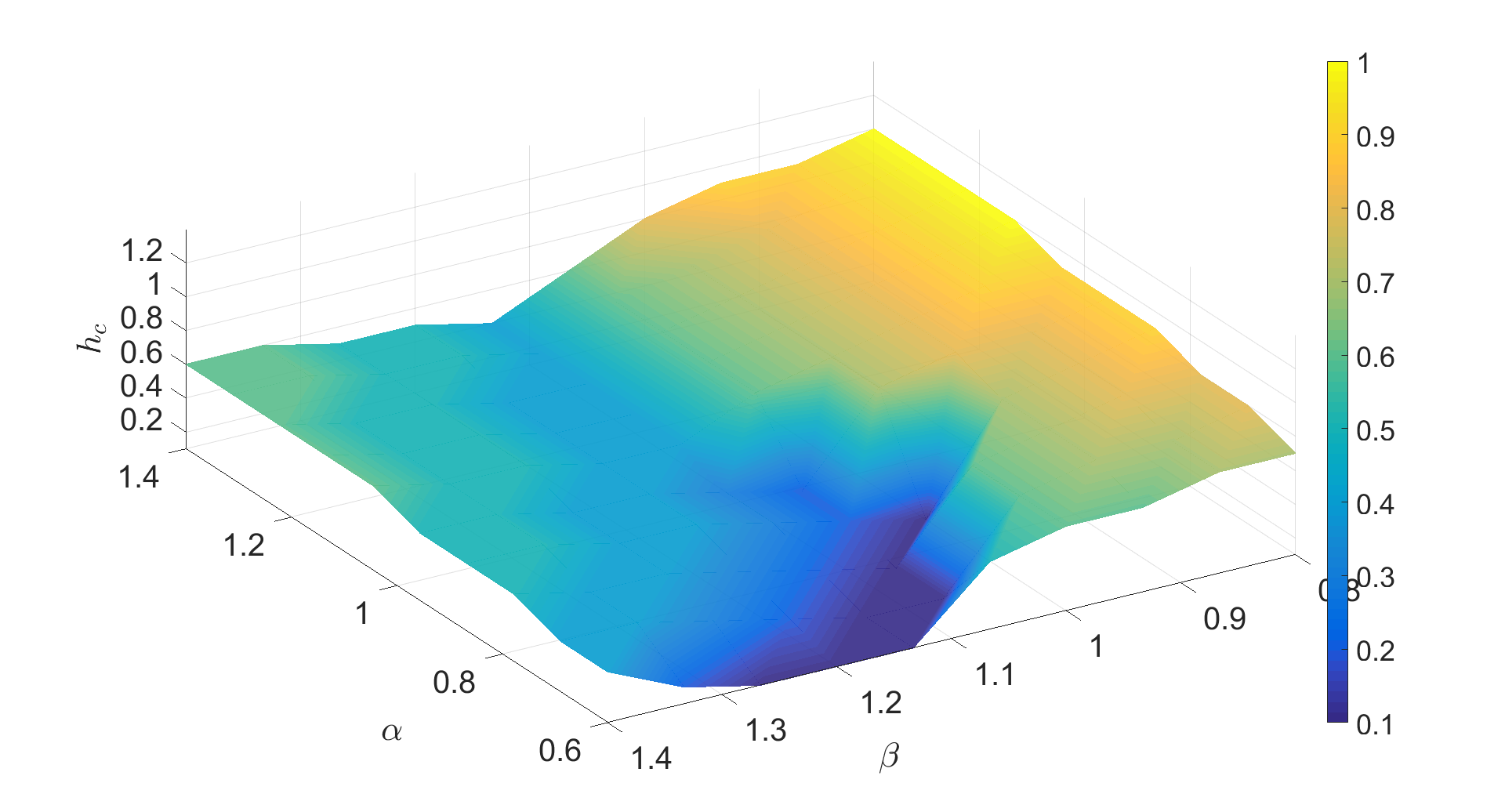}
\end{center}
\caption{Summary of numerical investigation for $c=0.25$ MP distribution with quartic potential. On the top, we have the evolution of the derivative of the potential $\bar{\mathcal{U}}_k^\prime(\bar{\chi})$. In the middle, the evolution of a specific value $\bar{\mathcal{U}}_k^\prime(\bar{\chi}_0)$. On the bottom, we can show the dependency of $h_c$ on the parameters $\alpha$ and $\beta$ defining the regulator family. On the left for $\tau(y)=1$ and on the right for $\tau(y) \neq 1$.}\label{figevolMP1}
\end{figure}

\begin{figure}
\begin{center}
\includegraphics[scale=0.15]{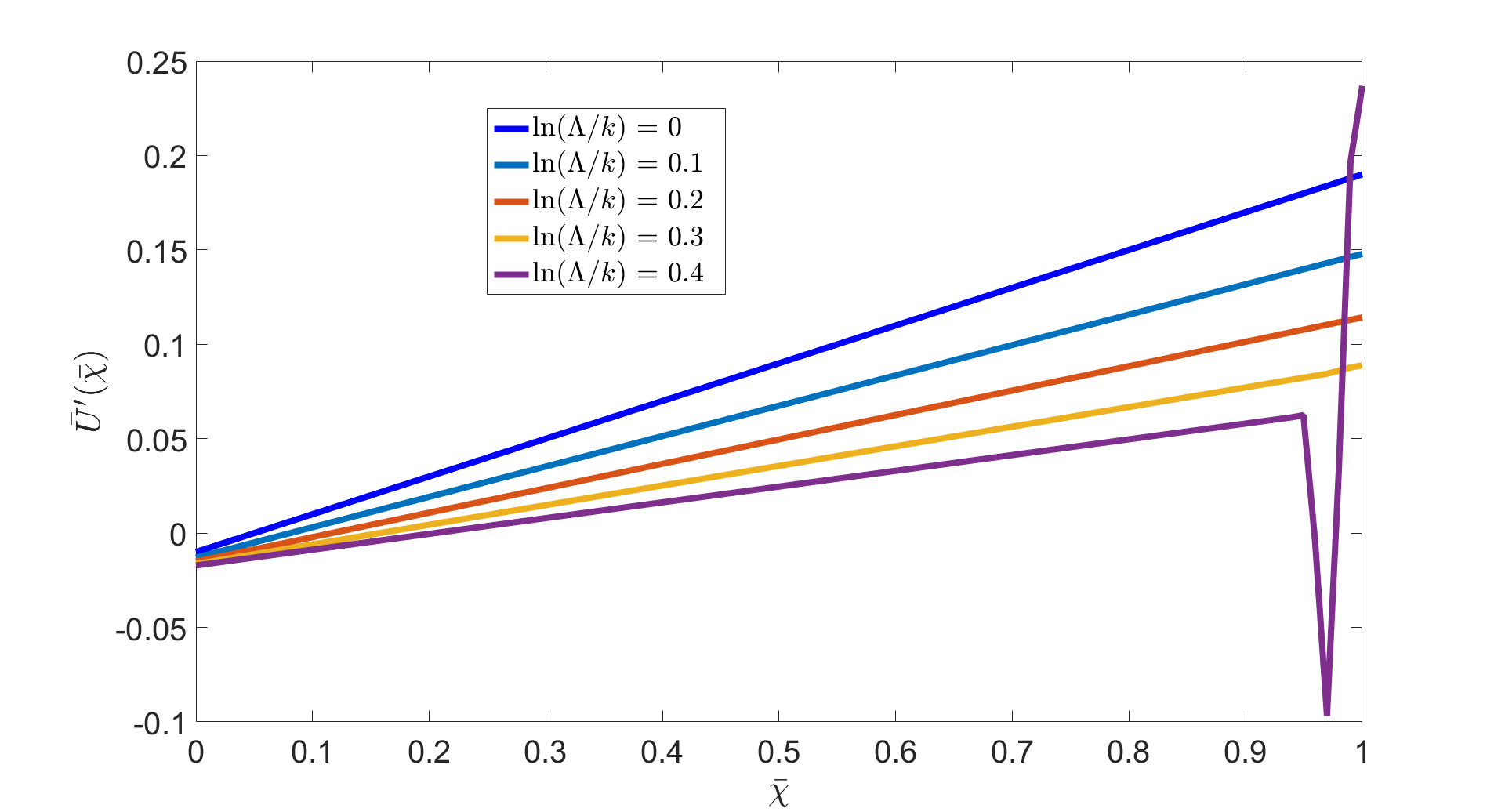}\,\includegraphics[scale=0.15]{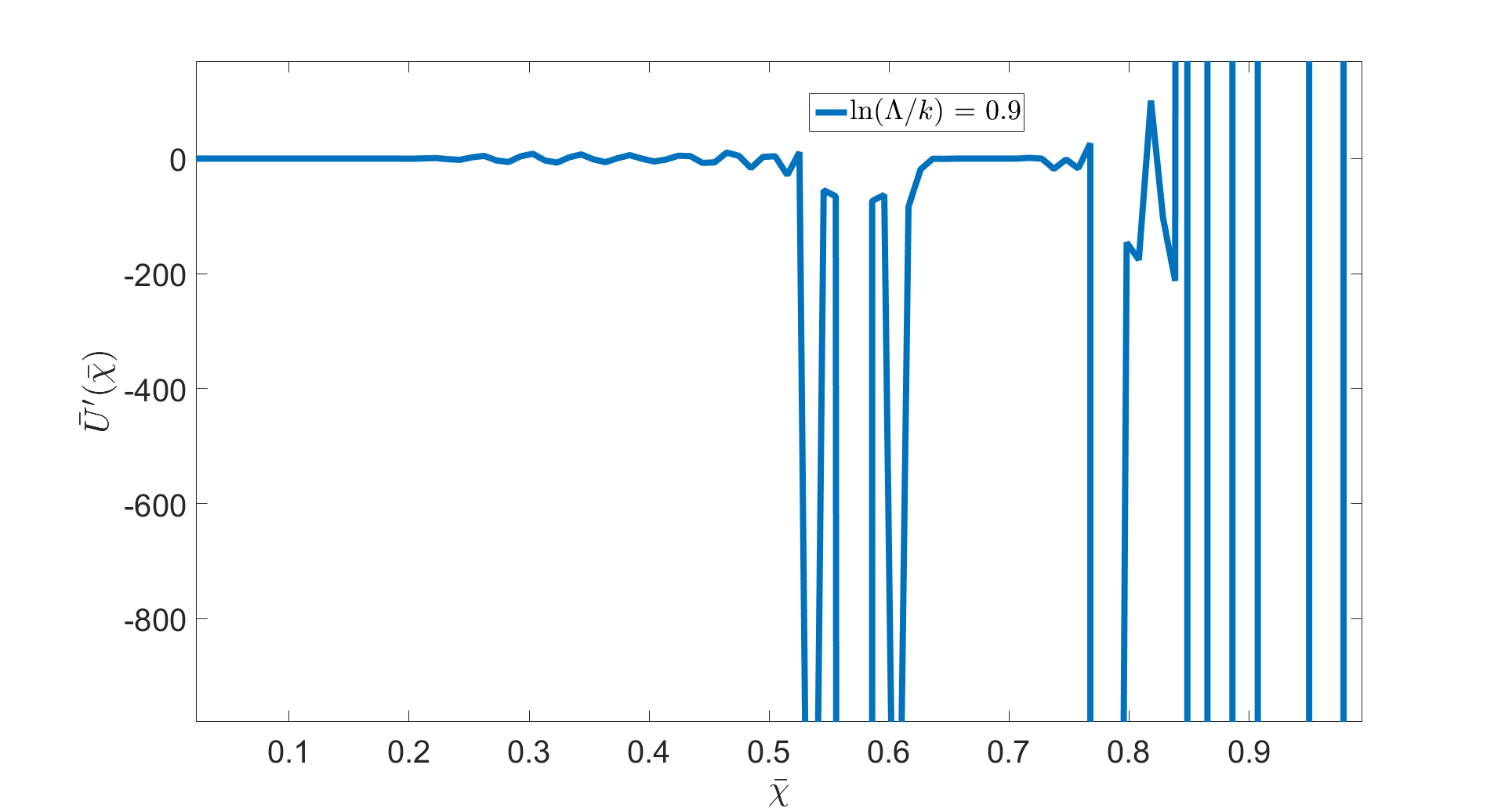}\\
\includegraphics[scale=0.15]{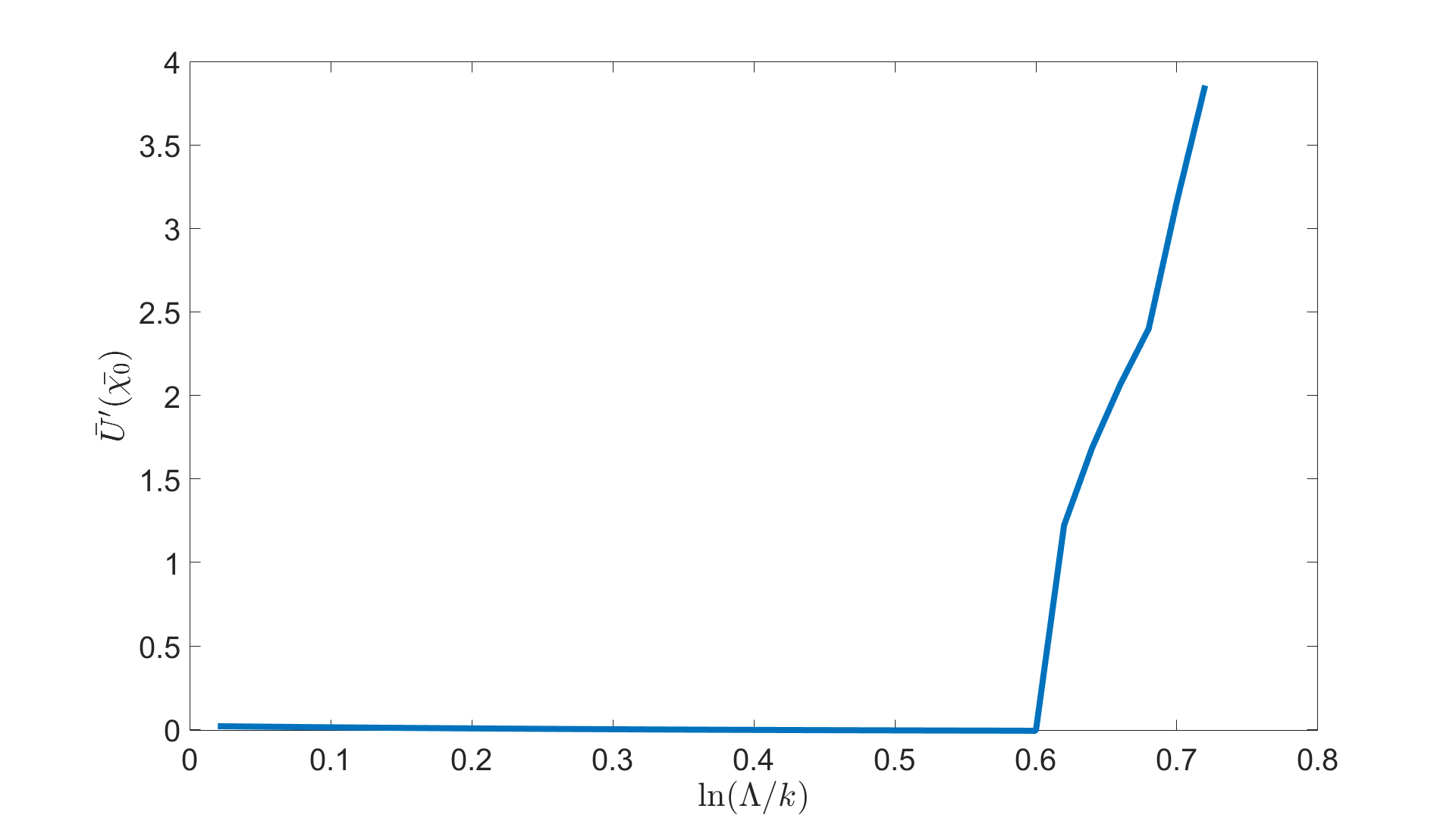}\,\includegraphics[scale=0.15]{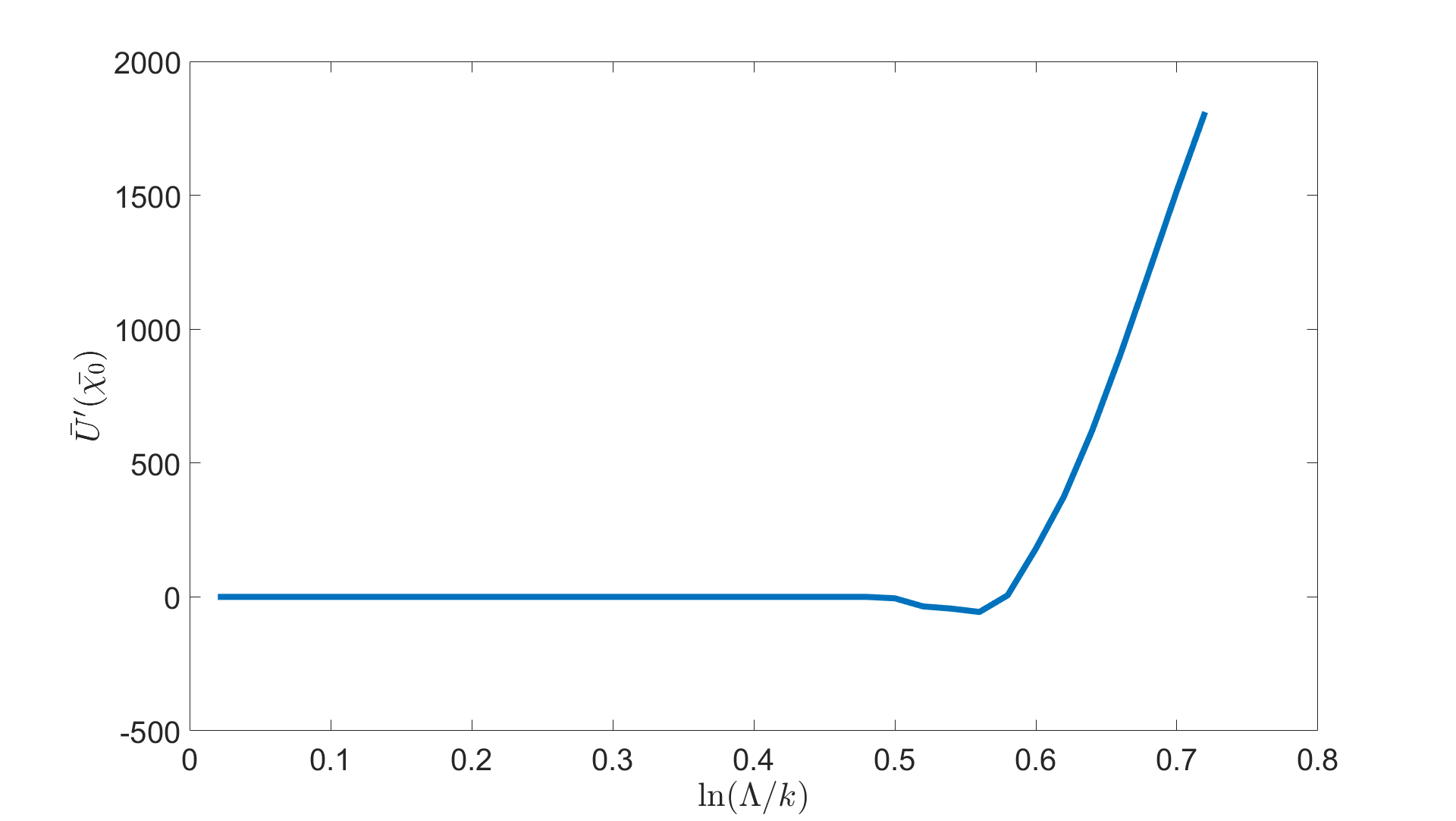}\\
\includegraphics[scale=0.15]{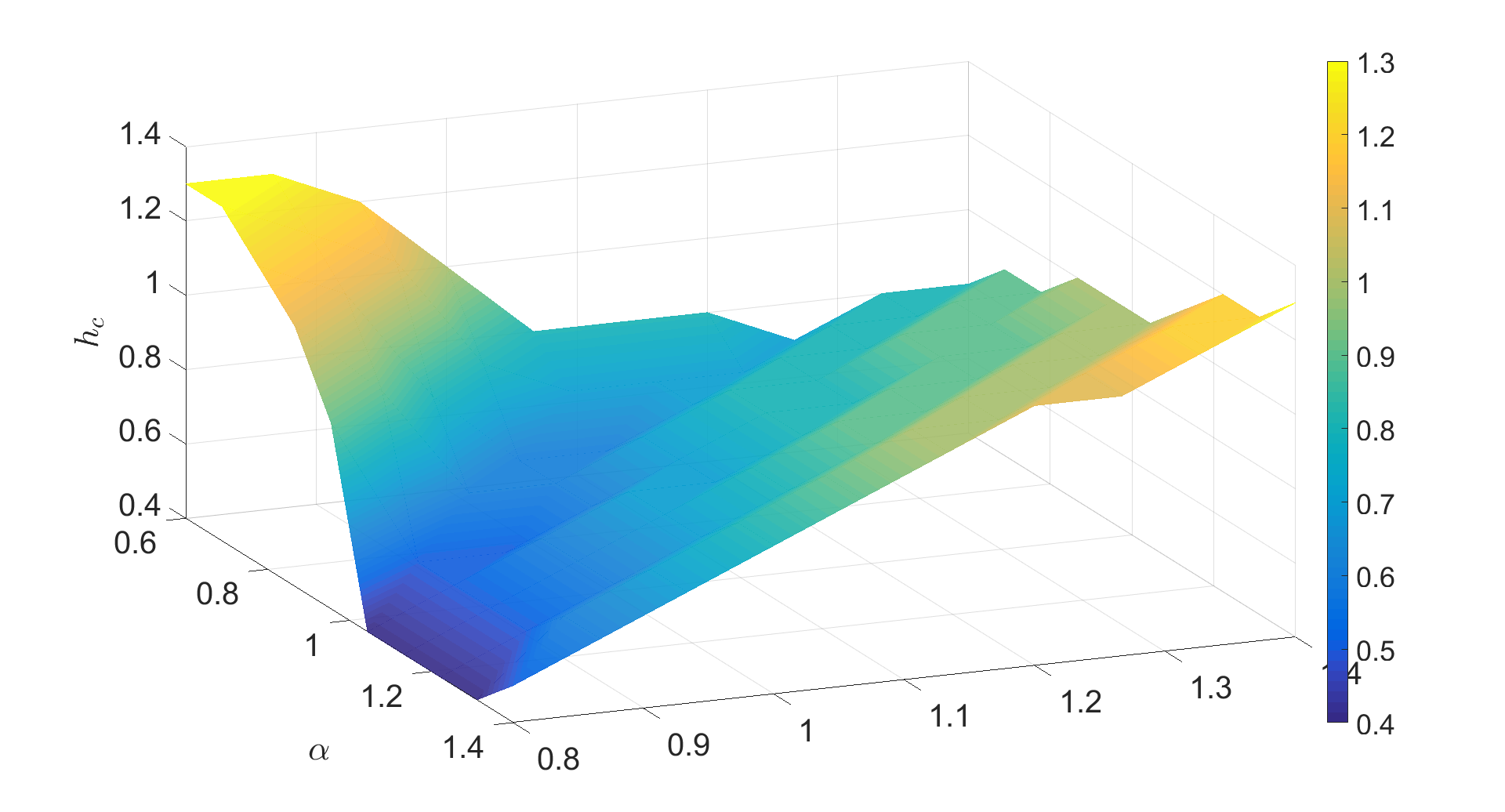}\,\includegraphics[scale=0.15]{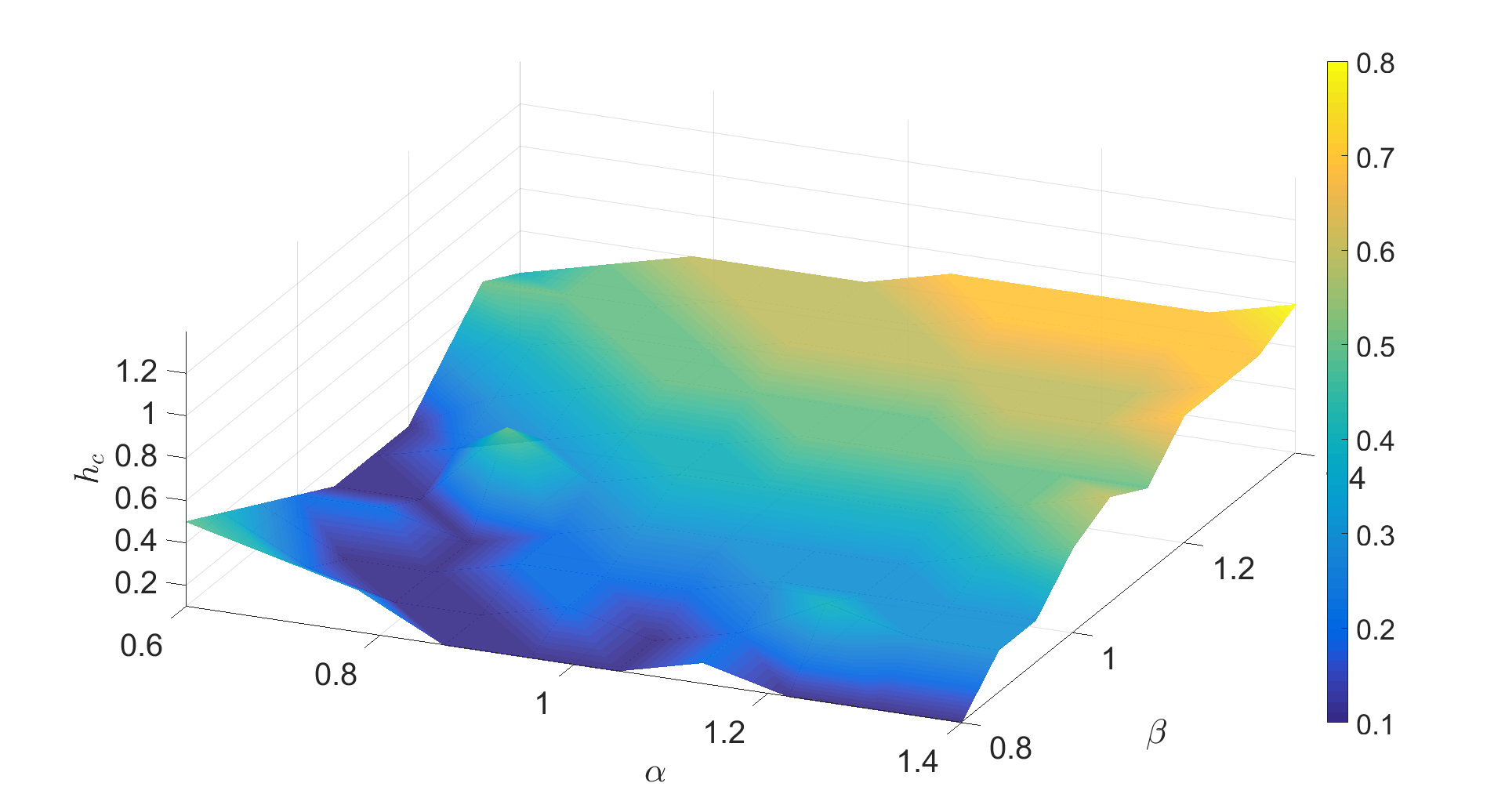}
\end{center}
\caption{Summary of numerical investigation for $c=1$ MP distribution with quartic potential. On the top, we have the evolution of the derivative of the potential $\bar{\mathcal{U}}_k^\prime(\bar{\chi})$. In the middle, the evolution of a specific value $\bar{\mathcal{U}}_k^\prime(\bar{\chi}_0)$. On the bottom, we can show the dependency of $h_c$ on the parameters $\alpha$ and $\beta$ defining the regulator family. On the left for $\tau(y)=1$ and on the right for $\tau(y) \neq 1$.}\label{figevolMP2}
\end{figure}

\medskip
Figures \ref{figevolMP1} and \ref{figevolMP2} summarize the results for MP law, respectively, for $c=0.25$ and $c=1$. We arrive at the same conclusions for MP as for Wigner distribution. In both cases, the flow becomes singular at a finite time below a critical surface, and the effective potential $U^\prime_k(\bar{\chi})$ diverges, and accordingly with the results of the previous sections, we interpret again the divergences as a breaking down of the equilibrium dynamics assumption. Once again, we show that a coarse-graining both in eigenvalues and frequency improves the reliability of the flow regarding the value of the critical mass $h_c$. However, the theoretical value $h_c=0$ (see Appendix \ref{appendix2}) is reached in smaller and more irregular regions than for Wigner law. 

\section{Two-particle irreducible formalism}\label{section2PI}

The 2PI formalism is especially relevant to address issues where the $2$-point function is a better order parameter than the $1$-point function. This is especially the case for disordered systems because in glassy phases the averaging of $q_i(t)$ generally vanishes. This formalism was first introduced by Luttinger and Ward \cite{kohn1960ground,luttinger1960ground}, see also \cite{benedetti20182pi} for a recent and pedagogical review and \cite{blaizot2021functional,dupuis2014nonperturbative} for self-consistent and comprehensives presentations of the formalism in the nonperturbative renormalization group framework.

\subsection{Formalism and $1/N$ expansion}

Let us provide a short presentation of formalism for our purpose. We use the convention of the section \ref{FRG}, and we denote as $\Phi=(\phi,\bar{\varphi})$ the doublet of fields, with components $\Phi_{\alpha,i}(t)$ at the time $t$. We denote as $S_k[\Phi]$ the classical action \eqref{classicalaction} which we added the regulator $\Delta S_k[\Phi]$ defined by \eqref{regulatordef}, and we introduce the generating functional of connected correlation functions:
\begin{equation}
W[J,\textbf{K}]=\ln \int d\Phi\, e^{- S_k[\Phi]+ \int dt \sum_{i,\alpha} \Phi_{i,\alpha}(t) J_{i,\alpha}(t)+\frac{1}{2} \int dt dt^\prime \sum_{i,j,\alpha,\beta} \Phi_{i,\alpha}(t) k_{i,\alpha;j,\beta}(t,t^\prime)  \Phi_{j,\beta}(t^\prime)}\,.\label{defW}
\end{equation}
In this equation the matrix $\textbf{K}$ with entries $k_{i,\alpha;j,\beta}(t,t^\prime)$ is assumed to be symmetric, both in the variables $t$ and $t^\prime$ and on the pairs $(i,\alpha)$ and $(j,\beta)$. To simplify the notations, we introduce the short notation $I=(i,\alpha)$, $J=(j,\beta)$ and so on. We thus have:
\begin{equation}
\frac{\delta k_{IJ}(t,t^\prime) }{ \delta k_{KL}(s,s^\prime)}= \frac{1}{2} \left( \delta_{IK} \delta_{JL} \delta(t-s)\delta(t^\prime-s^\prime) +\delta_{IL} \delta_{JK} \delta(t-s^\prime)\delta(t^\prime-s) \right)\,.
\end{equation}
\paragraph{2PI effective action.} As in section \ref{FRG} we denote as $\Xi_{a}(t)$ and $G_{k,ab}(t,t^\prime)$ the $1$ and $2$ point functions, obtained by deriving one and two times the generating functional $W[J,\textbf{K}]$. Hence, 
\begin{equation}
2\frac{\delta W}{\delta k_{ab}(t,t^\prime)}=G_{k,ab}(t,t^\prime)+\Xi_{a}(t)\Xi_{b}(t^\prime)\,.
\end{equation}
Now, we define the second Legendre transform, for fixed $1$ and $2$ point functions $\Gamma_{2,k}[\Xi,\textbf{G}]$ as:
\begin{align}
\nonumber \Gamma_{2,k}[\Xi,\textbf{G}]&:=-W_k[J,\textbf{K}]+\int dt \sum_a J_a(t) \Xi_a(t)\\
&+ \frac{1}{2}\int dt dt^\prime \sum_{a,b} \Xi_a(t) [k_{ab}(t,t^\prime)+R_{k,ab}(t-t^\prime)]\Xi_b(t^\prime) + \frac{1}{2} \Tr [\textbf{G}\textbf{K}]\,, \label{defLegendre}
\end{align}
and we have:
\begin{equation}
\frac{\delta \Gamma_{2,k}}{\delta \Xi_a(t)}=J_a(t)+\int dt^\prime \sum_b k_{ab}(t,t^\prime)  \Xi_b(t^\prime)\,,\quad \frac{\delta \Gamma_{2,k}}{\delta G_{ab}(t,t^\prime)}=\frac{1}{2} k_{ab}(t,t^\prime)\,.\label{quantummoveeq}
\end{equation}
\textit{on shell}, when the sources vanish, they give the effective dynamical equations of the theory, and causality must be restored as discussed in section \ref{Closed}. In \eqref{defW} we decompose $\Phi_{a}(t)=\Xi_a(t)+\xi_a(t)$ and expands the exponent in power of $\xi_a(t)$. This, integrating out the quadratic contribution, the definition \eqref{defLegendre} leads to (see Appendix \ref{appendix5}):
\begin{equation}
\boxed{\Gamma_{2,k}[\Xi,\textbf{G}]=S_k[\Xi]+\frac{1}{2}\Tr [\ln \textbf{G}^{-1}]+ \frac{1}{2}\Tr\, \textbf{G}\textbf{G}^{-1}_{0,k} + \Phi_k[\Xi,\textbf{G}]  \,,}\label{equationGamma2}
\end{equation}
where:
\begin{enumerate}
\item $\textbf{G}_{0,k}$ is the \textit{effective bare propagator} with entries:
\begin{equation}
(G_{0,k})_{ab}(t,t^\prime)=\frac{\delta^2 S_k}{\delta \Phi_{a}(t)\delta \Phi_{b}(t^\prime)}\,.
\end{equation}
\item $\Phi_k[\Xi,\textbf{G}]$ is the Ward-Luttinger functional (WLF), whose expansion start at two loops. Furthermore:
\begin{equation}
\boxed{\mathbf{\Sigma}:=2\frac{\delta \Phi_k}{\delta \textbf{G}}\,,} \label{equationselfenergy}
\end{equation}
is the self-energy, including only $1$PI contributions. 
\end{enumerate}
Taking the derivative with respect to the momentum, we recover \textit{on shell} the closed equation \eqref{closedSigma1}. In the same way, \textit{off shell}, we must have:
\begin{equation}
\boxed{\Sigma_{k,MM}\, \delta_{\lambda \lambda^\prime} \int dt =-\left(4\kappa_2 \,\int dt \int \mu(\lambda) d\lambda\, G_{k,\bar{\varpi} M}(\lambda,t,t)\right) \delta_{\lambda \lambda^\prime}\,.}
\end{equation}

\paragraph{$1/N$ expansion and quartic model.} As recalled in section \ref{largeN}, the vector field that we consider has a power counting and the perturbative series can be organized accordingly with a non-trivial $1/N$ expansion given with the scaling law $\sim N^{-\omega+1}$ with $\nu$ given by \eqref{scaling}. Hence, the WLF expands in power on $N$, $\Phi_k=\sum_{n=1}^\infty \Phi_k^{(1-n)}$, where $\Phi_k^{(1-n)}$ expands in diagrams involving Feynman diagrams with $\omega\equiv n$. At leading order we therefore retain:
\begin{equation}
\Phi_k = \Phi_{k,(1)}+\mathcal{O}(N^{-1})\,.
\end{equation}
Let us consider the quartic case. As recalled in section \ref{largeN}, the leading order diagrams are vacuum trees in the intermediate field formalism. But a tree is not 2PI, except for the primary of them, made of a single edge and two nodes, namely:
\begin{equation}
\Phi_{k,(1)}\equiv \vcenter{\hbox{\includegraphics[scale=1]{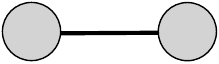}}}\,.
\end{equation}
Explicitly, in the large $N$ limit:
\begin{equation}
\Phi_{k,(1)}[\Xi,\textbf{G}]=2{\kappa_2} N \int \mu(\lambda) \mu(\lambda^\prime) d\lambda d\lambda^\prime\int dt \,G_{\bar{\varpi} M}(\lambda,t,t)G_{M M}(\lambda^\prime,t,t)\,,\label{WLF1}
\end{equation}
where $\kappa_2$ is defined in \eqref{potquart}. From the discussion given in the section \ref{Closed}, the WLF has to vanish \textit{on shell}, for $2$-point function solving the equation \eqref{quantummoveeq} for vanishing sources. 
\medskip

\paragraph{Renormalization group equation.} From the definition of $\Gamma_{2,k}$ given by equation \eqref{equationGamma2}, we have:
\begin{equation}
\frac{d}{ds}\Gamma_{2,k}[\Xi,\textbf{G}]=\frac{1}{2}\Tr\, \textbf{G}\dot{\textbf{G}}^{-1}_{0,k} +\frac{d}{ds} \Phi_k[\Xi,\textbf{G}]\,,
\end{equation}
where one more time, $s=\ln(k)$. Note that we keep both $\Xi$ and $\textbf{G}$ fixed in the derivation. On the other hand, from definition \eqref{defLegendre}, we get straightforwardly:
\begin{align}
\frac{d}{ds}\Gamma_{2,k}[\Xi,\textbf{G}]:=-\frac{\partial}{\partial s}W_k[J,\textbf{K}]+ \frac{1}{2}\int dt dt^\prime \sum_{a,b} \Xi_a(t) \frac{\partial R_{k,ab}}{\partial s}(t-t^\prime)\Xi_b(t^\prime)\,. \label{defLegendre2bis}
\end{align}
The partial derivative of $W_k[J,\textbf{K}]$ can be computed exactly as:
\begin{equation}
\frac{\partial}{\partial s}W_k[J,\textbf{K}]= \frac{1}{2} \Tr\frac{\partial}{\partial s} \textbf{R}_k (\textbf{G}+ \mathbf{\Xi}\otimes \mathbf{\Xi})= \frac{1}{2} \Tr \dot{\textbf{G}}^{-1}_{0,k}(\textbf{G}+ \mathbf{\Xi}\otimes \mathbf{\Xi})\,.
\end{equation}
Therefore:
\begin{equation}
\frac{d}{ds}\Gamma_{2,k}[\Xi,\textbf{G}]=\frac{1}{2}\Tr\, \textbf{G}\dot{\textbf{G}}^{-1}_{0,k}\,,
\end{equation}
and we conclude that the WLF is an RG invariant:
\begin{equation}
\boxed{\frac{d}{ds} \Phi_k[\Xi,\textbf{G}]=0\,.}
\end{equation}
Hence, on shell, the dependence on $k$ of $ \Phi_k[\Xi,\textbf{G}_{k}]$ is only through the dependency on $k$ of $\textbf{G}_{k}$:
\begin{equation}
\frac{d}{ds} \Phi_k[\Xi,\textbf{G}]= \Tr \frac{\delta \Phi_k}{\delta \textbf{G}} \frac{d \textbf{G}_k}{ds} = \frac{1}{2}\Tr  \mathbf{\Sigma}_k \frac{d \textbf{G}_k}{ds}\,,
\end{equation}
and generally:
\begin{equation}
\frac{d}{ds} \Phi_k^{(n)}[\Xi,\textbf{G}]=\Tr  \frac{\delta  \Phi_k^{(n)}}{\delta \textbf{G}_k}\frac{d \textbf{G}_k}{ds}=-\Tr  \frac{\delta  \Phi_k^{(n)}}{\delta \textbf{G}_k} \textbf{G}_k\frac{d \textbf{G}^{-1}_k}{ds}\textbf{G}_k\,,
\end{equation}
where $\Phi_k^{(n)}$ denotes the $n$-th functional derivative with respect to $\textbf{G}_k$. 

\subsection{Ginzburg-Landau expansion near critical temperature}

In this section we assume the following decomposition for the solution of the equation of motion:
\begin{equation}
G_{k,\bar{\varpi} M}(\lambda,t,t^\prime)=\bar{G}_{k,\bar{\varpi} M}(\lambda,t^\prime-t)+\Omega_k(\lambda,t,t^\prime)\,,\label{2ptsFunction}
\end{equation}
\begin{equation}
G_{k,MM}(\lambda,t,t^\prime)=\bar{G}_{k,MM}(\lambda,t^\prime-t)+\Phi_k(\lambda,t,t^\prime)\,,\label{2ptsFunction2}
\end{equation}
where $\bar{G}_{k,\bar{\varpi} M}\propto \theta(\tau)$ is the expected $2$-point function in the \textit{normal phase} (i.e., high-temperature phase), assumed to be translation invariant in time and satisfying fluctuation-dissipation theorem \cite{Aron_2010}. The contributions $\Omega_k$ and $\Phi_k$ are a bi-local time-dependent order parameters, assumed to be small, that breaks explicitly the time-translation symmetry. To the lowest order, one expects this is equivalent to write the self energy $\Sigma_{k,\bar{\varpi} M}$ as:
\begin{equation}
\Sigma_{k,\bar{\varpi} M}(\lambda,t,t^\prime)=\bar{\Sigma}_{k,\bar{\varpi} M}(\lambda, t^\prime-t)+\Delta_k(\lambda,t,t^\prime) \,,
\end{equation}
where $\Delta_k$, assumed to be small, is bi-local in time as well. Indeed, from the Dyson equation we have in Fourier space:
\begin{align}
\nonumber G_{k,\bar{\varpi} M}(\lambda,\omega,\omega^\prime)&=\bar{G}_{k,\bar{\varpi} M}(\lambda,\omega)\delta_{\omega,-\omega^\prime}\\\nonumber
&+ \bar{G}_{k,\bar{\varpi} M}(\lambda,\omega)\Delta_k (\lambda,-\omega,-\omega^\prime) \bar{G}_{k,\bar{\varpi} M}(\lambda,-\omega^\prime)+\mathcal{O}(\Delta_k^2)\,.
\end{align}
Hence:
\begin{equation}
\Omega_k(\lambda,\omega,\omega^\prime)=\bar{G}_{k,\bar{\varpi} M}(\lambda,\omega)\Delta_k (\lambda,-\omega,-\omega^\prime) \bar{G}_{k,\bar{\varpi} M}(\lambda,-\omega^\prime)+\mathcal{O}(\Delta_k^2)\,.
\end{equation}
%with:
%\begin{equation}
%F(\tau):=\int_0^\infty dt_1\, \bar{G}_{k,\bar{\varpi} M}(t_1)\int_0^{\tau-t_1} dt_2\,\bar{G}_{k,\bar{\varpi} M}(t_2)\,.
%\end{equation}
%Assuming $ \int dt\, \bar{G}_{k,\bar{\varpi} M}(t)<\infty$, we have for $\tau$ large enough:
%\begin{equation}
%\Omega_k= \Delta_k\, \left( \int dt\, \bar{G}_{k,\bar{\varpi} M}(t)\right)^2+\mathcal{O}(\delta_k^2)\,.
%\end{equation}
%The decomposition \eqref{2ptsFunction} agrees with the large-scale behavior expected for the spherical model -- see \eqref{asymptoticbehavior}. Indeed, for $t^\prime, t \to \infty$ such that $t/t^\prime =: \varepsilon < 1$, $\theta(t^\prime-t)=\theta(t^\prime(1-\varepsilon))=1$. In contrast, as $\varepsilon > 1$, $\theta(t^\prime-t)=0$. For $\varepsilon=1$, it depends on the convention for the Heaviside function and analytic augments of Appendices \ref{appendix1} and \ref{appendix2} show that it must be convenient to choose: $\theta(0)=0$. 

Near the transition, we expect that both $\Omega_k$ and $\Delta_k$ are small enough to make a power expansion valid. From equation \eqref{equationselfenergy}, we introduce the Legendre transform $F_k$ of $\Phi_k$:
\begin{equation}
F_k[\mathbf{\Sigma}]=\Phi_k[\mathbf{G}]+ \frac{1}{2}\Tr\, \textbf{G}\mathbf{\Sigma}\,,
\end{equation}
where we disregarded the classical field $\Xi$. We thus rewrite the effective action \eqref{equationGamma2} as (up to irrelevant constants):
\begin{align}
\nonumber\Gamma_{2,k}[\textbf{G}]&=\frac{1}{2}\Tr [\ln (\textbf{G}^{-1}_{0,k}-\mathbf{\Sigma})]+ \frac{1}{2}\Tr\, \textbf{G}\mathbf{\Sigma} + \Phi_k[\Xi,\textbf{G}]\\
&=\frac{1}{2}\Tr [\ln (\textbf{G}^{-1}_{0,k}-\mathbf{\Sigma})]+F_k[\mathbf{\Sigma}]\,,
\end{align}
And we define $\mathbf{\Sigma}:=\mathbf{\Sigma}_\text{eq}+\mathbf{\Delta}$ and  $\tilde{\Gamma}_{2,k}[\mathbf{\Sigma}]$ as:
\begin{equation}
\tilde{\Gamma}_{2,k}[\mathbf{\Sigma}]=\frac{1}{2}\Tr [\ln (\textbf{G}^{-1}_{0,k}-\mathbf{\Sigma})]+F_k[\mathbf{\Sigma}]\,,
\end{equation}
such that on shell we must have:
\begin{equation}
\frac{\delta \tilde{\Gamma}_{2,k}[\mathbf{\Sigma}]}{\delta \Sigma_{\alpha\beta}}=0\,.
\end{equation}
We assume that on shell function $\mathbf{\Sigma}_\text{k,eq}$ is known. Hence, defining $\tilde{\gamma}_{2,k}[\mathbf{\Delta}]:
=\tilde{\Gamma}_{2,k}[\mathbf{\Sigma}]-\tilde{\Gamma}_{2,k}[\mathbf{\Sigma}_\text{eq}]$, the equilibrium conditions for $\Delta$ is defining as:
\begin{equation}
\frac{\delta \tilde{\gamma}_{2,k}[\mathbf{\Delta}]}{\delta \mathbf{\Delta}}\big\vert_{\mathbf{\Sigma}_\text{eq}=\mathbf{\Sigma}_\text{k,eq}}=0\,.
\end{equation}
The bi-local matrix element $\mathbf{\Delta}\equiv \mathbf{\Delta}(\lambda,\omega,\omega^\prime)$ can be expanded in power of external frequencies and momentum $\lambda$. At zero order, it is a pure number, depending on $k$, and as a first approximation we can track the flow of this zero order contribution. We consider the following ansatz:
\begin{equation}
\mathbf{\Sigma}_{k}(\lambda,\omega,\omega^\prime)=\begin{pmatrix}
0 & \bar{\Sigma}_{k,\bar{\varphi}\phi}\\
\bar{\Sigma}_{k,\bar{\varphi}\phi} & 0
\end{pmatrix} \delta(\omega+\omega^\prime)+\begin{pmatrix}
0 & \Delta_k\\
\Delta_k & 0
\end{pmatrix}
\end{equation}
where $\bar{\Sigma}_{k,\bar{\varphi}\phi}$ is solution of the zero order closed equation \eqref{zeroorderclosed}, and matrix elements are in the ‘‘$\bar{\varphi} \phi$'' space. Focusing on the quartic theory, the closed equation for $\mathbf{\Sigma}_{k}$ is determined by the equations \eqref{equationselfenergy} and \eqref{WLF1}. Explicitly:
\begin{equation}
\Delta_k=-2\kappa_2 \, \int d\lambda \,\mu(\lambda)\int \frac{d\omega}{2\pi}\, \Phi_k(\lambda,\omega,-\omega)\,.
\end{equation}
Let us introduce the notation:
\begin{equation}
\bar{\mathbf{G}}_k(\lambda,\omega)\delta(\omega+\omega^\prime)=: \begin{pmatrix}
0 & A_k(-\omega)\\
A_k(\omega) & -A_k(\omega) (1+r_k(p)\rho^{(2)}_k(\omega)) A_k(-\omega) 
\end{pmatrix}\delta(\omega+\omega^\prime)\,,
\end{equation}
where $A_k$ is given by the solution of the closed $2$-point function equation \eqref{equationselfenergy} at order $0$ in $\Delta_k$. Furthermore, $\Omega_k$ can be computed as a perturbation series in $\Delta_k$ from the Dyson equation. Because we assume the condition \eqref{conditioncausal}, i.e. $\int d\omega\, A(\omega)=0$, we get:
\begin{align}
\nonumber -\Phi_k(\lambda,\omega,-\omega)=&  (A_k(\omega)A^2_k(-\omega)+A_k(-\omega)A^2_k(\omega))(1+r_k(p)\rho^{(2)}_k(\omega))\Delta_k\\
&+ 2\pi L_1(k) A_k(\omega)A_k(-\omega)\Delta_k^2+\mathcal{O}(\Delta_k^3)\,,
\end{align}
with:
\begin{equation}
L_1(k):=\int d\lambda \mu(\lambda)\int \frac{d\omega}{2\pi} A_k(\omega)(1+r_k(p)\rho^{(2)}_k(\omega))A_k(-\omega)\,.
\end{equation}
We furthermore define the integrals:
\begin{equation}
L_2(k):=\int d\lambda \mu(\lambda)\int \frac{d\omega}{2\pi} A_k(\omega)A_k(-\omega)\,,
\end{equation}
and
\begin{equation}
K(k):=\int d\lambda \mu(\lambda)\int \frac{d\omega}{2\pi} A_k(\omega)A_k^2(-\omega)(1+r_k(p)\rho^{(2)}_k(\omega))\,,
\end{equation}
such that the equation fixing the gap equation for $\Delta_k$ reads:
\begin{equation}
U^\prime(\Delta_k):=(a(k)-1) \Delta_k + b(k) \Delta_k^2=0\,, \label{eqdelta}
\end{equation}
with:
\begin{equation}
a(k):=4\kappa_2 K(k)\,,\qquad b(k):=4\pi \kappa_2 L_1(k)L_2(k)\,.
\end{equation}
The physical meaning of equation \eqref{eqdelta} can be analyzed numerically, and it is the topic of the next section. 
\medskip

\subsection{Numerical investigations}\label{numeric}
\begin{figure}
\begin{center}
\includegraphics[scale=0.5]{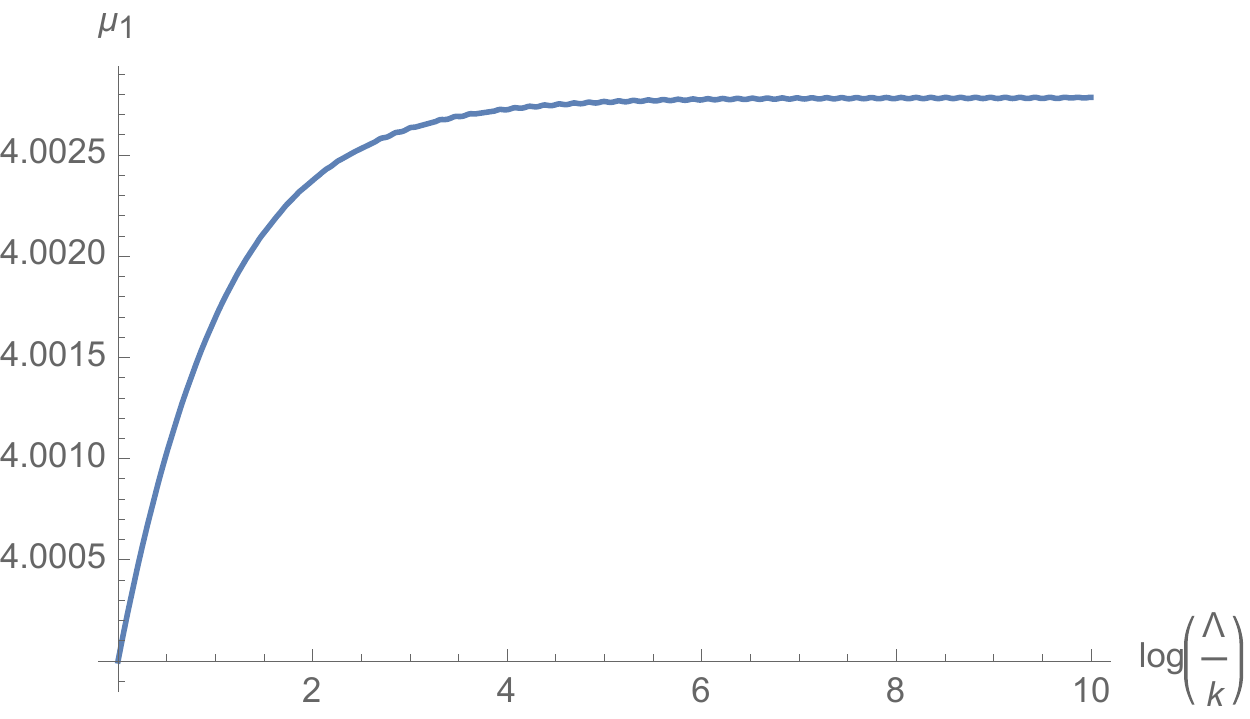}\qquad\includegraphics[scale=0.5]{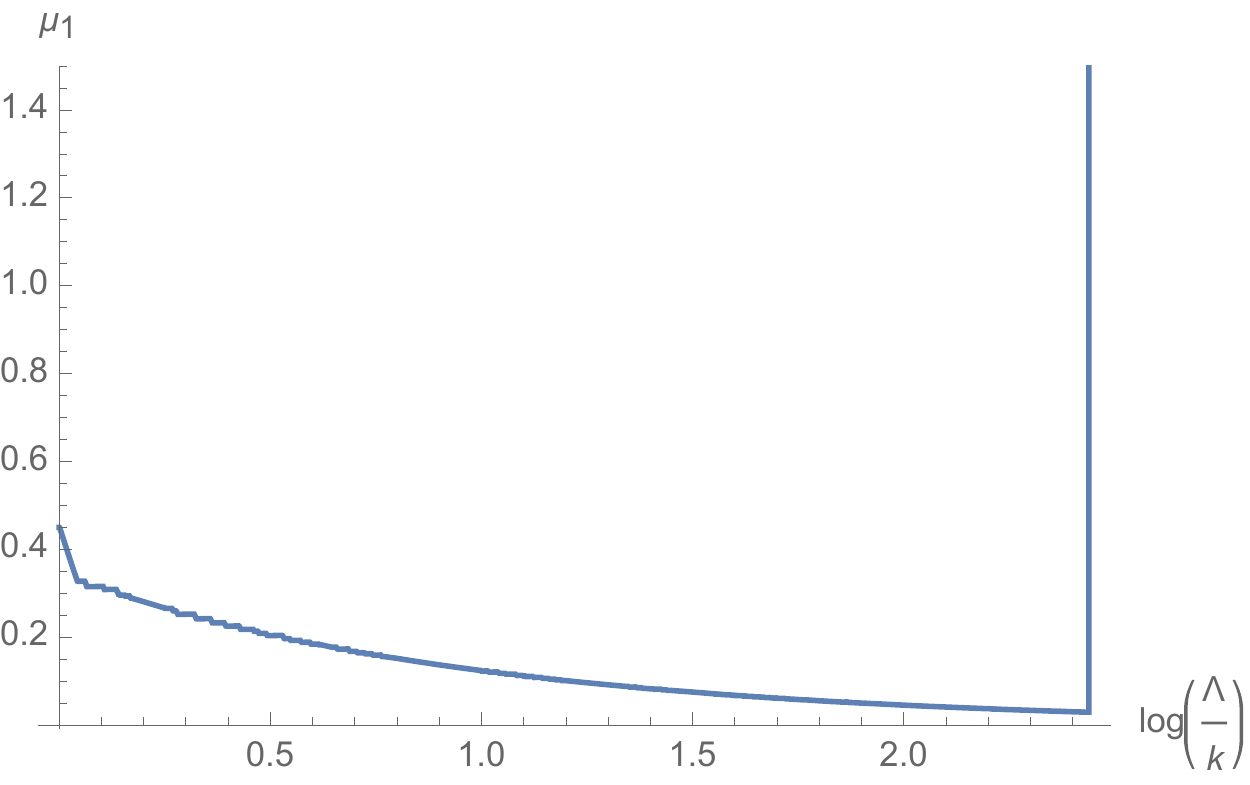}
\end{center}
\caption{Evolution of the mass parameter $\mu_1$ in the UV regime. Above the critical regime on the left and below the critical regime on the right.}\label{figevolmass}
\end{figure}

\begin{figure}
\begin{center}
\includegraphics[scale=0.5]{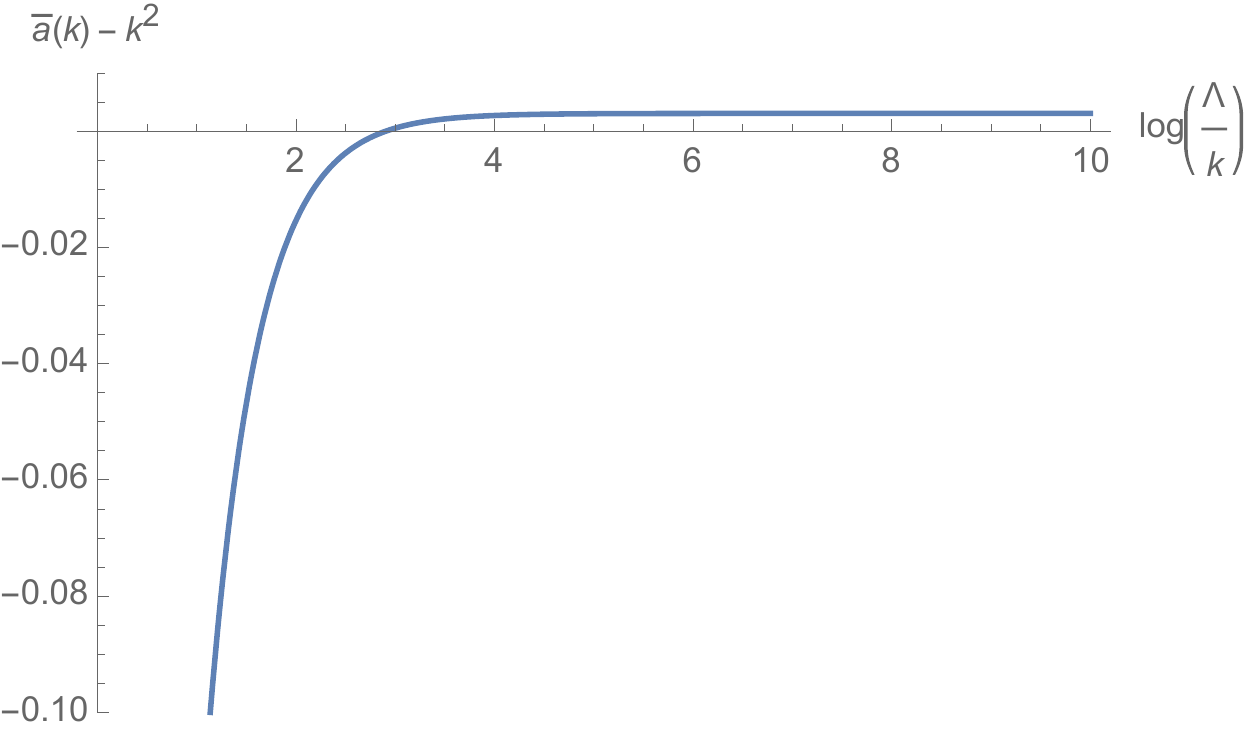}\qquad \includegraphics[scale=0.5]{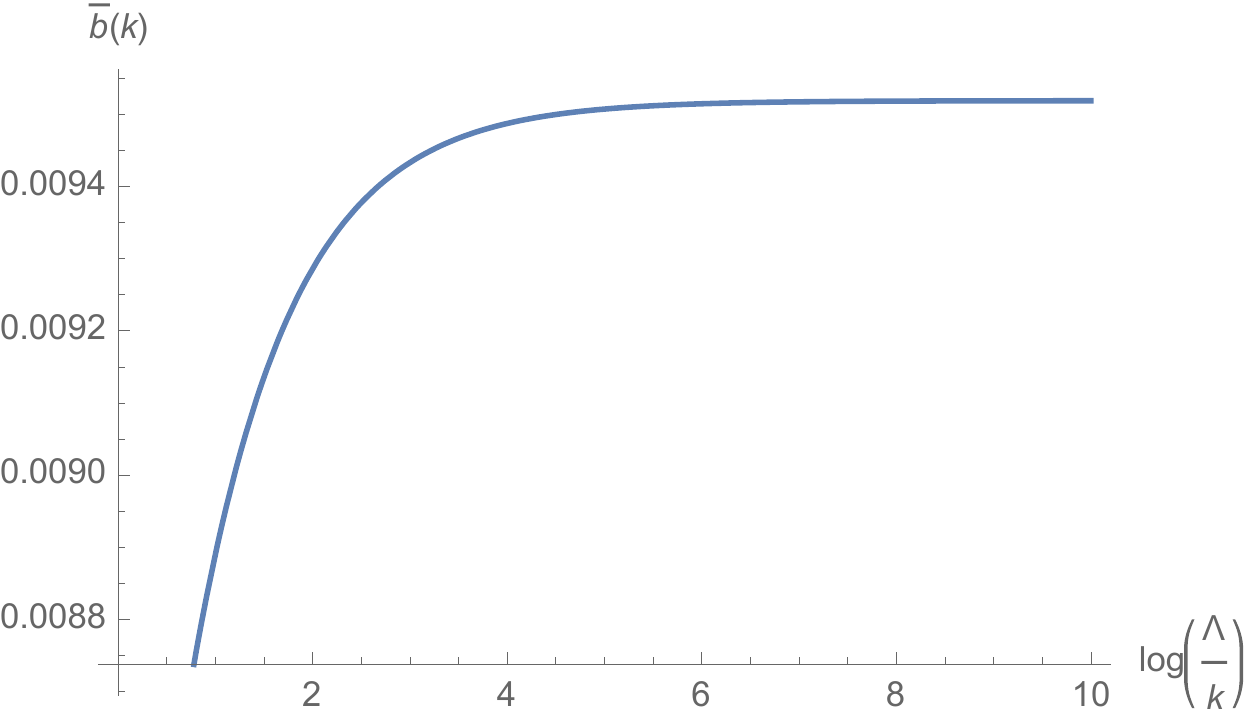}
\end{center}
\caption{Evolution of $\bar{a}(k)$ and $\bar{b}(k)$, the dimensionless versions of couplings involved in $U[\Delta_k]$ above the critical regime.}\label{figabTsupp}
\end{figure}

\begin{figure}
\begin{center}
\includegraphics[scale=0.8]{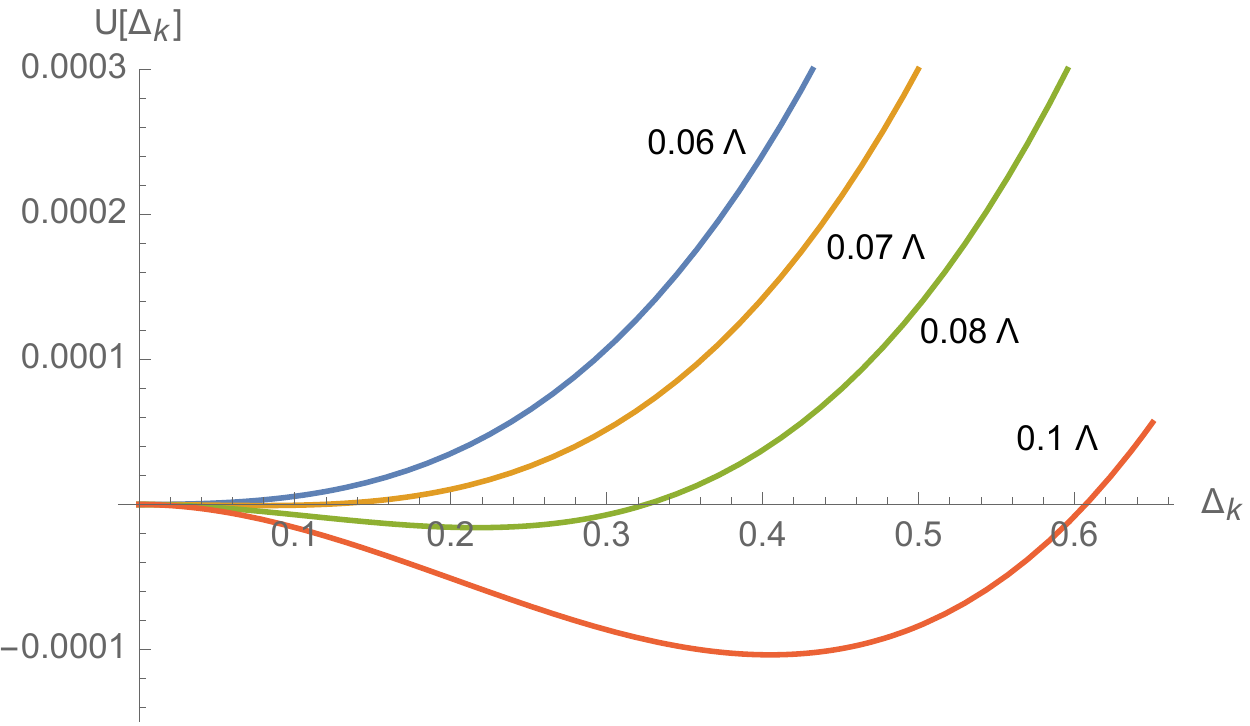}
\end{center}
\caption{Symmetry restoration in the equilibrium phase.}\label{figdeltaTsupp}
\end{figure}

\begin{figure}
\begin{center}
\includegraphics[scale=0.5]{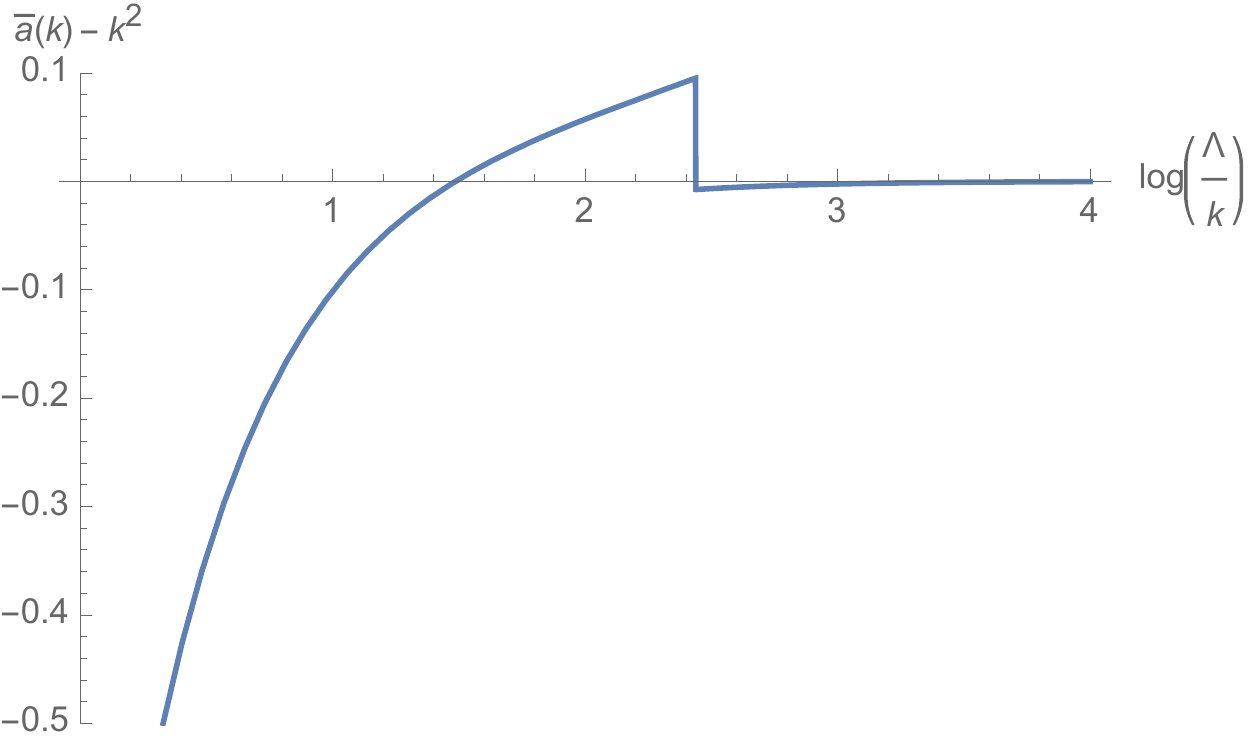}\qquad \includegraphics[scale=0.5]{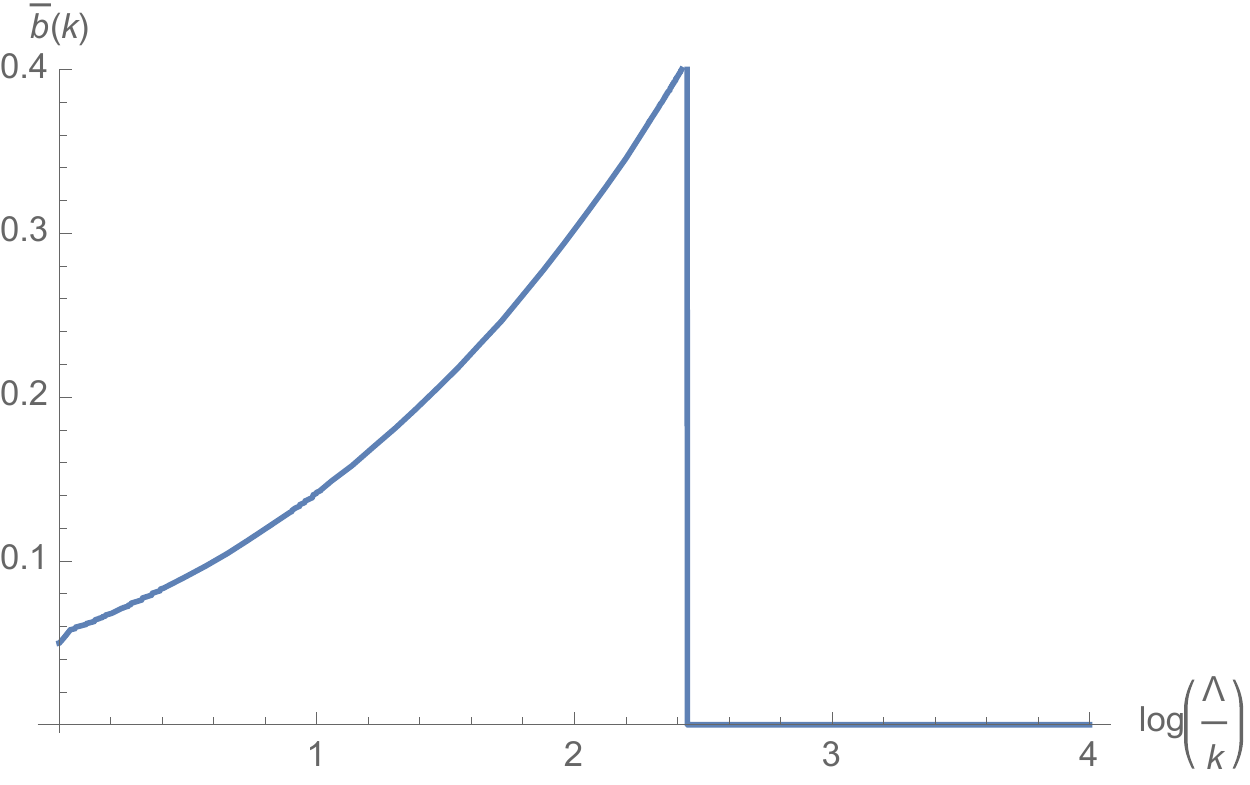}\\
\includegraphics[scale=0.5]{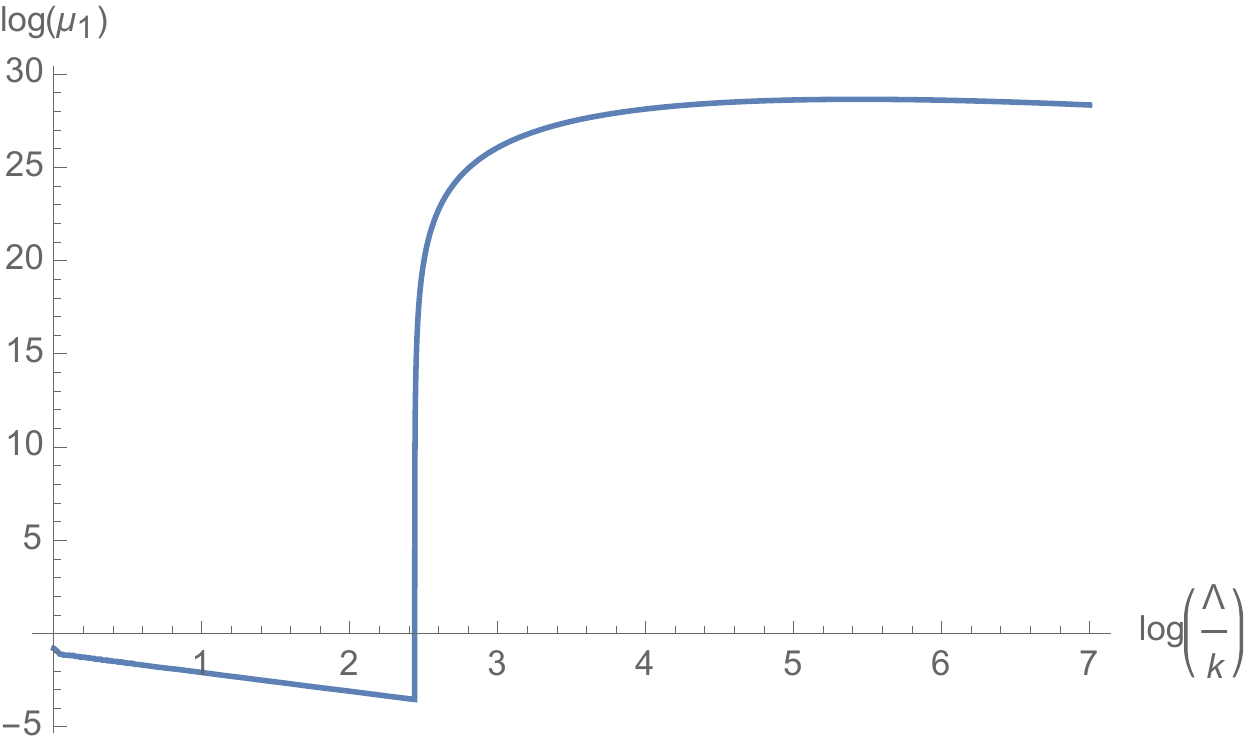}\qquad \includegraphics[scale=0.5]{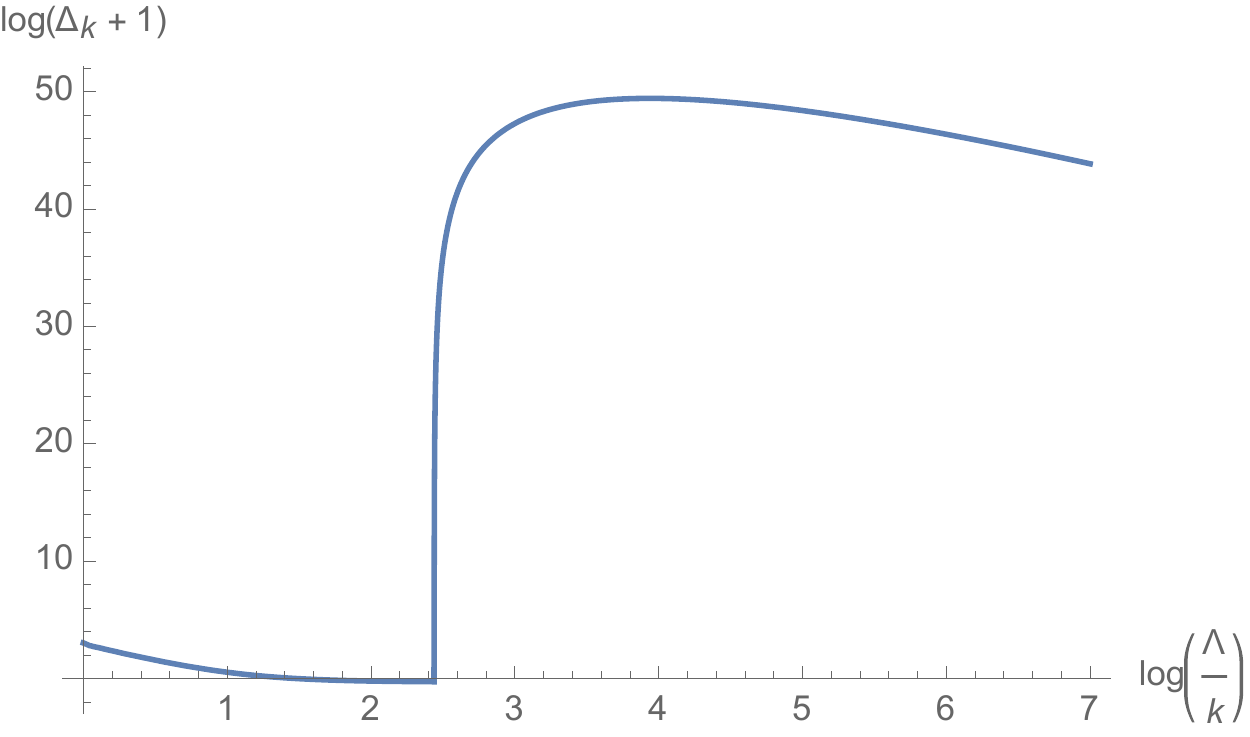}
\end{center}
\caption{Evolution of couplings $a$ and $b$ (on the top) and of $\ln (\Delta_k+1)$ and $\ln \mu_1$ (on the bottom).}\label{figdeltaTinf}
\end{figure}

The numerical results are summarized on Figures \ref{figevolmass}, \ref{figabTsupp}, \ref{figdeltaTsupp} and \ref{figdeltaTinf} to highlight the meaning of the divergences occurring at finite timescale, as Figure \ref{figflow} shows in the deep IR regime. On Figure \ref{figevolmass} one can show the same behavior in the deep UV. Above the critical regime, in the equilibrium phase, the mass $\mu_1(k)$ converges toward a finite value after a short transition time. In contrast, out of the equilibrium phase dynamics, the flow has an angular point (i.e., non-derivable) and diverges at finite timescale. 
In the equilibrium phase, the behavior of couplings $a(k)$ and $b(k)$ occurring on the definition of $U(\Delta_k)$, that reach a finite value after a transition regime also. The quantity $b(k)$ is positive, but $(a(k)-1)$ changes in sign along the trajectory. It is negative in the deep UV, but converges toward a positive value in the IR. This corresponds to a standard symmetry restoration scenario, and ultimately stability requires $\Delta_k=0$ for large scales, as Figure \ref{figdeltaTsupp} shows explicitly. 
Below the critical regime, where ergodicity is expected to be broken (i.e., out of equilibrium dynamics), things are significantly different, and Figure \ref{figdeltaTinf} summarizes what happens. At the beginning, the behavior for $a(k)$ and $b(k)$ is reminiscent of what happens for high temperature on Figure \ref{figabTsupp}. In particular, $b(k)$ is positive and $a(k)-1$ is negative, but reaches a positive value and symmetry is restored for a first time. However, at the point where the mass diverges, $a(k)-1$ reaches spontaneously a negative value again whereas $b(k)$ almost vanishes in the same time. This change makes the non-zero vacuum solution $\Delta_k\neq 0$ stable, and the Figure \ref{figdeltaTinf} shows the behavior of $\ln (\Delta_k+1)$. For $\log(\Lambda/k)$ small enough, the function is positive, and the symmetry is broken. It is restored for $\log(\Lambda/k)\approx 1.5$ until the point where the mass has a singular point, for $\log(\Lambda/k)\approx 2.4$. Until this point $\ln (\Delta_k+1)$ takes large and positives values, after what it reaches a smooth decreasing regime. 

\section{Concluding remarks}\label{conclusion}

In this paper, we introduced a nonperturbative renormalization group framework to investigate the large time equilibrium dynamics behavior of the $p=2$ soft spin model. We focused on the large $N$ limit where eigenvalues of the disorder matrix $J_{ij}$ follow the Wigner or Marchenko-Pastur distributions. The main originality of this approach is to avoid the inevitable bi-local interactions in time arising when the disorder is integrated out \cite{Lahoche:2021tyc}. Hence, local potential can be considered as a suitable approximation scheme to solve the exact RG flow. Because the eigenvalues of the disorder look like a fictitious momentum, we constructed RG based on a coarse-graining both in frequencies and eigenvalues through a choice of regulators compatible with causality and time reversal symmetry. We considered indeed many popular approximation schemes to solve exact flow equation in this paper: (i) The vertex expansion, (ii) the effective vertex expansions, (iii) the LPA around non-zero vacuum and (iv) the 2PI formulations were considered.   
For all these approximations, we investigated the reliability regarding analytic inputs, especially for the quartic potential. Our results, especially using local potential and non-zero vacuum expectation, are in qualitative agreements with analytical insights. Furthermore, we show that reliability is improved, regarding the choice of the regulator, for a coarse graining both in eigenvalues and frequency with respect to a coarse-graining only in frequency. 
\medskip

However, several aspects have been left out for future works. First, although we have essentially limited ourselves to quartic potentials in our numerical investigations in order to evaluate their reliability, our formalism is suitable for the study of any initial condition. Moreover, although we have limited ourselves to Wigner and Wishart ensembles, for which the inputs of the matrices are independent, this choice is also more restrictive than the formalism, and we could consider non-Gaussian distributions for the inputs of the matrices, or even distributions that do not follow an analytical law \cite{Potters1}. These aspects should be considered in forthcoming works, especially regarding signal detection issue discussed in the introduction, or for dynamics of disordered quantum systems \cite{rokni2004dynamical}. Finally, let us notice that our formalism neglect finite size effects \cite{fyodorov2015large}. These effects are totally blinded by the limit $N\to \infty$, which is assumed to justify the replacement of sums by integrals involving analytic distributions. 
\medskip

\paragraph{Acknowledgement:} VL addresses special thanks to “the little star” for his constant inspiration.

\pagebreak
\appendix

\begin{center}
\LARGE{Appendices}
\end{center}

\section{A short review on the $p=2$ spherical model's dynamics}\label{appendix1}

This section reviews relevant features of the $p=2$ spherical model in a pedagogical manner for readers who are not familiar with the field of spin glasses. The reader which would like to go further could consult the references \cite{Cugliandolo2,Cugliandolo3} and the book \cite{DeDominicisbook}. Mathematically, the model describes the behavior of the components $q_i(t)$ of a random vector of size $N$, whose dynamics obey equation:
\begin{equation}
\frac{d q_i}{dt}=-\sum_{j=1}^N [J_{ij}+\ell(t)\delta_{ij}]q_j(t)+\eta_i(t)\,,\label{sphericalmod}
\end{equation}
where $J_{ij}$ is a quenched random disorder whose entries are distributed accordingly with the Gaussian distribution of variance $\sigma^2/N$ and $\ell(t)$ is a Lagrange multiplier, ensuring that the constraint $\sum_{i=1}^N q_i^2=N$ holds $\forall \, t$. The noise field $\eta(t)$ is moreover assumed to be a Gaussian random field as well, accordingly to the equation \eqref{stateta}. Assuming $N$ large enough, this model provides an archetypal example of disordered dynamics and enjoys the property to be exactly solvable, including in the static limit, without requiring a replica method. Although, it does not exhibit a true glassy phase, but rather a ferromagnet behavior, it presents interesting characteristics, such as aging effects with a \textit{weak ergodicity breaking}, that is also found in systems with true  glassy dynamics. 
\medskip

\paragraph{Dynamical aspects in the eigenspace.} In the eigenspace, the model \eqref{sphericalmod} reads:
\begin{equation}
\frac{d q_\lambda}{dt}=-[\lambda+\ell(t)]q_\lambda(t)+\eta_\lambda(t)\,.
\end{equation}
This equation can be solved formally taking $t=0$ for the initial condition:
\begin{equation}
q_\lambda(t)=q_\lambda(0)\,e^{-(2\sigma + \lambda)t}\rho(t)+\int_0^t dt^\prime\, e^{-(2\sigma+\lambda)(t-t^\prime)}\,\eta_\lambda(t^\prime)\frac{\rho(t)}{\rho(t^\prime)}\,,\label{evolution}
\end{equation}
with:
\begin{equation}
\rho(t):=e^{2\sigma t - \int_0^t dt^\prime \ell(t^\prime)}\,.
\end{equation}
Furthermore, in the large $N$ limit, the spherical constraint $\sum_{i=1}^N q_i^2=N$ imposes :
\begin{equation}
\int_{-2\sigma}^{2\sigma} d\lambda \,\mu(\lambda) \langle q_\lambda^2(t) \rangle =1\,,
\end{equation}
which should be valid for each time $t\geq 0$ and whereas before the bracket notation means averaging over the noise. These equations translate as a self-consistent equation for $\Theta(t):=1/\rho^2(t)$:
\begin{equation}
1=\int_{-2\sigma}^{2\sigma} d\lambda \,\mu(\lambda) \left\{ q_\lambda^2(0) \,e^{-2(2\sigma + \lambda)t}\rho^2(t)+2D\int_0^t dt^\prime e^{-2(2\sigma+\lambda)(t-t^\prime)}\,\eta_\lambda(t^\prime)\frac{\rho^2(t)}{\rho^2(t^\prime)} \right\}\,,\label{self}
\end{equation}
depending on the choice of the initial conditions. One of the main statements of the analytic study in \cite{Cugliandolo3} is that the system is not able to reach an equilibrium regime in general, except for very special initial conditions. Indeed, if the system has a macroscopic occupation number for the smaller eigenvalue $\lambda=-2\sigma$, namely:
\begin{equation}
q_\lambda^2(0)=\frac{\delta(\lambda+2\sigma)}{\mu(2\sigma)}\,,
\end{equation}
hence, the self equation \eqref{self} shows that $q_\lambda(t\to \infty)\to 0$ excepts for boundary eigenvalue $\lambda=-2\sigma$, $q_{2\sigma}(t\to \infty) \to \sqrt{1-D/D_c}$ where the critical temperature $D_c$ is:
\begin{equation}
\frac{1}{D_c}:=\int_{-2\sigma}^{2\sigma} d\lambda\, \mu(\lambda)\frac{1}{2\sigma+\lambda}\,.\label{criticaltemp}
\end{equation}
On the contrary for the uniform initial condition $q_\lambda(0)=1$ $\forall\, \lambda$, the system fails to thermalize in a finite time. This condition is physically equivalent to a random initial condition for the variable $q_i(t)$, close to what we expect experimentally. Taking the Laplace transform of \eqref{self}, we obtain the self-consistent equation:
\begin{equation}
\tilde{\Theta}(z)=\int_{-2\sigma}^{2\sigma} d\lambda\, \mu(\lambda) \frac{1+2D \tilde{\Theta}(z)}{z+2(2\sigma+\lambda)}\,,
\end{equation}
where $\tilde{\Theta}(z)$ denotes the Laplace transform of ${\Theta}(t)$. The large-time behavior can be obtained by expanding the previous equation to the power of $z$. We thus obtain:
\begin{equation}
\gamma(t) \sim \left(1-\frac{D}{D_c} \right)\, t^{3/4}\,.
\end{equation}
Therefore, from \eqref{evolution}, it follows that $q_\lambda(t)$ vanishes exponentially as $t\to \infty$ with relaxation time $\tau^{-1}(\lambda)=2\sigma+\lambda$, excepts for the component $\lambda=-2\sigma$ which have an infinite relaxation time. The $2$-point correlation function $C(t,t^\prime)$, defined as:
\begin{equation}
C(t,t^\prime)=\int_{-2\sigma}^{2\sigma} d\lambda\, \mu(\lambda) \,\langle q_\lambda(t) q_{\lambda^\prime}(t^\prime) \rangle\,,
\end{equation}
and can be computed using \eqref{evolution} and $\rho(t)$ given by \eqref{self}. Setting $t^\prime=0$, and assuming random initial conditions, it is easy to check that for $t$ large enough:
\begin{equation}
C(t,0) \sim \left(1-\frac{D}{D_c}\right) t^{-3/4}\,,
\end{equation}
and thus that the correlation follows a power law. Most interesting is the behavior of the correlation as both $t$ and $t^\prime$ go to infinity. There are essentially two interesting limits:
\begin{enumerate}
\item $t,t^\prime \to \infty$ but $(t-t^\prime)/t \ll 1$. In that limit, the correlation function is almost translation invariant, i.e., it depends mainly on $\epsilon:=t-t^\prime$ and goes toward the self-overlapping $Q=1-D/D_c$ as $\epsilon\to\infty$ (after $t$ and $t^\prime$).
\item $t,t^\prime \to \infty$ but $(t-t^\prime)/t= \mathcal{O}(1)$. In this case, the translation invariance is lost, and the correlation function depends on $t$, $t^\prime$ and $\epsilon$. As $\epsilon/t$ be small enough, we have again $C\to 1-D/D_c$. But for $\epsilon \gg t$, $C\to 0$ and the correlation vanishes. 
\end{enumerate}
To summarize:
\begin{align}
\lim_{\epsilon\to \infty} \lim_{t\to \infty} C(t+\epsilon,t)&=Q\\
\lim_{t\to \infty} \lim_{\epsilon\to \infty} C(t+\epsilon,t)&=0\label{asymptoticbehavior}
\end{align}
Hence, if asymptotically the times remain arbitrarily close, the system seems to balance like a ferromagnetic system. However, the system fails to thermalize, and for time intervals much larger than the age of the system, the correlations tend to be zero.
\medskip

\paragraph{Static limit.} According to equation \eqref{eqmodel}, the static limit of the system is fully described by the partition function:
\begin{equation}
Z=\int \left(\prod_{i=1}^N dq_i \right) \, \exp \left(\frac{1}{2D} \left[ \sum_{i,j=1}^N -J_{ij} q_i q_j - \ell \sum_{i=1}^N (q_i^2-1) \right] \right)\,.
\end{equation}
The integral over $q$ is Gaussian, performing it, we get in the eigenspace:
\begin{equation}
Z= \exp \left(-\frac{1}{2} \sum_\lambda \ln \left(\frac{\ell+\lambda}{D}\right) +\frac{\ell N}{2D}\right)\,.
\end{equation}
The value of $\ell$ is fine-tuned to impose the spherical condition, which reads in the large $N$ limit:
\begin{equation}
\frac{1}{D}=\int_{-2\sigma}^{2\sigma}\, \frac{d\lambda}{2\pi \sigma^2}\sqrt{4\sigma^2-\lambda^2}\, \frac{1}{\ell+\lambda}\,.\label{conditionl}
\end{equation}
It is easy to check that $\ell$ decreases with $D$ to the critical value $\ell_c=2\sigma$, defining, in turn, the critical temperature $D_c$ by the same formula \eqref{criticaltemp} as for the dynamics. Hence, as for standard Bose-Einstein condensation, we show that the component $q_{-2\sigma}$, fails to be included in the integral approximation of the discrete sums. Indeed:
\begin{equation}
\sum_\lambda \langle q_\lambda^2 \rangle \to \int_{-2\sigma}^{2\sigma} d\lambda \,\mu(\lambda) \langle q_\lambda^2 \rangle + \frac{1}{N} \left( \langle q_{2\sigma}^2\rangle + \langle q_{-2\sigma}^2\rangle \right)\,.
\end{equation}
As all the components $\langle q_\lambda^2 \rangle $ remain of order $1$, the approximation works. But the failure of the approximation near $D=D_c$ means that it becomes wrong. The component $ \langle q_{-2\sigma}^2\rangle$ becomes of order $N$, more precisely from \eqref{conditionl}:
\begin{equation}
\langle q_{-2\sigma}^2\rangle = N \left(1-\frac{D}{D_c} \right)\,.
\end{equation}
As for the dynamics, we recover in the static limit the condensation property, meaning that the system behaves as a disguised ferromagnet at equilibrium, and the disorder is not strong enough to induce a glassy phase. 

\section{Large time behavior of the quartic model}\label{appendix2}
In this section, according to the method detailed in \cite{bray2002theory}, we study the large-time behavior of the Langevin dynamic, which solution is considered in the relation \eqref{evolution}. To be more precise, we consider the following definition:
\bea
g(t)=\int_0^t\, dt' \ell(t'), \quad a(t)=\sum_\lambda \frac{q^2_\lambda}{N}\rightarrow \int  \mu(\lambda) q_\lambda^2\,d\lambda,
\eea
where $q_\lambda(t)$ is  the time-dependent solution of the Langevin equation taken in the eigen-direction. The quentity $a(t)$ is assumed to be self-averaged for large $N$. Hence, we can easily show the solution for dynamics after the quench as:
\bea\label{xxmodel}
\langle q_\lambda^2\rangle=e^{-2(\lambda t+g(t))}+2D\int_0^t dt'\, e^{-2\lambda(t-t')-2(g(t)-g(t'))}.
\eea
The function $\ell(t)$ may be expanded as a power expansion of the mean value $a(t)$:
\beq
\ell(t)=2\sigma+\sum_{n\in\mathbb{N}}h_n a^n(t).
\eeq
In this note, we consider the quartic potential, such that $\ell(t)=h_0+2\sigma+h_1a(t)$. Assuming that dynamics is frozen on the minimum of the potential for $t$ large enough, namely $\ell(t)\to 2\sigma$, one obtain the asymptotic closed equation:
\bea
-\frac{h_1}{h_0}\int_{-2\sigma}^{2\sigma}\mu(\lambda) e^{-2\lambda t-2g(t)} d\lambda=1+2D\frac{h_1}{h_0}\int_{-2\sigma}^{2\sigma}\mu(\lambda) \int_0^t\, dt'e^{-2\lambda (t-t')-2(g(t)-g(t'))},
\eea
and we will check that our assumption for the asymptotic behavior of $\ell(t)$ for large time is self consistent. This expression can be rewritten as:
\bea\label{nounouse}
G(t)=-\frac{h_1}{h_0}\Big(H(t)+2D\int_0^t dt'\, H(t-t')G(t')\Big),
\eea
where we introduced the following definitions
\bea\label{momi}
G(t)=e^{2g(t)-m t},\quad H(t)=\int_{-2\sigma}^{2\sigma}\mu(\lambda) e^{-2\lambda t-m t}d\lambda,\quad m\in\mathbb{R} \label{defGH}
\eea
where $H(t)\sim 1/t^{3/2}$ for large $t$, and remains finite at the origin. Note that the relation \eqref{nounouse} is invariant under the choice of the parameter $m\in\mathbb{R}.$ To solve this closed equation we move to the Laplace transformation by integrated as $\bar G(p):=\int_0^\infty\, dt\, e^{-pt}G(t)$. Using the fact that $H(t-t')$ may be replaced by $H(t-t')\theta(t-t')$ in the integration domain,  we get  the simple relation 
\bea\label{vinco}
\bar G(p)=-\frac{h_1}{h_0}\Big(1+2D\bar G(p)\Big)\bar H(p)\Leftrightarrow \bar G(p)=-\frac{1}{2D+\frac{h_0}{h_1}\bar H^{-1}(p)}.
\eea
At this stage, we have to make a few remark. In contrast with the spherical model reviewed in the previous section, the closed equation does not holds for all time, but only asymptotically. Hence, one may be disappointed by the way we integrated over $t$ the closed relation from $0$ to $\infty$ by taking the Laplace transform. Indeed, we assume that the solution of the closed equation provides the true asymptotic behavior for $G(t)$, an assumption that can be motivated from the observation that, for $t$ large enough, $H(t-t^\prime)$ suppresses low time contributions provided that $G(t)$ has a finite limit for short times.
\medskip

Before determining the asymptotic expression of the function $G$, let us assume that $m=4\sigma$.  Integrating the function $\bar{H}(p$) we come to
\bea
\bar H(p)=\int_{-2\sigma}^{2\sigma}\mu(\lambda)\frac{1}{2(\lambda+2\sigma)+p}d\lambda.
\eea
Considering the Wigner distribution $\mu(x)=\frac{\sqrt{4\sigma^2-x^2}}{4\pi \sigma}$  and by replacing $\bar H(p)$ by $\bar H(0)+\bar H(p)-\bar H(0)$ we get
\bea
\bar H(p)&=&\bar H(0)+\int_{-2\sigma}^{2\sigma}\mu(\lambda)\frac{1}{2(\lambda+2\sigma)+p}d\lambda-\int_{-2\sigma}^{2\sigma}\mu(\lambda)\frac{1}{2(\lambda+2\sigma)}d\lambda\cr
&=& \bar H(0)-\frac{p}{2}\int_{-2\sigma}^{2\sigma}\mu(\lambda)\frac{1}{[2(\lambda+2\sigma)+p](\lambda+2\sigma)}d\lambda\,,
\eea
where $\bar{H}(0)\equiv 1/2\sigma$. Also, the critical temperature $D\equiv T_c$ is determined such that the denominator of the relation \eqref{vinco} vanish, and
\begin{equation}
T_c:=-\frac{h_0}{h_1}\frac{1}{2 \bar H(0)}.\label{criticaltempWing}
\end{equation}
Then by setting $u=\lambda+2\sigma$ we come to:
\bea
\bar H(p)=-\frac{h_0}{h_1}\frac{1}{2T_c}-\frac{p}{8\pi\sigma}\int_0^{4\sigma} \frac{\sqrt{4\sigma-u}}{\sqrt{u}(p+2u)}du.
\eea
In the above relation, the integral may be computed asymptotically as $p\to 0$ from:
\beq
\int_0^{\infty} \frac{du}{\sqrt{u}(p+2u)}\sim \frac{\pi}{\sqrt{2}}\frac{1}{p^{1/2}},
\eeq
and we get finally as $p\to \infty$:
\bea
\bar H(p)\approx -\frac{h_0}{h_1}\frac{1}{2T_c}-\frac{\sqrt{\sigma}}{4\sqrt{2}}\,p^{1/2}
\eea
Note that the large time limit corresponds to the small values of $p$ and then 
the function $G(p)$ can be deduced in the limit $p\rightarrow 0$ 
\bea
\bar G(p)=\frac{1}{2}\frac{1+\tilde{A}(\sigma) T_c p^{1/2}}{T_c-T-\tilde{A}(\sigma)TT_c p^{1/2}}\approx \frac{1}{2(T_c-T)}\Big[1+\frac{\tilde{A}(\sigma)T_c^2 p^{1/2}}{T_c-T}+\cdots]
\eea
where $\tilde{A}(\sigma)=\frac{h_0}{h_1}\frac{\sqrt{\sigma}}{4\sqrt{2}}$. The inverse transformation can be simply given by
\bea
G(t)=\frac{\delta(t)}{2(T_c-T)}+\frac{1}{2(T_c-T)^2}\frac{\tilde{A}(\sigma)T_c^2}{t^{3/2}}
\eea
which asymptotically at $t$ writes as 
\bea
G(t)\propto \frac{1}{t^{3/2}}.
\eea
In an attempt to implement the dependence of the distribution on the large-time behavior, let us consider the MP-distribution given in \eqref{mp}.
Then, if we set $m=-\lambda_-$, the expression of $\bar H(p)$ becomes
\bea
\bar H(p)=-\frac{h_0}{h_1}\frac{1}{2T_c}-\frac{p}{2}\int_{\lambda_{-}}^{\lambda_{+}}\,\frac{\sqrt{(\lambda-\lambda_{-})(\lambda_{+}-\lambda)}}{2\pi\sigma c\lambda}\frac{d\lambda}{[2(\lambda-\lambda_{-})+p](\lambda-\lambda_{-})}
\eea
Now by changing variable as $u=\lambda-\lambda_{-}$ we get
\bea
\bar{H}(p)=-\frac{h_0}{h_1}\frac{1}{2T_c}-\frac{p}{4\pi\sigma c}\int_{0}^{\lambda_{+}-\lambda_{-}}\,du\frac{\sqrt{\lambda_{+}-\lambda_{-}-u}}{(u+\lambda_-)(2u+p)\sqrt{u}}\,.
\eea
The integral diverges as $p\to 0$, but also as $\lambda_-\to 0$ and we will consider this case separately. 
\begin{itemize}
    \item For $\lambda_{-}\neq 0$ and $p\rightarrow 0$ the integral  behave as:
\bea\label{integronew}
\int_{0}^{\lambda_{+}-\lambda_{-}}\,du\frac{\sqrt{\lambda_{+}-\lambda_{-}-u}}{(u+\lambda_-)(2u+p)\sqrt{u}} \approx \frac{4\sqrt{2}}{\pi \sqrt{p}(2\lambda_--p)}
\eea
Therefore
\bea
\bar{H}(p)\approx -\frac{h_0}{h_1}\frac{1}{2T_c}-\frac{\sqrt{2(\lambda_+-\lambda_{-})}}{\pi^2\sigma c}\frac{p^{\frac{1}{2}}}{(2\lambda_--p)}\,,
\eea
and then the function $\bar G(p)$ becomes
\bea
\bar G(p)&\approx& \frac{1}{2(T_c-T)}\Big[1+\frac{\gamma(h_0,h_1)T_c^2}{(T_c-T)}\frac{p^{\frac{1}{2}}}{(2\lambda_{-}-p)}+\cdots\Big]\cr
&\approx& \frac{1}{2(T_c-T)}\Big[1+\frac{\gamma(h_0,h_1)T_c^2}{(T_c-T)}\frac{p^{\frac{1}{2}}}{2\lambda_{-}}+\cdots\Big],\\ &&\mbox{  with }\gamma(h_0,h_1)=\frac{2\sqrt{2h_1(\lambda_{+}-\lambda_{-})}}{h_0\pi^2c\sigma}.\nonumber
\eea
Finally the function $G(t)$ becomes for large $t$:
\bea
G(t)=\frac{\gamma(h_0,h_1)T_c^2}{4\lambda_{-}(T_c-T)^2}\frac{1}{t^{\frac{3}{2}}}+\cdots. 
\eea
Because of the definition of $G(t)$, we have:
\begin{equation}
\int_0^t \ell(t^\prime) dt^\prime - 2\sigma t = \ln\left(\frac{\gamma(h_0,h_1)T_c^2}{4\lambda_{-}(T_c-T)^2}\right)-\frac{3}{2}\ln t\,,
\end{equation}
hence: $\ell(t)$ behaves as $2\sigma+\mathcal{O}(t^{-1})$ for $t$ large enough, meaning that $a(t)$ converges toward the stable vacuum $-h_0/h_1$. 

\item The case $\lambda_-=0$ (namely $c=1$) is special. Indeed, in the limit $\lambda_-\to 0$, the integral over $\lambda$ diverges, and in particular $H(0)=\infty$. This seems to suggest that $T_c=0$, accordingly with the static limit, equation \eqref{conditionl}. Indeed, assuming $h_0<0$, we can set $m=-2h_0$, and:
\bea
\bar{H}(p)&=&\bar{H}(0)-\frac{p}{2}\int_0^{\lambda_+} \frac{\sqrt{\lambda(\lambda_{+}-\lambda)}}{2\pi\sigma \lambda}\frac{d\lambda}{[2(\lambda-h_0)+p](\lambda-h_0)}\cr&=&\bar H(0)-\frac{\sqrt{\frac{h_0-4}{h_0}}-\sqrt{\frac{8}{p-2h_0}+1}}{4}\,,\label{equationHlambda0}
\eea
where:
\begin{equation}
\bar{H}(0)\equiv \int_0^{\lambda_+} \frac{\sqrt{\lambda_{+}-\lambda}}{2\pi\sigma \sqrt{\lambda}}\frac{d\lambda}{2(\lambda-h_0)}=\frac{1}{4} \left(\sqrt{\frac{h_0-4}{h_0}}-1\right)\,,
\end{equation}

which again defines the critical temperature $T_c:=-h_0/(2 h_1 \bar{H}^{-1}(0))$. For large $t$, using standard results about the asymptotic behavior of Bessel functions, it is easy to check that $H(t)$ behaves as:
\begin{equation}
{H}(t)\sim \frac{e^{2h_0 t}}{\sqrt{2\pi t}}\,.
\end{equation}
In the low-temperature limit, $G(t)\sim -(h_1/h_0) H(t)$, and we conclude again that $a(t)$ goes toward the vacuum $-h_0/h_0$ as $1/t$, with infinite relaxation time. Note that transition temperature seems to have a different interpretation, as in the case $c<1$. Indeed, it looks rather as the boundary of the validity domain of the solution, which becomes singular at $h_0=0$ where the transition temperature vanishes formally. For more detail about the computation of the critical temperature for $p=2$ solf spin dynamics, see \cite{Lahoche:2022lmf}.
\end{itemize}

\section{Correlations function for response field}\label{appendix3}

In this section, we sketch the proof that correlation functions for the response field have to vanish, and in particular $G_{k,\bar{\varphi}\bar{\varphi}}=0$. The proof works as follows. Adding a linear driving force to the potential $V^\prime(Q^2)$ as $U^\prime(\phi^2) \phi_\lambda \to U^\prime(\phi^2) \phi_\lambda+ \int dt \sum_\lambda k_\lambda(t) \phi_\lambda(t)$, we modify the path integral \eqref{expressionZ} accordingly. However, the transformation is equivalent to a translation of the $\lambda$-th component of $\tilde{j}$ accordingly: $\tilde{j}_\lambda\to \tilde{j}_\lambda-ik_\lambda$. Therefore:
\begin{equation}
Z[j,\tilde{j}]\big\vert_{U^\prime} = Z[j,\tilde{j}-ik]\big\vert_{U}\,.
\end{equation}
For vanishing sources, $Z$ must be equal to $1$ by construction, and in fact that the generating functional $Z[j]$, setting $\tilde{j}=0$ in the previous expression, reads formally:
\begin{equation}
Z[j,\tilde{j}=0]= \left\langle \exp \left(\int dt \sum_{\lambda=1}^N j_\lambda(t) \phi_\lambda(t) \right) \right\rangle_\eta \,,
\end{equation}
the bracket meaning averaging over noise field. Therefore: $Z[0,-ik]\vert_{U}=1$, and 
\begin{equation}
\langle \prod_{p=1}^P \bar{\varphi}_{\lambda_p} \rangle \propto \frac{\delta^PZ[0,-ik]\vert_{U}}{\delta k_{\lambda_1}\cdots \delta k_{\lambda_P}}  =  \frac{\delta^P\, 1}{\delta k_{\lambda_1}\cdots \delta k_{\lambda_P}} =0\,.
\end{equation}
Hence we have $G_{k,\bar{\varphi}\bar{\varphi}}=0$ but also $\langle \bar{\varphi}_\lambda \rangle \equiv \bar{\varpi}_\lambda=0$. 

\section{Effective equation for $\Gamma_k^{(6)}$}\label{EVEAdendeum}

\begin{figure}
\begin{center}
\includegraphics[scale=0.8]{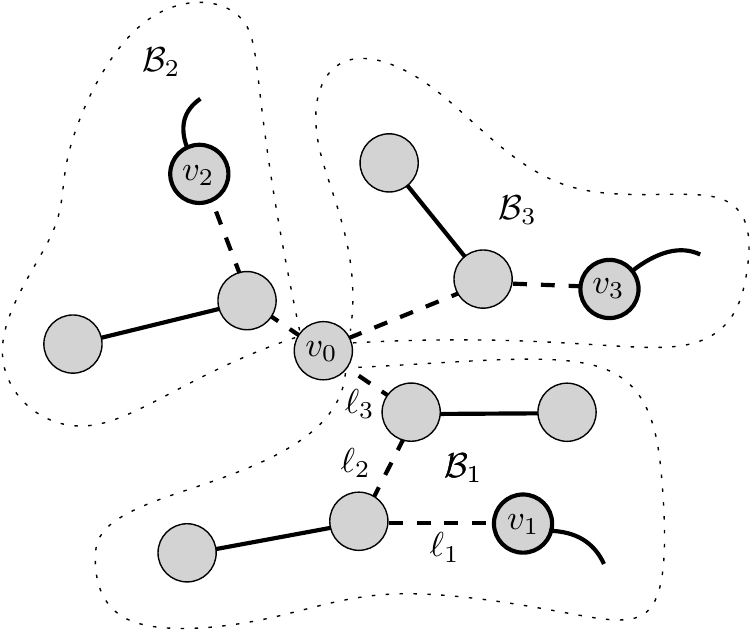}\qquad \includegraphics[scale=1]{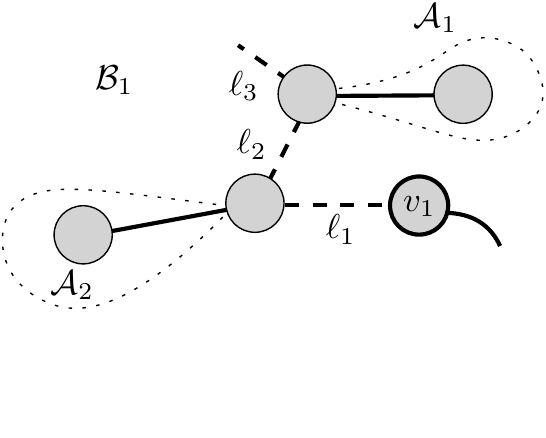}
\end{center}
\caption{On left: A typical graph contributing to $\Gamma_k^{(6)}$. On right: Structure of the arm $\mathcal{P}_1$.}\label{contributionphi6}
\end{figure}

Accordingly, with our derivation of effective $4$-point function in section \ref{largeN}, let us consider the $1$PI $6$-point function $\Gamma_k^{(6)}$ for the equilibrium theory \eqref{eqmodel}. From section \ref{largeN}, it follows that Feynman leading order diagrams contributing to its perturbative expansion have to be deduced from vacuum trees by deleting three loop vertices on leaves. Figure \ref{contributionphi6} provides an example, where cancelled leaves are pictured as ciliated loop vertices in the LVR, and denoted as $v_1$, $v_2$ and $v_3$. Let $\mathcal{P}$ the smallest path connecting the three boundaries $v_1$, $v_2$, $v_3$. The path $\mathcal{P}$ decomposes as $\mathcal{P}=\mathcal{P}_1\cup \mathcal{P}_2\cup\mathcal{P}_3$, where $\mathcal{P}_i$ is connected with the vertex $v_i$ and such that $\mathcal{P}_1 \cap \mathcal{P}_2 \cap \mathcal{P}_3= \{v_0 \}$. For instance $\mathcal{P}_1=\{\ell_1,\ell_2,\ell_3 \}$, all the edges building $\mathcal{P}$ being drawn as dotted edges. Along the path $\mathcal{P}_i$, there are some branches connected to the effective vertices, as pictured on the right of Figure \ref{contributionphi6} for the arm $\mathcal{P}_1$. Along the path, there are two connected components, say $\mathcal{A}_1$ and $\mathcal{A}_2$. It is not hard to check that these contributions are involved in the perturbation series for the full $2$-point function. Hence, the formal sum of these diagrams keeping the lengths $\vert \mathcal{P}_i \vert$ fixed leads to an effective tree with $L= \sum_i \vert \mathcal{P}_i \vert=7$ edges and $V=L+1=8$ vertices. Following section \ref{largeN}, it is moreover not hard to check that the branches of this effective tree are nothing but a formally resumed version of $4$-point function $1PI$ functions $\Gamma_k^{(4)}$, the Figure \ref{contributionphi6} (on right) providing a typical graph involved in the perturbative expansion of it.

\section{LPA=LPA$^\prime$ in the large N limit}\label{appendix4}
\begin{figure}
\begin{center}
\includegraphics[scale=0.8]{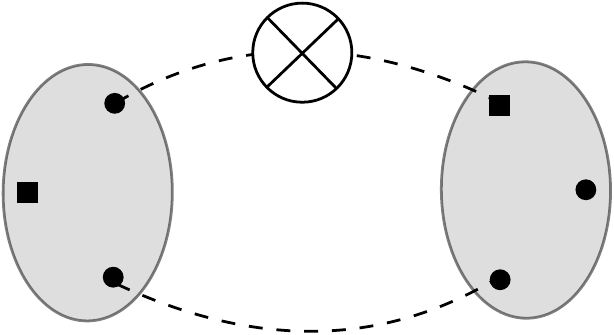}\qquad \includegraphics[scale=0.8]{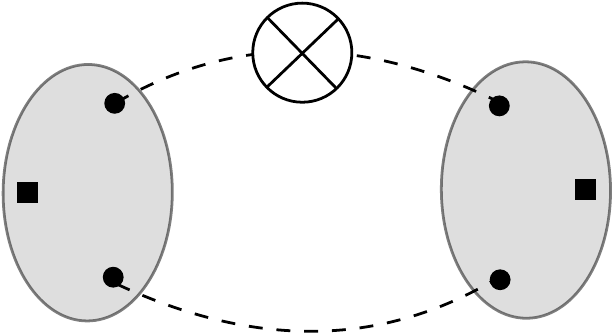}
\end{center}
\caption{Effective diagrams contributing to the computation of anomalous dimensions $\eta_Z$ and $\eta_Y$. The gray discs materialize effective $3$-point vertices.}\label{figureeta}
\end{figure}

In this section, we sketch the proof that the anomalous dimensions $\eta_Z$ and $\eta_Y$:
\begin{equation}
\eta_Z:= k \frac{d}{dk} \ln (Z_k)\,,\qquad \eta_Y:= k \frac{d}{dk} \ln (Y_k)\,,
\end{equation}
are both of order $1/N$, and can be discarded in the limit $N\to \infty$. Because of the $O(N)$ invariance, the contribution involving $\Gamma_k^{(4)}$ in the left-hand side of the flow equation for $\Gamma_k^{(2)}$ does not contribute for $\eta_Z$ and $\eta_Y$ in the large $N$ limit, because the relevant loop creating one face (see section \ref{vertexexp}) does not depend on the externals $\lambda$ and $\omega$ (see the discussion below equation \eqref{flow2points}). The remaining contribution to the left-hand side involves $\Gamma^{(3)}_k$ vertices, which are non-vanishing in the broken phase. Accordingly, with the graphical notation of the section \ref{LPAformalism}, this contribution takes the form pictured in Figure \ref{figureeta}. It is easy to check that $\Gamma_k^{(3)} \sim 1/N$. Indeed, setting $\chi=\kappa$, we have formally:
\begin{align}
\vcenter{\hbox{\includegraphics[scale=0.7]{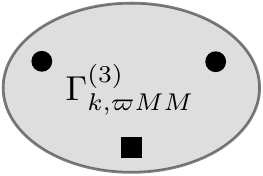}}}&= \frac{2\mu_2\sqrt{\kappa}}{N}\Bigg(\vcenter{\hbox{\includegraphics[scale=0.7]{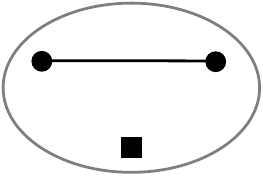}}}+\vcenter{\hbox{\includegraphics[scale=0.7]{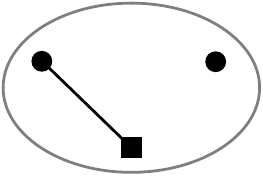}}}+\vcenter{\hbox{\includegraphics[scale=0.7]{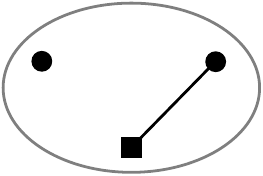}}}\Bigg)+\frac{4\mu_3(\kappa)^{3/2}}{N^2}\, \vcenter{\hbox{\includegraphics[scale=0.7]{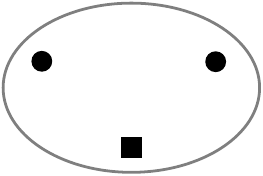}}}\label{exp3pts}
\end{align}
Let us focus on the connected components of the effective propagator. Hence, the leading order contributions to the graph of Figure \ref{figureeta} create one face and looks for instance as:
\begin{equation}
\vcenter{\hbox{\includegraphics[scale=0.7]{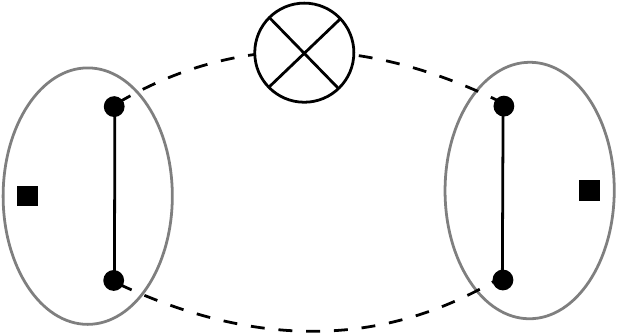}}}\,,
\end{equation}
which contributes to the flow of $Y_k$, but scales as $1/N$. The contributions of disconnected pieces for the propagator do not change this analysis. Indeed, the disconnected pieces arise with a factor $1/N$, and sums created per disconnected piece do not compensate the additional $1/N$ factors, and contributions to $Y_k^\prime$ and $Z_k^\prime$ have to be of order $1/N$ at best. Therefore, $\text{LPA}=\text{LPA}^\prime$ in the large $N$ limit.  

\section{Details about the derivation of \eqref{equationGamma2}}\label{appendix5}
From \eqref{defW}, we must have:
\begin{align}
\nonumber-\Gamma_{2,k}[\Xi,\textbf{G}]&+\int dt \sum_a J_a(t)\, \Xi_a(t)
 + \frac{1}{2} \Tr [\textbf{G}\textbf{K}]\\\nonumber
&+\frac{1}{2} \int dt dt^\prime \sum_{a,b} \Xi_a(t) [k_{ab}(t,t^\prime)+R_{k,ab}(t-t^\prime)](t,t^\prime)\Xi_b(t^\prime)\\
&=\ln \int d\Phi\, e^{- S_k[\Phi]+ \int dt \sum_{a} \Phi_{a}(t) J_{a}(t)+\frac{1}{2} \int dt dt^\prime \sum_{a,b} \Phi_{a}(t) k_{ab}(t,t^\prime)  \Phi_{b}(t^\prime)}\,.
\end{align}
On the left-hand side, we decompose $\Phi_{a}(t)=\Xi_a(t)+\xi_a(t)$, 
\begin{align}
\nonumber S_k[\Xi+\xi]&=S_k[\Xi]+\int dt \frac{\delta S_k}{\delta \Phi_{a}(t)} \xi_a(t)\\
&+\frac{1}{2} \int dt dt^\prime \frac{\delta^2 S_k}{\delta \Phi_{a}(t)\delta \Phi_{b}(t^\prime)} \xi_a(t)\xi_b(t^\prime)+\mathcal{W}[\Xi+\xi]
\end{align}
where we assumed to sum over repeated indices. The remaining contribution, $\mathcal{W}[\Xi+\xi]$ including non-Gaussian contributions. After some algebraic manipulations, we thus obtain:
\begin{align}
\nonumber&\exp \left(\Gamma_{2,k}[\Xi,\textbf{G}]+\Tr \left[\textbf{G}\frac{\delta \Gamma_{2,k}}{\delta \textbf{G}}\right]\right)\\
&=\int d\xi\, e^{- S_k[\Xi]+ \int dt \sum_{a} \left(\frac{\delta \Gamma_{2,k}}{\delta \Xi_a(t)}-\frac{\delta S_k}{\delta \Phi_{a}(t)}\right)\xi_a(t)-\frac{1}{2}\int dt dt^\prime \sum_{a,b} \left(\frac{\delta^2 S_k}{\delta \Phi_{a}(t)\delta \Phi_{b}(t^\prime)}-2\frac{\delta \Gamma_{2,k}}{\delta G_{ab}(t,t^\prime} \right) \xi_a(t) \xi_b(t)-\mathcal{W}[\Xi+\xi]}\,.
\end{align} 
Now we can expand the left-hand side in a number of loops to recover the expansion \eqref{equationGamma2}.

\printbibliography[heading=bibintoc]
\end{document}